%% file: main.tex
\documentclass[a4paper,11pt]{article}
\pdfoutput=1 %
\makeatletter
\let\NNWZWarticlecaption\@makecaption
\makeatother
\usepackage{jheppub-compat}
\makeatletter
\let\@makecaption\NNWZWarticlecaption
\makeatother %

\usepackage{amsmath}
\usepackage{amssymb}
\usepackage{mathrsfs}
\usepackage{graphicx}
\usepackage[font=small,labelfont=bf,labelsep=period]{caption}
\usepackage{subcaption}

\usepackage{hyperref}
\usepackage{bm}
\usepackage{dsfont}
\usepackage{xcolor}
\usepackage{multirow}
\usepackage{booktabs}
\usepackage{comment}
\usepackage{enumitem}
\usepackage{algorithm} 
\usepackage{algpseudocode} 
\newtheorem{theorem}{Theorem}[section]
\newtheorem{proposition}[theorem]{Proposition}

\newcommand{\bC}{\mathbb{C}}
\newcommand{\bE}{\mathbb{E}}

\newcommand{\bZ}{\mathbb{Z}}

\newcommand{\cD}{\mathcal{D}}

\newcommand{\cH}{\mathcal{H}}

\newcommand{\cK}{\mathcal{K}}

\newcommand{\cW}{\mathcal{W}}

\newcommand{\mf}[1]{\mathfrak{#1}}

\newcommand{\OPE}{\mathrel{\leadsto}}

\definecolor{cardinal}{rgb}{0.6,0,0}
\definecolor{darkgreen}{rgb}{0,0.5,0}

\hypersetup{
 pdftitle={Learning from Duality in Neural Network Field Theories},
 pdfauthor={Pietro Capuozzo, Brandon Robinson, Benjamin Suzzoni},
 pdfsubject={Neural field theories, level--rank duality and charged preparations},
 pdfkeywords={Conformal and W Symmetry, Field Theories in Lower Dimensions, Renormalization and Regularization}
}
\begin{document}

\title{Learning from Duality in Neural Network Field Theories
}
\author[a]{Pietro Capuozzo,}
\emailAdd{pietro.capuozzo17@alumni.imperial.ac.uk}
\author[b]{Brandon Robinson,}
\emailAdd{b.j.robinson@uva.nl}
\author[c]{and Benjamin Suzzoni}
\emailAdd{b.suzzoni@benterre.com}

\affiliation[a]{STAG Research Centre, University of Southampton, Southampton, SO17 1BJ, UK}
\affiliation[b]{Institute of Physics, University of Amsterdam, 904 Science Park, 1098 XH Amsterdam, Netherlands}
\affiliation[c]{Department of Mathematical Sciences, Ulsan National Institute of Science and Technology, 50 UNIST-gil, Ulsan 44919, South Korea}

\abstract{We develop neural realizations of positive-level Wess--Zumino--Witten models using bosonic Wakimoto and fermionic Fourier--Berezin fields. Level--rank duality connects the charged sectors of the fermionic construction, where explicit neural observables recover conformal-block data and enable the transfer of information about unknown correlators between paired theories. We also show that distinct neural presentations of the same continuum current algebra exhibit different finite-width behavior, with approximation accuracy depending on architecture, regulator, and parameter distribution. These results connect the exact mathematical structure of conformal field theory with the expressive and computational properties of neural field theories.
}

\maketitle

\section{Introduction}\label{sec:introduction}

Duality is a central organizing principle in theoretical and mathematical physics.  A suitable duality frame can simplify a calculation by reorganizing the degrees of freedom, symmetries or couplings used to describe a theory \cite{Seiberg:1994rs,Giveon:1994fu}.  At the interface of physics and machine learning such a non-trivial relation in the space of field theories poses a challenge: if two neural architectures realize (quantum) field-theoretic structures related by an exact duality, what does that relation tell us about their finite approximations, computational costs and learning behavior?  %

Neural Network Field Theory (NNFT) provides a setting for investigating these questions. It associates field-theoretic correlation functions with expectations over ensembles of neural parameters.  Appropriate choices of architecture and parameter law can reproduce free-field and conformal-field-theory kernels in a regulated infinite-width limit \cite{Halverson:2020trp,Halverson:2021aot,Halverson:2024axc,Robinson:2025ybg}.  However, a limiting covariance is not a complete description of the finite ensemble: higher connected moments, shared-parameter correlations, regulator dependence and the cost of evaluating composite readouts can all depend on how the same kernel is realized \cite{Yaida:2020,Roberts:2021principles,Demirtas:2023fir,Ferko:2026luj,Harvey:2026iqd}.  This provides a particularly sharp setting for studying duality.  An exact relation between selected continuum observables supplies a fixed physical reference, while the underlying neural implementations can still be varied and compared.  Throughout this work, neural feature width, regulator removal, finite operator-mode rank and statistical estimation are treated as distinct operations.

Symmetry-based machine learning often builds group actions into neural features, for example through equivariant architectures, Lie-algebra-valued layers or learned symmetry generators \cite{Cohen:2016,Kondor:2018,Finzi:2020,Bogatskiy:2020,Dehmamy:2021,Lin:2024LieNeurons}.  Our question concerns the operator algebra generated by composite neural observables. Its affine level, normal ordering and descendant states require information beyond the transformation of features under a finite-dimensional group. We calculate these structures from specified neural parameter laws and then examine the finite-ensemble corrections.

We study level--rank duality of Wess--Zumino--Witten (WZW) current algebras as a controlled example. Our central question is which features of the neural realizations are constrained by exact algebraic data and which depend on the finite presentations. WZW models are interacting two-dimensional conformal field theories that have applications in string theory, representation theory and three-dimensional topological field theory \cite{Wess:1971yu,Witten:1983tw,Moore:1988qv,Gepner:1986wi}.  Their holomorphic currents generate affine Kac--Moody algebras $\widehat{\mathfrak g}_k$, and the Sugawara construction produces the corresponding stress tensor \cite{Sugawara:1968,Kac:1990gs}.  At positive integral level, the integrable highest-weight sectors and fusion rules are controlled by the affine symmetry; chiral correlation functions decompose into conformal blocks constrained by current Ward identities and Knizhnik--Zamolodchikov (KZ) equations \cite{Kac:1990gs,Knizhnik:1984nr}.  In this setting, current operator product expansions (OPEs), central charges, representation branching and block transport provide different, independently checkable levels of agreement.  %

Level--rank duality gives a precise relation between two such current algebras.  For $\widehat{\mathfrak{su}}(N)_K$ and $\widehat{\mathfrak{su}}(K)_N$, integrable sectors are paired through Young-diagram transposition together with identifications of appropriate charges, currents, fusion data and spaces of conformal blocks \cite{Naculich:1990pa,Naculich:1990hg,Nakanishi:1990hj,Ostrik:2014LevelRankTensorCategories}.  The free-fermion conformal embedding makes the shared structure explicit:
\begin{equation}
\widehat{\mathfrak{su}}(N)_K \oplus \widehat{\mathfrak{su}}(K)_N \oplus \widehat{\mathfrak{u}}(1)_{NK} \subset \widehat{\mathfrak{u}}(NK)_1.
\label{eq:intro_level_rank_embedding}
\end{equation}
The two non-Abelian current algebras act on the row and column indices of an $N\times K$ array of complex fermions, while the additional number-current sector accounts for their common charged parent \cite{Nakanishi:1990hj,Naculich:1990pa}. This is a correspondence between specified chiral structures, not an equivalence of the complete $\mf{su}(N)_K$ and $\mf{su}(K)_N$ WZW theories: their central charges generally differ, and paired conformal blocks need not be equal as functions.  Nor does the embedding identify the parameters or finite approximation schemes of the two neural current readouts.  

To study level--rank duality in NNFT, we construct affine currents from composite observables of regulated neural free fields. The neural architectures and their parameter laws determine the elementary correlations, while normal ordering and the associated operator constructions determine the current algebras. We evaluate composite correlators at finite width and fixed regulator, and then take the Gaussian limit before removing the regulator. This procedure allows us to study finite-width effects while retaining a well-defined continuum theory.

We develop two complementary constructions of positive-level affine currents. The bosonic construction combines neural scalar fields and commuting first-order $\beta\gamma$ fields through the nonlinear and derivative terms of the Wakimoto construction \cite{Wakimoto:1986gf,deBoer:1996ti}. The fermionic construction uses Fourier--Berezin fields whose regulated contractions approach the Cauchy kernel. Bilinears of these fields generate commuting $\widehat{\mathfrak{su}}(N)_K$ and $\widehat{\mathfrak{su}}(K)_N$ current algebras, with positive levels determined by the flavor multiplicities. Both constructions recover the prescribed affine structures in the continuum limit and allow finite-width corrections to be calculated from the underlying neural ensembles.

The fermionic construction also provides access to charged sectors. Angular projections of the Fourier features, together with a Neveu--Schwarz polarization, lead to the canonical anticommutation relations (CAR) and their Fock-space realization. Exterior powers of the $N\times K$ channel space determine the charged lowest-grade sectors, while negative current modes generate their descendants. For $NK=6$, Grassmann antisymmetry gives the two charge-two channels $(\mathbf3,\overline{\mathbf3})$ and $(\mathbf1,\mathbf6)$, relating the intermediate sectors of $SU(2)_3$ and $SU(3)_2$. Using the corresponding Ward identities and sewing data, we reconstruct their descendant expansions and find agreement with independent WZW calculations through the tested grades.

We next examine how different neural constructions approach the same continuum theory. A spectral-tilt parameter changes the distribution of sampled frequencies while preserving the limiting elementary correlations. The finite-width correlators admit inverse-width expansions whose coefficients depend on the spectral distribution, regulator and observable. We compare current, stress-tensor and higher-point correlations in direct $SU(N)_K$ constructions and their $SU(NK)_1/SU(K)_N$ coset counterparts. The preferred spectral distribution varies with the observable and neural width, while the choice of regulator affects the accuracy relative to the continuum CFT. We also find that changing how parameters are shared across fields can alter the relative accuracy of the direct and coset constructions.

These choices also affect the computational resources. The bosonic $SU(6)_1$ parent contains more elementary fields and requires more complicated composite observables than the direct $SU(2)_3$ construction. Their evaluation costs differ according to the correlators being computed and the intermediate calculations that can be reused. We further examine the sensitivity of the correlators to changes in the elementary field amplitudes, and test whether these adjustments improve their agreement with the continuum theory. Finally, we consider sampling uncertainty and the cost of estimating the neural ensemble averages. Together, these calculations describe how the choice of neural construction affects accuracy and computational efficiency. 

We also investigate finite-dimensional approximations of the fermionic state spaces. Retaining selected one-particle modes preserves their canonical anticommutation relations, while modifying the composite current algebra and its sewing amplitudes. We compare direct and neural optimization of polarized subspaces against conventional mode truncation. Optimizing the retained modes improves the approximation in two of the five settings studied, with similar performance from the direct and neural methods. The results illustrate how the choice of finite-dimensional subspace affects the approximation of WZW amplitudes.

Finally, we investigate how level--rank duality can be used to transfer information between paired theories. We begin with fusion coefficients, using Kac--Peterson modular data, the Verlinde formula and Young-diagram transposition to relate integrable representations grouped into simple-current orbits \cite{KacPeterson:1984,Verlinde:1988}. In the two prediction tasks, the correspondence can already be determined from  representation-theoretic data, giving the same predictions as a model supplied with the exact matching. We then consider charged correlators whose coefficients must be learned from observations. The level--rank pairing relates the corresponding source and target observables, allowing measurements in one sector to predict the response in the other. We construct these observables explicitly from charged angular polynomials of the finite Fourier--Berezin fields and examine their accuracy against WZW transport. Source-trained predictors recover the paired preparation responses without additional target-preparation measurements.

These results demonstrate how the algebraic relations of WZW models can guide the construction and study of neural field theories. Level--rank duality supplies exact relations between charged sectors, while the finite neural realizations provide a setting for investigating approximation, computational resources and the transfer of learned information.

\paragraph{Summary of contributions.}
Our main results can be organized into five contributions:
\begin{itemize}
\item \textbf{Neural realizations of positive-level affine algebras.} We construct bosonic Wakimoto and fermionic Fourier--Berezin architectures whose composite observables realize positive-level affine currents and their stress tensors in controlled continuum limits. We derive the limiting current algebras and calculate finite-width corrections directly from the underlying neural parameter laws.

\item \textbf{Charged sectors and conformal-block transport.} Angular projections of the fermionic fields yield the limiting CAR pairing and its Fock-space realization. We identify paired charged sectors and reconstruct their affine descendants, obtaining agreement with independent WZW/KZ calculations through the tested grades. Using physical Ward couplings, we also construct charged angular observables in the finite neural fields and establish controlled approximation limits for their sewing amplitudes.

\item \textbf{Presentation-dependent approximation and computational resources.} We compare direct Wakimoto and coset constructions of the same limiting current algebra, finding different finite-width errors, regulator dependence and computational costs. An explicit calculation of the coset stress-tensor correlator identifies the effects of regulator corrections in the parent theory. We also examine how parameter sharing and amplitude calibration affect correlation functions, and how sampling methods and computational workloads influence the cost of evaluating them. 

\item \textbf{Finite-mode compression of fermionic state spaces.} We study finite-dimensional approximations of fermionic state spaces through polarized mode selection and the resulting sewn amplitudes. Optimized subspaces improve on conventional mode truncation in two of the five conditions studied. Direct and neural optimization give comparable results, with approximation accuracy depending on the retained subspace and observable. 

\item \textbf{Learning from level--rank duality.} Modular and fusion data determine correspondences between simple-current orbits in paired WZW theories. We study their use in fusion prediction, finding that the map computed from representation data gives the same predictions as a supplied exact correspondence. We then construct finite-neural charged observables for unknown preparations and use source-sector measurements to predict their paired-sector responses. The transfer is verified in both directions against finite-neural observables and independent KZ transport, without measurements of the target preparations.  
\end{itemize}

\paragraph{For physics readers.}
WZW models provide an exact algebraic setting in which to study neural realizations of conformal field theories. We construct positive-level affine currents using bosonic Wakimoto and fermionic Fourier--Berezin fields. In the bosonic construction, a direct $SU(2)_3$ presentation and the $SU(6)_1/SU(3)_2$ coset recover the same continuum current sector, while displaying different finite-width corrections and regulator dependence. The fermionic construction realizes paired current algebras within a common free-fermion theory. Its charged sectors admit a CAR/Fock description, with descendant expansions that agree with independent WZW transport calculations. We extend this construction to charged angular observables of the finite neural fields, obtaining controlled approximations to the corresponding sewing amplitudes. Level--rank duality then relates preparations inferred from measurements in one sector to observables in the other. We also examine the approximation of sewn amplitudes using selected fermionic modes and the transfer of fusion data between integrable sectors. Together, these calculations connect the algebraic structure of WZW models with the finite-dimensional and statistical properties of their neural realizations.

\paragraph{For machine learning readers.}
Our results show how mathematical relations between field theories can inform the design and comparison of neural architectures. Networks with the same limiting current algebra can have substantially different approximation errors at finite width. We find that changing the frequency distribution can improve accuracy without increasing network width, with the preferred distribution depending on the observable and regulator. Different architectures also have different computational costs, and averaging over selected neural parameters can reduce sampling uncertainty. We examine learned compression of fermionic state spaces, finding improvements from the choice of retained modes and comparable performance from direct and neural optimization. Finally, we use level--rank duality in two learning problems. Fusion-label matching can be determined from public representation data, while charged preparations contain unknown information that must be learned from observations. Using explicit finite-neural charged observables, source-trained predictors transfer this information between paired WZW sectors without target-preparation measurements. The experiments also compare these predictions with exact source estimators and direct target training, providing a basis for assessing the benefits and costs of incorporating mathematical structure into neural learning.\bigskip

The remainder of the paper is organized as follows. Section~\ref{sec:background} establishes the bosonic and fermionic neural current constructions; Section~\ref{sec:quantized_neural_transport} develops charged-state geometry and four-point transport; Section~\ref{sec:finite_width_regulator_design} analyzes the direct/coset finite-width and resource comparisons; Section~\ref{sec:finite_resource_geometry} studies mode compression and transfer learning; and Section~\ref{sec:discussion} discusses the outstanding mathematical and computational questions.

\section{Neural architectures and affine current algebras}
\label{sec:background}

In this section, we introduce the ingredients for neural-network (NN) realizations of the current algebras appearing in Wess--Zumino--Witten (WZW) models. Our construction uses the neural-network field-theory (NNFT) correspondence, which associates field-theoretic correlation functions with expectation values over ensembles of network parameters \cite{Halverson:2020trp,Halverson:2021aot}. By choosing architectures and priors whose kernels reproduce the required free-field propagators in a certain limit, we are able to use standard free-field representations to construct WZW currents as composite observables of the neural fields and their derivatives.

We consider two complementary constructions. First, we introduce a bosonic substrate consisting of neural realizations of free scalar fields and commuting first-order $\beta\gamma$ ghost fields. Combined with the appropriate nonlinear readouts, derivative corrections, and normal-ordering prescriptions, these implement the Wakimoto representation \cite{Wakimoto:1986gf,deBoer:1996ti}. Second, we introduce a fermionic substrate consisting of Grassmann-valued neural fields equipped with a Gaussian Berezin functional \cite{Frank:2025zuk}. In the corresponding operator construction, signed Wick contractions of the elementary fields allow normally ordered fermion bilinears to generate the desired affine current sectors. The two constructions differ both in their elementary statistics and in the composite readouts required to obtain positive level.

Throughout, we distinguish the particular neural substrate specified by the parameterized random fields from the readouts, vacuum, and normal ordering that define the operator construction. This distinction is crucial because, beyond covariance kernels, the exchange statistics and the operator prescription are necessary to specify the algebra of composite operators.

Our target is the positive-level affine current algebra $\widehat{\mathfrak{su}}(N)_K$, with $N\geq2$ and $K\in\mathbb{Z}_{>0}$. For convenience, we will fix a set of Hermitian generators $t^a$ satisfying
\begin{align}
    \operatorname{tr}_N(t^at^b)     &=\frac{\delta^{ab}}{2},     &     [t^a,t^b]     &=i f^{ab}{}_{c}t^c,     \qquad a,b,c=1,\ldots,N^2-1.
\end{align}
This infinite-dimensional algebra can be described by an algebra of holomorphic currents $\mathcal{J}^a(z)$,
\begin{align}
    \mathcal{J}^a(z)\mathcal{J}^b(w)     \sim     \frac{K\delta^{ab}}{2(z-w)^2}\mathbf{1}     +     \frac{i f^{ab}{}_{c}\mathcal{J}^c(w)}{z-w}.
    \label{eq:target-current-ope}
\end{align}
The simple pole encodes the underlying $\mathfrak{su}(N)$ Lie algebra, while the double pole determines its central extension. The coefficient $K$ is the affine level; the factor of $1/2$ follows from the normalization of the generators. The Laurent modes of these currents generate the corresponding centrally extended loop algebra. 

As in the usual OPE notation, $\sim$ denotes equality up to terms regular as $z\to w$. In the neural realization, this relation is understood through expectation values over the network parameters in the infinite-width limit: first taken at separated points, after which $z\to w$ while the other insertion points remain separated. Thus \eqref{eq:target-current-ope} is a statement about ensemble correlation functions, and provides our local criterion for realizing the target current algebra.

\subsection{Bosonic substrate and the Wakimoto construction}
\label{sec:bosonicsubstrate}

The Wakimoto construction uses $N-1$ Cartan scalars and a commuting $\beta\gamma$ pair for each positive root of $\mathfrak{su}(N)$ \cite{Wakimoto:1986gf,deBoer:1996ti},
\begin{align}
    \phi^A, & \qquad A=1, \ldots,N-1,     &     (\beta^a,\gamma^a), & \qquad a=1,\ldots,\frac{N(N-1)}{2}.
\end{align}
The first-order fields have conformal weights $(1,0)$ with respect to the Wakimoto stress tensor, and are defined through their OPE,
\begin{align}\label{eq:phibetagamma_ope_definition}
    \partial\phi^A(z)\partial\phi^B(w) \sim & -\frac{\delta^{AB}}{(z-w)^2}, & \beta^a(z)\gamma^b(w) \sim & -\frac{\delta^{ab}}{z-w},
\end{align}
with the reverse-ordered singular product $\gamma^a(z)\beta^b(w)\sim+\delta^{ab}/(z-w)$; all remaining elementary products are regular. The affine currents are then ``normal-ordered'' polynomials in $\beta^a$, $\gamma^a$, and $\partial\phi^A$, together with terms proportional to $\partial\gamma^a$. Their coefficients depend on the Lie algebra and the level. We will come back to their exact definitions at the end of this subsection.

We first define neural architectures with the required elementary correlation functions. We denote the number of independently sampled frequency features by $L$.

Both the scalar and $\beta\gamma$ networks use complex frequencies drawn from the same distribution. At width $L$, we independently sample complex frequencies $W_j=p_j e^{i\varphi_j}$, $j=1,\ldots,L$, from the rotationally invariant law on the annulus $\mathcal A=\{W=pe^{i\varphi}:p_-<p<p_+\}$,
\begin{align}
\label{eq:bosonic-frequency-prior}
     d\nu_\alpha^{(\Lambda_B)}(W) &= \frac{1}{Z_{\alpha}^{(\Lambda_B)}}\frac{R_{\epsilon_B}(p)}{p^{1-2\alpha}}dp\frac{d\varphi}{2\pi},
    & Z_{\alpha}^{(\Lambda_B)} &= \int_{p_-}^{p_+}\frac{R_{\epsilon_B}(p)}{p^{1-2\alpha}}\,dp,
\end{align}
The parameter $\alpha\in\mathbb R$ controls the spectral tilt of the distribution with every real value admissible on a finite annulus. We denote the bosonic regulator data collectively by $\Lambda_B=(p_-,p_+,\epsilon_B)$, with momenta expressed in fixed reference units. The regulator $R_{\epsilon_B}(p)$ suppresses the IR and UV portions of the spectrum, allowing us to take a controlled limit as the annulus fills out the complex plane.

The order of limits is important: we first take the infinite-width limit $L\to\infty$ at fixed regulator parameters, then remove the annular cutoffs $p_-\to0$ and $p_+\to\infty$ at fixed $\epsilon_B>0$, and finally consider $\epsilon_B\to0$ for observables with well-defined continuum limits. A convenient choice for the bosonic construction is the symmetric regulator\footnote{Other regulator choices are studied in Section~\ref{sec:finite_width_regulator_design}.} 
\begin{align}
\label{eq:bosonic-regulator}
    R_{\epsilon_B}(p) &= e^{-\epsilon_B (p + 1/p)}.
\end{align}
At fixed $\epsilon_B$, this suppresses both small and large Fourier features as desired, so the annular cutoffs can be removed while retaining finite correlation functions. The subsequent $\epsilon_B\to0$ limit is an ordered limit of regulated correlators. In particular, the undifferentiated scalar covariance retains an additive IR divergence, while the IR regulated scheme has well-behaved finite limits. On the full annulus, the normalization is exactly
\begin{align}
\label{eq:bosonic-frequency-normalization}
      Z_\alpha^{(0,\infty,\epsilon_B)}&=\int_0^\infty p^{2\alpha-1}e^{-\epsilon_B(p+p^{-1})}\,dp=2\mathrm{K}_{2\alpha}(2\epsilon_B), 
\end{align}
where $\mathrm{K}_\nu$ denotes the modified Bessel function of the second kind. The leading small-$\epsilon_B$ behavior is 
\begin{align}
    Z_\alpha^{(0,\infty,\epsilon_B)}&=\begin{cases}
    \Gamma(2|\alpha|)\epsilon_B^{-2|\alpha|}[1+o(1)], &\alpha\neq0\\
    -2\ln(\epsilon_B) - 2\gamma_E + o(1),& \alpha = 0.
    \end{cases}
\end{align}
These asymptotics are nonuniform near $\alpha=0$, so we use the exact normalization at finite regulator.

The spectral tilt $p^{2\alpha}$ together with the compensating feature amplitude $p^{-\alpha}$ below realizes part of the architectural freedom studied in \cite{Zhang:2026tss}. Changing the spectral prior without compensating the amplitude generally changes the elementary kernel, but with the compensation, the elementary two-point kernels and the resulting Gaussian width limit are unchanged for any $\alpha$. Finite-width higher moments, however, generally depend on $\alpha$. The untilted choice, $\alpha=0$, matches the spectral weighting of \cite{Robinson:2025ybg}, and $\alpha=1$ recovers that in \cite{Frank:2026bui}. We will study this freedom in Section~\ref{sec:finite_width_regulator_design}, by looking at finite-width correlators of composite neural affine currents at different values of $\alpha$.

\paragraph{Free bosons.}

Following the logarithmic-kernel construction of \cite{Robinson:2025ybg}, we represent the scalar fields by single-hidden-layer random Fourier feature (RFF) networks,
\begin{align}
\label{eq:bosonic-scalar-architecture}
    \phi^A(z,\bar z) = \frac{C_\phi}{\sqrt L} \sum_{j=1}^{L} \frac{a_j^A}{|W_j|^\alpha} \cos\left(W_j z+\bar W_j\bar z+\delta_j\right).
\end{align}
The factor $|W_j|^{-\alpha}$ compensates the spectral tilt of the sampling law. The output weights and phases are Gaussian and uniform random variables respectively,
\begin{align}
    a_j^A &\overset{\mathrm{i.i.d.}}{\sim} \mathcal N(0,\sigma^2), & \delta_j &\overset{\mathrm{i.i.d.}}{\sim} \operatorname{Unif}_{[0,2\pi)},
\end{align}
with $\sigma>0$. The output weights ($a_j^A$), frequencies ($W_j$), and phases ($\delta_j$) are mutually independent parameter families. To reduce the number of neural parameters, we use the same frequencies and phases for all scalar fields. The Gaussian output weights remain independent across species, which ensures that the scalar two-point functions are diagonal in the ensemble.

We choose the normalization 
\begin{align}
\label{eq:boson-normalisation}
     C_\phi^2=\frac{4Z_\alpha^{(\Lambda_B)}}{\sigma^2}.
\end{align}
Writing $r=|z-w|$, the exact two-point function at any finite width $L$ is the expectation value over all network parameters, i.e. sampled frequencies, phases, and Gaussian output weights collectively denoted $\Theta$,
\begin{align}
\label{eq:bosonic-scalar-covariance}
    \bE_\Theta\left[\phi^A(z,\bar z)\phi^{B}(w,\bar w)\right] &= \delta^{AB}\mathcal K_\phi(r),\qquad 
    \mathcal K_\phi(r) \equiv 2\int_{p_-}^{p_+} \frac{R_{\epsilon_B}(p)}{p}{\mathrm J}_0(2pr)\,dp,
\end{align}
where ${\mathrm J}_\nu$ is the Bessel function of the first kind. 

Removing the annular cutoffs at fixed ${\epsilon_B}$ and then expanding for small ${\epsilon_B}$ gives, at fixed $r>0$,
\begin{align}
\label{eq:bosonic-scalar-log-limit}
    \lim_{\substack{p_-\to0\\ p_+\to\infty}} \cK_\phi(r) =-2\ln\left(r{\epsilon_B}\right) - 4\gamma_E +\frac{\epsilon_B}{r}\left(4r^2 - 1\right) +\mathcal O\left({\epsilon_B}^2\ln\left(\frac{1}{\epsilon_B}\right)\right),
\end{align}
where $\gamma_E$ is the Euler--Mascheroni constant. The position-independent infrared divergence drops out of expectations of gradients.  Hence, the limiting contraction
\begin{align}
\label{eq:bosonic-scalar-gradient-ope}
    \partial\phi^A(z)\,\partial\phi^{B}(w)
    \OPE
    -\frac{\delta^{AB}}{(z-w)^2},
\end{align}
as expected of holomorphic weight-one free-boson derivatives. Here and below, the notation ${\OPE}$ refers to the Gaussian theory obtained after taking the infinite-width limit and removing the regulators, and holds for arbitrary insertions of additional neural observables in the expectation value. We also suppress the antiholomorphic arguments when discussing its holomorphic correlation functions.

\paragraph{$\boldsymbol{\beta\gamma}$ ghosts.}

For each positive root, we introduce a pair of complex-valued neural fields,
\begin{subequations}
\label{eq:bosonic-ghost-architecture}
\begin{align}
    \beta^a(z,\bar z)&=-\frac{C_\beta}{\sqrt L}\sum_{j=1}^{L}\left[ b_j^a v_j(z,\bar z) - c_j^a u_j(z,\bar z) \right],\\
    \gamma^a(z,\bar z) &= \frac{C_\gamma}{\sqrt L} \sum_{j=1}^{L} \left[ b_j^a u_j(z,\bar z) + c_j^a v_j(z,\bar z) \right].
\end{align}
\end{subequations}
With $W_j$ drawn from $d\nu_\alpha^{(\Lambda_B)}$ as above, the ghost features are
\begin{subequations}
\label{eq:bosonic-ghost-features}
\begin{align}
    u_j(z,\bar z) &= \sqrt{2}\,p_j^{1/2-\alpha}e^{i\varphi_j/2} \cos\left(W_j z+\bar W_j\bar z\right), \\
    v_j(z,\bar z) &= i\sqrt{2}\,p_j^{1/2-\alpha}e^{i\varphi_j/2} \sin\left(W_j z+\bar W_j\bar z\right).
\end{align}
\end{subequations}
The real output weights satisfy $b_j^a,\ c_j^a\overset{\mathrm{i.i.d.}}{\sim}\mathcal N(0,\sigma^2)$. The same weights and features appear in both members of each $\beta\gamma$ pair, producing nontrivial mixed two-point functions.

The required continuum $\beta\gamma$ contraction fixes the product of the normalization constants to be of the form
\begin{align}
\label{eq:bosonic-ghost-normalization-product}
     C_\beta C_\gamma=\frac{Z_\alpha^{(\Lambda_B)}}{i\sigma^2}
\end{align}
which we use to fix the remaining rescaling freedom by choosing
\begin{align}
\label{eq:bosonic-ghost-normalizations}
    C_\beta=-i\sqrt{\frac{Z_\alpha^{(\Lambda_B)}}{\sigma^2}},
  \qquad
  C_\gamma=\sqrt{\frac{Z_\alpha^{(\Lambda_B)}}{\sigma^2}}.
\end{align}

For $z-w=re^{i\theta}$, the exact mixed two-point function is
\begin{align}
\label{eq:bosonic-ghost-covariance}
    \bE\!\left[\beta^a(z)\gamma^{b}(w)\right] &= \delta^{ab}\partial_z\cK_\phi(r)= -2\delta^{ab}e^{-i\theta} \int_{p_-}^{p_+} R_{\epsilon_B}(p){\mathrm J}_1(2pr)\,dp.
\end{align}
Thus, the scalar and first-order kernels are related by a holomorphic derivative, for any width $L$ and distribution parameter/regulator. Rotational invariance also gives
\begin{align}
\label{eq:bosonic-ghost-diagonal-covariances}
    \bE\!\left[\beta^a(z)\beta^{b}(w)\right] = \bE\!\left[\gamma^a(z)\gamma^{b}(w)\right] = 0.
\end{align}

In the regulator-free Gaussian limit, the nontrivial contractions are
\begin{align}
\label{eq:bosonic-ghost-opes}
    \beta^a(z)\gamma^{b}(w) &\OPE -\frac{\delta^{ab}}{z-w}, & \gamma^a(z)\beta^b(w) &\OPE \frac{\delta^{ab}}{z-w},
\end{align}
as expected from the continuum limit \eqref{eq:phibetagamma_ope_definition}. The opposite signs in \eqref{eq:bosonic-ghost-opes} are consistent with commuting fields: exchanging the insertions also exchanges $z$ and $w$.

\paragraph{Wakimoto currents.}

 With Gaussian Wick factorization, \eqref{eq:bosonic-scalar-gradient-ope} and \eqref{eq:bosonic-ghost-opes} (or equivalently \eqref{eq:phibetagamma_ope_definition}) supply the Wakimoto free-field contractions. Those can then be used to construct holomorphic currents for the affine extension of any simply-laced Lie algebra, $\widehat{\mathfrak{g}}_K$.

We will keep the present exposition simple and only illustrate the construction for $\widehat{\mathfrak{su}}(2)_K$, which requires one scalar field $\phi$ and one commuting pair $(\beta,\gamma)$. A sketch of the $\widehat{\mathfrak{su}}(N)_K$ construction can be found in Appendix~\ref{app:sunk_wakimoto}. Choosing the Hermitian generators $t^a=\sigma^a/2$, the three affine currents are
\begin{subequations}\label{eq:wakimoto-su2-currents}
\begin{align}
  \mathcal{J}^1(z) &= \frac{1}{2}:\beta(1-\gamma^2):-i\sqrt{\frac{K+2}{2}}:\gamma\partial\phi: - \frac{K}{2}\partial\gamma,\\
  \mathcal{J}^2(z) &= -\frac{i}{2}:\beta(1+\gamma^2):+\sqrt{\frac{K+2}{2}}:\gamma\partial\phi: - \frac{iK}{2}\partial\gamma,\\
  \mathcal{J}^3(z) &= i\sqrt{\frac{K+2}{2}}:\partial\phi: + :\beta\gamma:.
\end{align}
\end{subequations}

The colons denote elementary Wick ordering, which, at finite regulator, is defined by subtracting the elementary pair contractions. For example
\begin{subequations}
\label{eq:wakimoto-elementary-wick-products}
\begin{align}
    :\!\gamma\partial\phi\!: &= \gamma\partial\phi - \bE[\gamma\partial\phi],\\
    :\!\beta\gamma^2\!: &= \beta\gamma^2 - 2\bE[\beta\gamma]\gamma - \bE[\gamma^2]\beta.
\end{align}
\end{subequations}
At finite width, this prescription defines the readout but does not imply Gaussian factorization of its correlation functions. However, for this rotationally invariant regulated ensemble and our choice of network architectures,
\begin{align}
\label{eq:wakimoto-vanishing-coincident-contractions}
    \bE[\beta(z)\gamma(z)]=\bE[\gamma(z)^2] = \bE[\gamma(z)\partial\phi(z)] = 0.
\end{align}
Hence, the subtractions in \eqref{eq:wakimoto-su2-currents} vanish and the currents can be evaluated directly as the displayed neural polynomials. We retain the normal-ordering symbols to specify the corresponding free-field operators in the continuum limit.

In the Gaussian continuum limit, these currents satisfy
\begin{align}
  \mathcal{J}^a(z)\mathcal{J}^b(w)\OPE \frac{K\delta^{ab}}{2(z-w)^2} + \frac{i\epsilon^{ab}{}_c\mathcal{J}^c(w)}{z-w},
\end{align}
as required by \eqref{eq:target-current-ope}. 

\paragraph{Stress tensor.}

The Sugawara readout requires affine, rather than merely elementary, normal ordering of the composite currents.  In the $\widehat{\mathfrak{su}}(2)_K$ Wakimoto construction detailed above, the diagonal affine products are
\begin{subequations}
\label{eq:su2-neural-affine-products}
\begin{align}
    :\!\mathcal{J}^1\mathcal{J}^1\!:\!_{\mathrm{aff}} &= (\mathcal{J}^1)^2 - \frac{1-\gamma^2}{2}\,\partial \mathcal{J}^3 - \gamma(\partial\gamma)\mathcal{J}^3 - \frac{K}{4}(\partial\gamma)^2,\\
    :\!\mathcal{J}^2\mathcal{J}^2\!:\!_{\mathrm{aff}} &= (\mathcal{J}^2)^2 - \frac{1+\gamma^2}{2}\,\partial \mathcal{J}^3 + \gamma(\partial\gamma)\mathcal{J}^3 + \frac{K}{4}(\partial\gamma)^2,\\
    :\!\mathcal{J}^3\mathcal{J}^3\!:\!_{\mathrm{aff}} &= (\mathcal{J}^3)^2 + \gamma\,\partial\beta - \beta\,\partial\gamma.
\end{align}
\end{subequations}
All quantities on the right-hand sides are evaluated at the same input, and products denote ordinary multiplication of the regulated neural outputs. In particular, $\partial \mathcal{J}^3$ is obtained by differentiating the current readout,
\begin{align}
    \partial \mathcal{J}^3=i\sqrt{\frac{K+2}{2}}\,\partial^2\phi+\gamma\,\partial\beta+\beta\,\partial\gamma.
\end{align}
Thus the affine products require only the original neural fields and their derivatives.

The correction terms in \eqref{eq:su2-neural-affine-products} are fixed by the elementary continuum contractions and the specified Wakimoto current construction.\footnote{In the Gaussian continuum theory, these expressions reproduce the conventional affine normal products $:\mathcal{J}^a\mathcal{J}^a:\!_{\mathrm{aff}}(z)=\lim_{\zeta\to z}\left[\mathcal{J}^a(\zeta)\mathcal{J}^a(z) -\frac{K}{2(\zeta-z)^2}\mathbf{1}\right]$ with no sum over $a$. The subtraction removes the second-order pole. There is no first-order pole for equal current components. Expanding the remaining elementary fields about $z$ gives the derivative corrections in \eqref{eq:su2-neural-affine-products}. This relation is understood inside correlation functions, with the infinite-width and regulator limits taken before the coincident-point limit \cite{deBoer:1996ti}.} At finite width and fixed regulators, \eqref{eq:su2-neural-affine-products} specifies our choice of composite neural readouts without assuming Gaussian factorization of the finite-width statistics.

The stress-energy neural field is defined by the normalized sum of the diagonal affine products,
\begin{align}
\label{eq:su2-wakimoto-stress-tensor}
    T_{\mathrm W}(z) &= \frac{1}{K+2}\sum_{a=1}^{3}:\!\mathcal{J}^a\mathcal{J}^a\!:\!_{\mathrm{aff}}(z)\nonumber\\
    &= -\frac12\bigl(\partial\phi(z)\bigr)^2 - \beta(z)\partial\gamma(z) - \frac{i}{\sqrt{2(K+2)}}\,\partial^2\phi(z).
\end{align}
The second line follows by substituting the explicit current readouts and combining the derivative corrections. As for the currents, the elementary coincident contractions entering this expression vanish for our rotationally invariant regulated ensemble. The affine corrections fix the coefficient of $\beta\partial\gamma$ and supply the scalar background-charge term.

The two-point function of this readout can be evaluated directly from the neural parameter ensemble. Taking the infinite-width limit at fixed regulators, followed by removal of the annular cutoffs at fixed $\epsilon_B$ and finally $\epsilon_B\to0$, gives
\begin{align}
\label{eq:su2-neural-stress-two-point}
    \bE\!\left[T_{\mathrm W}(z)T_{\mathrm W}(w)\right] \leadsto \frac{c_K}{2(z-w)^4}, \qquad c_K=\frac{3K}{K+2},
\end{align}
for fixed $z\ne w$. This reproduces the stress-tensor two-point function of the $SU(2)_K$ WZW model and verifies the central term of its Virasoro algebra, in agreement with the Wakimoto construction \cite{deBoer:1996ti}. Checking the remaining Virasoro terms requires correlators with further insertions in the continuum theory and is not carried out here. A derivation of this two-point function, including the finite-width and regulator-dependent contributions, is detailed in Appendix~\ref{app:finite_width_TT}.

\subsection{Fermionic substrate, CAR and the level--rank embedding}
\label{sec:explicit-fermionic-WZW}

The fermionic route replaces commuting first-order fields by Grassmann-valued neural outputs with a Gaussian Berezin functional. We first define the finite parameter ensemble before passing to its canonical anticommutation relation (CAR)/Fock operator completion.

The fermionic construction uses its own regulated Fourier prior.  At width $L$, the frequency parameters $W_j=p_je^{i\varphi_j}$ ($j=1,\ldots,L$) are sampled independently from
\begin{align}
\label{eq:fermionic_frequency_prior}
 d\nu_{\alpha}^{(\Lambda_F)}(W)
 &=\frac{e^{-\epsilon_Fp}}{Z_{\alpha}^{(\Lambda_F)}p^{1-2\alpha}}\,dp\,\frac{d\varphi}{2\pi},
 &Z_{\alpha}^{(\Lambda_F)}
 &=\int_{p_-}^{p_+}e^{-\epsilon_Fp}\,\frac{dp}{p^{1-2\alpha}},
\end{align}
where $0<p_-<p_+<\infty$, $\epsilon_F>0$ and $\Lambda_F=(p_-,p_+,\epsilon_F)$. The regulated annulus makes this a probability law for any fixed $\alpha$. The large-annulus limit $p_-=0$, $p_+=\infty$, used later, requires $\alpha>0$ for normalization. 

For every channel $A=1,\ldots,2NK$ and neuron $j$, introduce Grassmann generators $\xi_{Aj}$ and $\bar\xi_{Aj}$. Let $\mathcal W=(W_1,\ldots,W_L)$ be the collection of sampled frequencies. For any polynomial observable $\mathcal O_L(\xi,\bar\xi;\mathcal W)$, we write $\langle\mathcal O_L\rangle_{\mathrm{Gr}\mid\mathcal W}$ when emphasizing that its Grassmann coefficients are contracted at fixed $\mathcal W$.  The full neural expectation is the successive operation
\begin{align}
 \label{eq:fermionic_full_expectation_definition}
 \bE_{\Theta_F}[\mathcal O_L] &:=\int\left(\prod_{j=1}^L d\nu_{\alpha}^{(\Lambda_F)}(W_j)\right)\langle\mathcal O_L\rangle_{\mathrm{Gr}\mid\mathcal W}=\bE_{\mathcal W}\!\left[ \langle\mathcal O_L\rangle_{\mathrm{Gr}\mid\mathcal W}  \right].
\end{align} 
Here $\Theta_F$ collects the Grassmann coefficients and Fourier parameters, and $\bE_W$ below denotes averaging one frequency under $d\nu_{\alpha}^{(\Lambda_F)}$ rather than over all $\mathcal W$. %

The normalized conditional contraction satisfies $\langle1\rangle_{\mathrm{Gr}\mid\mathcal W}=1$ and vanishes on odd monomials. Its higher even moments obey the signed (Pfaffian) Wick rule with the following ordered pairings:
\begin{align}
\label{eq:fermionic_parameter_pairing}
 \langle\xi_{Aj}\bar\xi_{B\ell}\rangle_{\mathrm{Gr}\mid\mathcal W} &=\delta_{AB}\delta_{j\ell},&\langle\bar\xi_{Aj}\xi_{B\ell}\rangle_{\mathrm{Gr}\mid\mathcal W}&=-\delta_{AB}\delta_{j\ell}, 
\end{align}
with $\langle\xi_{Aj}\xi_{B\ell}\rangle_{\mathrm{Gr}\mid\mathcal W} =\langle\bar\xi_{Aj}\bar\xi_{B\ell}\rangle_{\mathrm{Gr}\mid\mathcal W} =0$.  All higher even moments of Grassmann-linear combinations $g_1,\ldots,g_{2m}$ are specified by 
\begin{equation}
\langle g_1\cdots g_{2m}\rangle_{\mathrm{Gr}\mid\mathcal W} = \operatorname{Pf}\!\left(\bigl[\langle g_r g_s\rangle_{\mathrm{Gr}\mid\mathcal W}\bigr]_{r,s=1}^{2m} \right).
 \label{eq:fermionic_parameter_wick_rule}
\end{equation}
These algebraic rules determine the contraction of every polynomial neural observable.

Setting 
$\vartheta_{W_j}(z)=W_jz+\bar W_j\bar z$, we define the two fixed Fourier feature functions
\begin{align}
\label{eq:fermionic_features_recall}
 U_j(z,\bar z)&=\sqrt{2W_j}\cos\vartheta_{W_j}(z),&
 V_j(z,\bar z)&=i\sqrt{2W_j}\sin\vartheta_{W_j}(z).
\end{align}
Here the branch of $\sqrt{2W_j}$ is fixed identically in both features. Since every nonzero contraction contains two factors of this common square root \(\sqrt{2W_j}\), the result is independent of its branch. The $2NK$ neural fermions are thus constructed from the architecture

\begin{align}
\label{eq:fermionic_architecture_recall}
 \chi_L^{A}(z,\bar z;\Theta_F)  &=e^{i\pi/4}\sqrt{\frac{Z_\alpha^{(\Lambda_F)}}{L}}  \sum_{j=1}^L p_j^{-\alpha}  \left[\xi_{Aj}U_j(z,\bar z)+\bar\xi_{Aj}V_j(z,\bar z)\right].
\end{align}
The $p_j^{-\alpha}$ factor compensates the spectral tilt of the sampling law. At a fixed frequency configuration, \eqref{eq:fermionic_parameter_pairing} gives the empirical covariance 
\begin{align}
\label{eq:fermionic_empirical_kernel} \left\langle\chi_L^A(z)\chi_L^B(w)\right\rangle_{\mathrm{Gr}\mid\mathcal W}  &=\delta^{AB}\frac{2Z_\alpha}{L}\sum_{j=1}^L  p_j^{-2\alpha}W_j  \sin\!\left(\vartheta_{W_j}(z)-\vartheta_{W_j}(w)\right)\\\nonumber&=:\delta^{AB}S_{L,\mathcal W}(z,w)%
\end{align}
Only after averaging the frequency parameters does this become the rotationally covariant regulated Cauchy kernel, 
\begin{align}
\label{eq:fermionic_ensemble_kernel}
 \bE_{\Theta_F}\!\left[\chi_L^A(z)\chi_L^B(w)\right]  &=2\delta^{AB}\frac{|z-w|}{z-w} \int_{p_-}^{p_+}e^{-\epsilon_Fp}{\mathrm J}_1(2p\vert z-w\vert)\,dp\\\nonumber
&=: \delta^{AB}\frac{F_1^{(\Lambda_F)}(|z-w|)}{z-w}\OPE \frac{\delta^{AB}}{z-w},
\end{align}
Here $Z_\alpha$ and $Z_{\alpha,F}$ both abbreviate $Z_\alpha^{(\Lambda_F)}$, and $\OPE$ denotes our regulated continuum limit ($L\to\infty$, then $p_-\to0$, $p_+\to\infty$, followed by $\epsilon_F\to0$).
For $r>0$, $F_1^{(\Lambda_F)}(r)=2r\int_{p_-}^{p_+}e^{-\epsilon_Fp}\mathrm J_1(2pr)\,dp$. The regulated two-point function is independent of $L$, while higher-point correlations receive finite-width corrections from the shared frequency samples.

\paragraph{Complex fermions.} %

We relabel the $2NK$ Grassmann-valued neural channels of \eqref{eq:fermionic_architecture_recall} as $\chi_L^{ja,s}$, with $j=1,\ldots,N$, $a=1,\ldots,K$, and $s=1,2$. Their continuum free-fermion operator interpretation is developed below. We combine them into $NK$ pairs of complex fermionic fields, 
\begin{align}
\label{eq:explicit-complex-fermions}
        \psi_{ja,L}(z)     &=     \frac{\chi_L^{ja,1}(z)+i\chi_L^{ja,2}(z)}{\sqrt2},     &     \widetilde\psi_{ja,L}(z)     &=     \frac{\chi_L^{ja,1}(z)-i\chi_L^{ja,2}(z)}{\sqrt2}.
\end{align}
At finite width we display the subscript $L$ where it matters; in subsequent composite formulas we suppress it when the same regulated width and parameters are understood. The indices $j$ and $a$ label the rows and columns of the fermionic channel array. The tilde denotes the chiral conjugate species; it does not denote an antiholomorphic field or pointwise complex conjugation of the regulated Fourier output.

With $\bE_{\Theta_F}$ denoting the full expectation over both the Grassmann generators and the shared frequencies, the elementary ensemble-averaged two-point functions are 
\begin{align}
\label{eq:fermionic-elementary-contractions}
    \bE_{\Theta_F}[\psi_{ja}(z)\widetilde\psi_{kb}(w)]    &=    \bE_{\Theta_F}[\widetilde\psi_{ja}(z)\psi_{kb}(w)]    =    \delta_{jk}\delta_{ab}    \frac{F_1^{(\Lambda_F)}(|z-w|)}{z-w},   
\end{align}
and $ \bE_{\Theta_F}[\psi_{ja}\psi_{kb}] = \bE_{\Theta_F}[\widetilde\psi_{ja}\widetilde\psi_{kb}] =0$. Here $F_1^{(\Lambda_F)}$ is the regulator-dependent radial factor of the frequency-averaged fermion kernel from \eqref{eq:fermionic_ensemble_kernel}. Its continuum limit is $1$ at separated insertion points, giving the required Cauchy kernel. Fermionic exchange reverses both the field order and the insertion positions. The equal signs of the two displayed mixed contractions are therefore consistent with exchange antisymmetry. 

The signed fermionic contraction rule fixes the non-Abelian current terms.

\paragraph{Currents.}

We denote elementary fermionic Wick ordering by $:\,\,:_0$. For a bilinear readout, it is defined by subtracting its Berezin contraction at fixed frequency, keeping the ordering identical:
\begin{align}
\label{eq:fermionic-elementary-wick-product}
    :\!\widetilde\psi_{ja}\psi_{kb}\!:_0(z)    =    \widetilde\psi_{ja}(z)\psi_{kb}(z)    -    \left\langle        \widetilde\psi_{ja}(z)\psi_{kb}(z)    \right\rangle_{\mathrm{Gr}\mid\mathcal W}.
\end{align}
The contraction on the right-hand side is evaluated with the same sampled frequencies and regulator as the composite; only afterwards is the frequency ensemble averaged.  Higher products are defined by the corresponding signed Wick subtractions.
This prescription allows us to specify the composite readouts at finite width without assuming Gaussian factorization of the full ensemble.
For the undifferentiated bilinear, displayed above, the conditional subtraction vanishes because $S_{L,\mathcal W}(z,z)=0$. For derivative bilinears it need not: at fixed sampled frequencies, for example,
\begin{equation}
\left\langle\widetilde\psi_{ja}(z)\partial\psi_{ja}(z)\right\rangle_{\mathrm{Gr}\mid\mathcal W}=-\frac{2Z_\alpha^{(\Lambda_F)}}{L}\sum_{k=1}^Lp_k^{-2\alpha}W_k^2.\label{eq:fermionic-derivative-wick-subtraction}
\end{equation}
This term vanishes after isotropic frequency averaging, but must be subtracted conditionally when defining the finite-$L$ stress readout. In the Gaussian continuum limit, the prescription agrees with normal ordering in the chiral fermion vacuum.

Choose Hermitian generators $t^A$ of $\mathfrak{su}(N)$ and $T^a$ of $\mathfrak{su}(K)$ satisfying
\begin{align}
    \operatorname{tr}_N(t^At^B) &= \frac{\delta^{AB}}{2}, &\operatorname{tr}_K(T^aT^b) &= \frac{\delta^{ab}}{2}.
\end{align}
Their commutators define the structure constants $f^{AB}{}_{C}$ and $\widetilde f^{ab}{}_{c}$ in the Hermitian convention used in \eqref{eq:target-current-ope}.

The row, column, and number current readouts are
\begin{subequations} \label{eq:fermionic-row-column-currents}
\begin{align}
    \mathcal{J}^A(z) &= \sum_{a=1}^{K}\sum_{j,k=1}^{N} :\!\widetilde\psi_{ja}(t^A)_{jk}\psi_{ka}\!:_0(z), \\
    \widetilde{\mathcal{J}}^a(z) &= \sum_{j=1}^{N}\sum_{b,c=1}^{K} :\!\widetilde\psi_{jb}(T^a)_{bc}\psi_{jc}\!:_0(z), \\
    \mathring{\mathcal{J}}(z) &= \sum_{j=1}^{N}\sum_{a=1}^{K} :\!\widetilde\psi_{ja}\psi_{ja}\!:_0(z).
\end{align}
\end{subequations}

The row generators act on the $N$ channels at fixed column index, while the column generators act on the $K$ channels at fixed row index. The number current sums over all $NK$ channels.

In the Gaussian continuum limit, signed Wick contractions  give
\begin{subequations}\label{eq:fermionic-positive-affine-levels}
    \begin{align}
    \mathcal{J}^A(z)\mathcal{J}^B(w) &\sim \frac{K\delta^{AB}}{2(z-w)^2} + \frac{i f^{AB}{}_{C}\mathcal{J}^C(w)}{z-w},\\
    \widetilde{\mathcal{J}}^a(z)\widetilde{\mathcal{J}}^b(w) &\sim \frac{N\delta^{ab}}{2(z-w)^2} +  \frac{i\widetilde f^{ab}{}_{c}\widetilde{\mathcal{J}}^c(w)}{z-w},\\
    \mathcal{J}^A(z)\widetilde{\mathcal{J}}^a(w) &\sim0, \\ \mathring{\mathcal{J}}(z)\mathring{\mathcal{J}}(w) &\sim \frac{NK}{(z-w)^2}.
\end{align}
\end{subequations}

The mixed products of $\mathring{\mathcal{J}}$ with either traceless-current sector are also regular. The factors $K$ and $N$ count the independent flavor contributions to the respective double contractions. The fermionic contraction signs give positive level in both sectors.

For one sampled frequency define the scalar (one-channel) Berezin contraction
\begin{equation}
 s_W(z,w)=2Z_\alpha p^{-2\alpha}W\sin\!\bigl(\vartheta_W(z)-\vartheta_W(w)\bigr),\qquad S_L(z,w)=\frac1L\sum_{j=1}^Ls_{W_j}(z,w).
 \label{eq:neural_fermionic_one_feature_audit}
\end{equation}
where the frequency law is \eqref{eq:fermionic_frequency_prior} and $C_\Lambda(z,w)=\mathbb E_W s_W(z,w)$.
Let $E=\mathbb C^N\otimes\mathbb C^K$. For $M,P\in\operatorname{End}(E)$, we set $\mathcal{J}^M=:\!\widetilde\psi M\psi\!:_0$,
where the pair of fermion indices $(j,a)$ is combined into a single composite index $I=(j,a)$, so that explicitly $\mathcal{J}^M=:\!\widetilde\psi_I M_{IJ}\psi_J\!:_0$. Conditional Berezin contraction of the actual neural outputs then gives $\langle \mathcal{J}^M(z)\mathcal{J}^P(w)\rangle_{\mathrm{Gr}\mid\mathcal W}
=\operatorname{tr}_E(MP)S_L(z,w)^2$. Averaging the independent frequency samples gives the exact finite-width result
\begin{equation}
 \mathbb E_{\Theta_F}[\mathcal{J}^M(z)\mathcal{J}^P(w)]=\operatorname{tr}_E(MP)\left[\left(1-\frac1L\right)C_\Lambda(z,w)^2 +\frac1L\mathbb E_W\bigl[s_W(z,w)^2\bigr]\right].
 \label{eq:neural_fermionic_jj_finite_width}
\end{equation}
In particular, for $z-w=re^{i\theta}$, the collision term is an explicit Bessel integral of the sampled Fourier features: 
\begin{equation}
 \mathbb E_W[s_W(z,w)^2]=2Z_\alpha^{(\Lambda_F)} e^{-2i\theta}\int_{p_-}^{p_+}p^{1-2\alpha}e^{-\epsilon_F p}\mathrm{J}_2(4pr)\,dp.
 \label{eq:neural_fermionic_jj_collision_bessel}
\end{equation}
Thus the positive $K/2$ double-pole coefficient for a row current is obtained by first taking $L\to\infty$ at fixed regulator, then $C_\Lambda\to1/(z-w)$, and finally using $\operatorname{tr}_E[(t^A\otimes I_K)(t^B\otimes I_K)]=K\delta^{AB}/2$.  The same direct calculation gives column level $N$ and number-current level $NK$.
The $L^{-1}$ term shows why the exact affine OPE is not a finite-width statement; its inclusion spoils the algebraic relation expected in the continuum limit. The absence of the first-order pole is explained by the fact that the expectation value of a single neural current vanishes, and thus never contributes to the two-point function above.\bigskip

The antisymmetric Lie bracket of the simple pole in \eqref{eq:fermionic-positive-affine-levels} can be checked independently by studying the correlation functions of three neural currents. With $s_{ab}=s_W(z_a,z_b)$ and $S_{ab}=L^{-1}\sum_j s_{W_j}(z_a,z_b)$, their exact conditional Berezin expectation is
\begin{equation}
 \langle \mathcal{J}^M(z_1)\mathcal{J}^P(z_2)\mathcal{J}^Q(z_3)\rangle_{\mathrm{Gr}\mid\mathcal W} = -\operatorname{tr}_E(M[P,Q])S_{12}S_{23}S_{31}.
 \label{eq:neural_fermionic_jjj_conditional}
\end{equation}
Writing $\mu_{ab}=\mathbb E_Ws_{ab}$, $\mu_{ab,cd}=\mathbb E_W(s_{ab}s_{cd})$, and $\mu_{12,23,31}=\mathbb E_W(s_{12}s_{23}s_{31})$, independent feature indices give the exact expansion
\begin{align}
 \mathbb E_\mathcal W[S_{12}S_{23}S_{31}]={}&\frac{(L-1)}{L^2} \bigl(\mu_{12,23}\mu_{31} +\mu_{12,31}\mu_{23}+\mu_{23,31}\mu_{12}\bigr) \nonumber\\ &+\frac1{L^2}\mu_{12,23,31}+\frac{(L-2)(L-1)}{L^2}\mu_{12}\mu_{23}\mu_{31}.
 \label{eq:neural_fermionic_jjj_frequency_expansion}
\end{align}
The ordered limit is
\[
 -\frac{\operatorname{tr}_E(M[P,Q])}{z_{12}z_{23}z_{31}}.
\]
For row generators it equals $-iKf^{ABC}/(2z_{12}z_{23}z_{31})$, fixing the non-Abelian sign from the neural Berezin pairing.

Thus the ordered Gaussian limit yields $\widehat{\mathfrak{su}}(N)_K$, with commuting $\widehat{\mathfrak{su}}(K)_N$ and $\widehat{\mathfrak{u}}(1)_{NK}$ sectors. We emphasize once more that this only holds in the ordered limit considered above. The finite-width algebra deviates substantially from the continuum affine one.\bigskip

Collecting our observations, we can conclude that the neural current sectors give the standard free-fermion conformal embedding in \eqref{eq:intro_level_rank_embedding} with the standard branching qualifications \cite{Nakanishi:1990hj, Ostrik:2014LevelRankTensorCategories}. This identifies the commuting current subalgebras. Charged readouts and the additional data needed for descendant sewing are developed in Section~\ref{sec:quantized_neural_transport}.

\paragraph{Stress tensor.}

The full fermionic stress-energy readout is quadratic in the elementary neural fields:
\begin{align}
\label{eq:fermionic-neural-stress-tensor}
    T_F(z) =-\frac12\sum_{j=1}^{N}\sum_{a=1}^{K}\left(:\!\widetilde\psi_{ja}\partial\psi_{ja}\!:_0(z) + :\!\psi_{ja}\partial\widetilde\psi_{ja}\!:_0(z)\right).
\end{align}
As for the current bilinears, the elementary normal products are defined by subtraction of the regulated contractions, with the fermionic ordering kept fixed. \eqref{eq:fermionic-neural-stress-tensor} defines the stress-energy readout at finite width and fixed regulator.

In the continuum limit, this field admits the Sugawara decomposition
\begin{align}
    T_F &=\frac{1}{N+K}\sum_A:\!\mathcal{J}^A\mathcal{J}^A\!:\!_{\mathrm{aff}} + \frac{1}{N+K}\sum_a:\!\widetilde{\mathcal{J}}^a\widetilde{\mathcal{J}}^a\!:\!_{\mathrm{aff}} + \frac{1}{2NK}:\!\mathring{\mathcal{J}}\mathring{\mathcal{J}}\!:\!_{\mathrm{aff}},
    \label{eq:fermionic-stress-decomposition}    \\
    c_F &=\frac{K(N^2-1)}{N+K} +\frac{N(K^2-1)}{N+K} +1 = NK.
\end{align}
Here $:\!\cdot\!:\!_{\mathrm{aff}}$ denotes the affine normal product of currents, distinct from elementary fermion Wick ordering. The three terms are the row, column and number-current stresses. In particular, the stress tensor associated with the $SU(N)_K$ target is the first term, with central charge $K(N^2-1)/(N+K)$.

In the same continuum limit, a direct signed Wick-contraction calculation gives
\begin{align}
\label{eq:fermionic-stress-ope-verification}
T_F(z)\psi_{ja}(w)&\sim\frac{\psi_{ja}(w)}{2(z-w)^2}+\frac{\partial\psi_{ja}(w)}{z-w},\nonumber\\
T_F(z)T_F(w)&\sim\frac{NK}{2(z-w)^4}+\frac{2T_F(w)}{(z-w)^2}+\frac{\partial T_F(w)}{z-w},
\end{align}
showing that $\psi$ has weight $1/2$. A similar formula with $\widetilde\psi$ shows that it also has weight $1/2$.\bigskip

The finite-$L$ stress readout in \eqref{eq:fermionic-neural-stress-tensor} can likewise be contracted in the conditional Berezin functional.  For the same $S_L$ as \eqref{eq:neural_fermionic_one_feature_audit}, 
\begin{equation}
 \langle T_F(z)T_F(w)\rangle_{\mathrm{Gr}\mid\mathcal W} =\frac{NK}{2}\left[ (\partial_zS_L)(\partial_wS_L) -S_L\partial_z\partial_wS_L\right].
 \label{eq:neural_fermionic_tt_conditional}
\end{equation}
Define $\mathcal B[F]=(\partial_zF)(\partial_wF) -F\partial_z\partial_wF$; after averaging sampled frequencies,
\begin{equation}
 \mathbb E_{\Theta_F}[T_F(z)T_F(w)] =\frac{NK}{2}\left[\left(1-\frac1L\right)\mathcal B[C_\Lambda]+\frac1L\mathbb E_W\mathcal B[s_W]\right].
 \label{eq:neural_fermionic_tt_finite_width}
\end{equation}
The collision term is an explicitly computable single-frequency neural integral.  Since $C_\Lambda\to(z-w)^{-1}$ in the ordered limit, $\mathcal B[C_\Lambda]\to(z-w)^{-4}$ and hence the coefficient of the limiting fourth-order pole is $NK/2$. The full $T_FT_F$ operator-product identity still requires the Gaussian CAR/Fock Wick theorem and its normal-product prescription.

The neural Berezin functional specifies the ordered contractions, but a CAR representation additionally requires a mode polarization, vacuum and adjoint. Choose the Neveu--Schwarz vacuum and introduce operator fields
\begin{align}
    \widehat\psi_{ja}(z) &=\sum_{r\in\mathbb Z+1/2}\widehat\psi_{ja,r}z^{-r-1/2},&
    \widehat{\widetilde\psi}_{ja}(z)=\sum_{r\in\mathbb Z+1/2}\widehat{\widetilde\psi}_{ja,r}z^{-r-1/2}.
\end{align}

The hats distinguish these operators from the Grassmann-valued neural outputs. Their modes satisfy the anticommutation relations
\begin{align}
    \{ \widehat\psi_{ja,r},\widehat{\widetilde\psi}_{kb,s}\}&=\delta_{jk}\delta_{ab}\delta_{r+s,0},\qquad\{ \widehat\psi_{ja,r},\widehat\psi_{kb,s}\} = \{ \widehat{\widetilde\psi}_{ja,r},\widehat{\widetilde\psi}_{kb,s}\} =0.
    \label{eq:fermionic-CAR-not-Grassmann}
\end{align}
Positive modes annihilate the vacuum, and the adjoint operation is $\bigl(\widehat\psi_{ja,r}\bigr)^\dagger=\widehat{\widetilde\psi}_{ja,-r}$.

We note here that the CAR operators above are not the Grassmann integration generators: Berezin integration plays the role of ordered Wick contractions, whereas the vacuum and adjoint define the additional operator completion.\bigskip

Finally, we conclude this section with the following remark. Finite mode truncation provides a useful comparison with the annular Fourier regulator. For $|z|>|w|$, retaining $M$ positive-frequency mode pairs gives the contraction 
\begin{align}
\label{eq:Hardy-versus-RFF-regulator}
    G_M(z,w) = \sum_{n=0}^{M-1}z^{-n-1}w^n =\frac{1-(w/z)^M}{z-w}\longrightarrow\frac1{z-w}.
\end{align}
This is the finite Hardy contraction. It can be used in a CAR construction based on a finite one-particle subspace, with more general choices described by the corresponding projectors. The Hardy mode cutoff $M$ and the neural annular-frequency regulator have different finite kernels, even though both approach the Cauchy contraction.  Finite-mode CAR therefore need not preserve the affine algebra or WZW blocks. The additional affine modules and charged-sector maps needed for conformal blocks are developed in Section~\ref{sec:quantized_neural_transport}.

\section{Neural readouts, fermionic state spaces and four-point transport}
\label{sec:quantized_neural_transport}

We now construct angular and charged readouts from the fermionic neural architecture. Their limiting correlations provide the starting point for a CAR/Fock reconstruction of descendant propagation, which we compare with WZW four-point transport.

Throughout, we distinguish the finite neural observables from the CAR/Fock completion of their limiting correlations. The latter provides the state spaces and pairings used in the sewing calculation. Identifying its individual block outputs directly with limits of finite-neural charged observables requires additional probe constructions and convergence control, specified below.

Using the two-stage expectation defined in~\eqref{eq:fermionic_full_expectation_definition}, we can calculate the finite-width corrections to the Gaussian Wick factorization explicitly. At fixed $\mathcal W$, higher field moments are Pfaffians of the empirical kernel~\eqref{eq:fermionic_empirical_kernel}. Averaging a Pfaffian over frequencies generally differs from taking the Pfaffian of the averaged kernel. Consequently, a two-point output that is exact at every width does not imply exact Gaussian higher-point outputs. 

For a single channel, each sampled frequency contributes a rank-two kernel. Its $2r$-point neural moment equals the Gaussian/Wick moment built from the averaged two-point kernel, multiplied by $s_r(L):=\frac{L^{(r)}}{L^r}$, where the factor $L^{(r)} = \tfrac{\Gamma(L+1)}{\Gamma(L-r+1)}$ counts ordered choices of $r$ distinct features.  Equivalently, the single-channel source functional is $[1+Q(\eta)/L]^L$, where $Q(\eta)$ is its quadratic source term. Non-Gaussian connected source cumulants begin at order $L^{-1}$. In particular, the four-point output is suppressed by $s_2(L) = 1-1/L$ even though the two-point output is independent of $L$. In the Cauchy limit this gives an order-$L^{-1}$ mismatch with the standard free-fermion OPE normalization. The exact identity, proof and separate symmetry checks are given in Appendix~\ref{app:finite_width_cft_constraints}.

The shared frequency pool also couples the finite-width corrections across channels. When two or more factors depend on the same sampled frequency, the resulting terms modify direct and exchanged channel contractions differently. Their dependence on insertion points generally cannot be absorbed into a single normalization.  On the UV-removal path $p_-=0$, $p_+=\infty$, $\epsilon_F\downarrow0$ with $0<\alpha<1$, the generic mixed-channel corrections scale as $(L\epsilon_F^{2\alpha})^{-1}$ and yield a nonzero regulator-subtracted dilation Ward defect at leading order. Thus $L\epsilon_F^{2\alpha}\to\infty$ is necessary for the bare Gaussian parent four-point limit along this path. Appendix~\ref{app:neural_width_projective_uv} gives the precise domain and formulas. The condition applies to the four-point correlations of the parent neural ensemble along this joint width--regulator limit.

\subsection{Angular readouts and the limiting fermionic state space}
\label{sec:fermionic_state_geometry}

The first extension of the output is a fixed linear projection onto angular harmonics. At $z=Re^{i\theta}$, define
\begin{align}
    \widehat\chi_{L,n}^A(R;\Theta_F)    &=\int_0^{2\pi}\frac{d\theta}{2\pi}\,    e^{in\theta}\chi_L^A(Re^{i\theta};\Theta_F),\qquad n\in\bZ.
    \label{eq:fermionic_neural_angular_readout}
\end{align}
This readout uses the same feature frequencies and Grassmann coefficients as the original field. Applying the Jacobi--Anger expansion to \eqref{eq:fermionic_features_recall} gives
\begin{subequations}  
\label{eq:fermionic_feature_bessel_modes}
    \begin{align}
    \widehat U_{j,n}(R) &=\sqrt{2W_j}\,i^n \mathrm{J}_n(2p_jR)e^{-in\varphi_j}\mathbf 1_{\{n\ \mathrm{even}\}},\\\widehat V_{j,n}(R)&=\sqrt{2W_j}\,i^n \mathrm{J}_n(2p_jR)e^{-in\varphi_j}\mathbf 1_{\{n\ \mathrm{odd}\}},
  \end{align}
\end{subequations}
and therefore the explicit mode network
\begin{align}
    \widehat\chi_{L,n}^A(R;\Theta_F) &=\frac{\sqrt{iZ_{\alpha,F}}}{\sqrt L}\sum_{j=1}^{L}p_j^{-\alpha}\left[\xi_{Aj}\widehat U_{j,n}(R)+\bar\xi_{Aj}\widehat V_{j,n}(R)\right].
    \label{eq:fermionic_neural_mode_network}
\end{align}
Thus selecting an angular mode replaces the trigonometric features by Bessel features with a specified angular phase. For a fixed network sample, $n$ selects the angular
mode and $R$ sets the radius of the projection circle.

We compute the correlations by the same two-stage average as for the original fields. Berezin contraction at fixed frequencies gives
\begin{align}
    &\left\langle\widehat\chi_{L,n}^A(R)\widehat\chi_{L,m}^B(S)   \right\rangle_{\mathrm{Gr}\mid\mathcal W}=\delta^{AB}\frac{iZ_{\alpha,F}}{L}\sum_{j=1}^{L}p_j^{-2\alpha}\left[\widehat U_{j,n}(R)\widehat V_{j,m}(S)-\widehat V_{j,n}(R)\widehat U_{j,m}(S)\right].
    \label{eq:fermionic_conditional_mode_covariance}
\end{align}
The even/odd harmonic split makes the antisymmetry explicit. The phases in \eqref{eq:fermionic_feature_bessel_modes}, together with $W_j=p_je^{i\varphi_j}$, produce $e^{i(1-n-m)\varphi_j}$. Averaging over the uniformly distributed frequency angle therefore selects $n+m=1$. With the regulated spectral prior, the complete two-point output is 
\begin{align}
&\bE_{\Theta_F}\!\left[\widehat\chi_{L,n}^A(R)\widehat\chi_{L,m}^B(S)\right]=-2\delta^{AB}(-1)^n\delta_{n+m,1}\int_{p_-}^{p_+}e^{-\epsilon_F p} \mathrm{J}_n(2pR)\mathrm{J}_m(2pS)\,dp.
\label{eq:fermionic_angular_bessel_covariance}
\end{align}
This separates two architectural consequences: rotational invariance fixes which mode pairs survive the ensemble average, while the radial integral determines their regulator-dependent amplitudes. The selection rule is exact at finite width after averaging.

Upon regulator removal, the averaged field kernel becomes $\delta^{AB}/(z-w)$.
For distinct radii $R,S>0$, removing the regulator gives
\begin{align}
\lim_{\mathrm{reg.}\to0}\bE_{\Theta_F}\!\left[\widehat\chi_{L,n}^A(R)\widehat\chi_{L,m}^B(S)\right]
=\delta^{AB}\delta_{n+m,1}
\begin{cases}
S^{n-1}/R^n, & R>S,\quad n\geq1,\\
-R^{-n}S^{n-1}, & R<S,\quad n\leq0,\\
0, & \text{otherwise}.
\end{cases}
\label{eq:fermionic_outer_angular_limit}
\end{align}

These formulas describe outputs of the same angular readout evaluated in the two radial orderings. For the symmetric weight-$\frac12$ assignment, set $r=n-\frac12$ and $s=m-\frac12$. At finite width, introduce the radially rescaled angular readout
\begin{align}
    a_{L,r}^A(R;\Theta_F)&=R^{r+\frac12}\widehat\chi_{L,r+\frac12}^A(R;\Theta_F),\qquad r\in\mathbb Z+\tfrac12.
    \label{eq:fermionic_rescaled_angular_mode}
\end{align}
The difference between its two limiting cross-radius pairings is then
\begin{align}
    \eta_{rs}^{AB}&=\delta^{AB}\delta_{r+s,0}.    \label{eq:fermionic_architecture_mode_pairing}
\end{align}
Thus, the elementary fermionic pairing has been recovered from the parameter-space correlations of explicit neural outputs. The half-integer labels use the untwisted, symmetric weight-$\frac12$ mode convention.

To obtain the many-particle state space, we also need higher mode correlations. At fixed frequencies these are Pfaffians of \eqref{eq:fermionic_conditional_mode_covariance}. Taking $L\to\infty$ at fixed regulator makes the empirical pairing converge to its spectral expectation and yields Gaussian Grassmann correlators. After pairing the real channels into complex fields, the pairings between negative-mode states and their positive-mode duals are determinants. The resulting exterior algebra supplies the CAR/Fock completion described in Section~\ref{sec:explicit-fermionic-WZW}. Its vacuum is the degree-zero operator reference state. This construction organizes the limiting correlations; it introduces no additional sampled neural parameter and does not impose Gaussian factorization at finite $L$.

\subsection{Charged readouts and channel resolution}
\label{sec:architecture_charged_outputs}

We next construct outputs that carry fermion number and resolve the row and column channel content. Use the complementary complex fields $\psi_L^I,\widetilde\psi_{L,I}$ of \eqref{eq:explicit-complex-fermions}, with $I=(j,a)$ and $E=\mathbb C^N\otimes\mathbb C^K$. A fixed tensor $v\in\Lambda^qE$ specifies the degree-$q$ polynomial readout
\begin{align}
\mathcal V^+_{q,L}(v;z) &=\frac{1}{q!}\,v^{I_1\cdots I_q}\widetilde\psi_{L,I_1}(z)\cdots\widetilde\psi_{L,I_q}(z),
\label{eq:transport_finite_charged_readout}
\end{align}
The complementary negative-charge readout $\mathcal V^-_{q,L}$ uses the $\psi_L$ fields and dual channel tensors. These products require no self-contraction subtraction as same-species contractions vanish, while mixed coincident composites require separate prescriptions. 

The sign of the charge follows from the current convention $\mathcal{J}^M=:\!\widetilde\psi M\psi\!:_0$. In the continuum, \begin{equation}
 \mathcal{J}^M(z)\widetilde\psi_I(w)\sim \frac{M^J{}_I\widetilde\psi_J(w)}{z-w},\qquad \mathcal{J}^M(z)\psi^I(w)\sim -\frac{M^I{}_J\psi^J(w)}{z-w}.
 \label{eq:transport_charged_orientation}
\end{equation}
Thus $\mathcal V^+$ has positive fermion number and transforms in $\Lambda^qE$. The complementary negative-charge readout transforms in the dual space.

For $(N,K)=(2,3)$, the six complex field channels form a $2\times3$ array. Antisymmetry of two charge-one insertions places their channel tensor $v_1\wedge v_2$ in
\begin{align}
    \Lambda^2(\mathbb C^2\otimes\mathbb C^3)&\cong\bigl(\operatorname{Sym}^2\mathbb C^2\otimes\Lambda^2\mathbb C^3\bigr) \oplus \bigl(\Lambda^2\mathbb C^2\otimes\operatorname{Sym}^2\mathbb C^3\bigr)\cong(\mathbf3,\overline{\mathbf3})\oplus(\mathbf1,\mathbf6).
    \label{eq:charge_two_architecture_decomposition}
\end{align}
Thus the architecture already supplies two allowed row--column channels at the lowest grade. They can be separated by fixed linear maps: the row $SU(2)$ quadratic Casimir $\mathsf C_{2,R}$ has eigenvalues $2$ and $0$ on the two summands, giving the projectors
\begin{align}
    \Pi_{(\mathbf3,\overline{\mathbf3})} &=\tfrac12\mathsf C_{2,R}, &\Pi_{(\mathbf1,\mathbf6)} &=\mathbf1-\tfrac12\mathsf C_{2,R},
    \label{eq:architecture_channel_projectors}
\end{align}
The projected channel tensors are
\begin{align}
    s_\mu(v_1,v_2) &=\Pi_\mu(v_1\wedge v_2),\qquad\mu\in\{(\mathbf3,\overline{\mathbf3}), (\mathbf1,\mathbf6)\}.
    \label{eq:architecture_channel_seed}
\end{align}
These fixed tensors define the charge-two neural readouts
\begin{align}
    \mathcal V^+_{2,L}(s_\mu(v_1,v_2);z) &=\frac12\,s_\mu(v_1,v_2)^{IJ}\widetilde\psi_{L,I}(z)\widetilde\psi_{L,J}(z).
    \label{eq:section3_projected_charged_readout}
\end{align}
The projectors follow directly from the row $SU(2)$ action on the fermion channels. The four-point amplitude also requires the descendant couplings and propagation constructed below.

The external charges $(1,1,2,2)$ used below sum to six. Their product has zero parameter-space expectation unless a compensating charge-$-6$ probe is included. Denote a chosen probe by $\mathcal B^-_{6,L}$; for each such choice, the finite-neural observable is
\begin{align}
    &\mathscr C_{L,\Lambda_F}[\mathcal B^-_6;\mathbf v,\mathbf z] =\int dP_{\mathcal W}\,\Bigl\langle\mathcal B^-_{6,L}\,\mathcal V^+_{1,L}(v_1;z_1)\mathcal V^+_{1,L}(v_2;z_2)    \mathcal V^+_{2,L}(v_3;z_3)\mathcal V^+_{2,L}(v_4;z_4)\Bigr\rangle_{\mathrm{Gr}\mid\mathcal W}.
    \label{eq:transport_finite_neural_four_point}
\end{align}
The probe's spatial or mode dependence is part of the definition of this correlator. For each specified probe, we contract using $S_{L,\mathcal W}$ at fixed $\mathcal W$ and then average over frequencies. Charge balance alone does not select an individual conformal block.

A local $SU(6)$-singlet probe at a fixed point $y$ is
\begin{align}
    \mathcal B^-_{6,L}(y)&=\frac1{6!}\varepsilon_{I_1\cdots I_6}\psi_L^{I_1}(y)\cdots\psi_L^{I_6}(y), \qquad \varepsilon_{12\cdots6}=1.   \label{eq:section3_local_neural_singlet_probe}
\end{align}
The probe is invariant under the $SU(6)$ action on the channel array. With the displayed field ordering and $\varepsilon_{12\cdots6}=1$, its conditional contraction is
\begin{align}
&\left\langle\mathcal B^-_{6,L}(y)\prod_{a=1}^{4}\mathcal V^+_{q_a,L}(v_a;z_a)\right\rangle_{\mathrm{Gr}\mid\mathcal W}\nonumber\\
&\qquad=-\varepsilon(v_1\wedge v_2\wedge v_3\wedge v_4)\prod_{a=1}^{4}S_{L,\mathcal W}(y,z_a)^{q_a},\qquad(q_1,q_2,q_3,q_4)=(1,1,2,2).
\label{eq:section3_local_probe_conditional_factorization}
\end{align}
The minus sign comes from the factor $(-1)^{6(6-1)/2}$: all six negative-charge fields precede the six positive-charge fields in the Berezin contraction. The factorials in the charged readouts give the standard exterior-product normalization.

At fixed $\mathcal W$, let $H_{L,\mathcal W}\cong\mathbb C^{2L}$ be the formal coefficient space of the two Fourier features at each sampled frequency. Define the feature-evaluation vector and skew form 
\begin{align}
    f_{\mathcal W}(z) &=\sqrt{\frac{i Z_{\alpha,F}}{L}}\, \bigl(p_j^{-\alpha}U_j(z),p_j^{-\alpha}V_j(z)\bigr)_{j=1}^{L}, &\mathsf J&=\bigoplus_{j=1}^{L}
      \begin{pmatrix}0&1\\-1&0\end{pmatrix}.
    \label{eq:section3_feature_coefficient_space}
\end{align}
The field coefficients in this basis are linear combinations of the existing Grassmann generators, not additional random variables. By \eqref{eq:fermionic_empirical_kernel}, their conditional pairing is exactly
\begin{equation}
    S_{L,\mathcal W}(z,w)=f_{\mathcal W}(z)^{\mathsf T}\mathsf Jf_{\mathcal W}(w).
    \label{eq:section3_feature_skew_pairing}
\end{equation}
This coefficient space has dimension $2L$ before any linear-dependence reduction and is distinct from the Hardy mode space of Section~\ref{sec:explicit-fermionic-WZW}.  Let us define $\kappa_{\cW}\in H_{L,\cW}^*$ by  $\kappa_{\cW}(z)(h)=h^{\mathsf T}\mathsf Jf_{\cW}(z)$.  The negative-charge fields transform in $E^*$, with $E=\bC^6$, so the $SU(6)$-invariant six-field backgrounds belong to
\begin{equation}
    [\Lambda^6(H_{L,\mathcal W}\otimes E^*)]^{SU(6)}\cong\operatorname{Sym}^6(H_{L,\mathcal W})\otimes\det(E^*),
    \label{eq:section3_local_probe_invariant_space}
\end{equation}
by skew Howe duality.  The chosen determinant covector $\varepsilon\in\det(E^*)$ identifies an invariant background with a symmetric mod tensor $b\in\operatorname{Sym}^6(H_{L,\cW})$.  We normalize this identification so that $b = f_{\cW}(y)^{\odot 6}$ gives the local probe in \eqref{eq:section3_local_neural_singlet_probe}. Its conditional contraction with the four external charged readouts is then 
\begin{align}
    &-\varepsilon(v_1\wedge v_2\wedge v_3\wedge v_4)\big\langle b\,\kappa_{\cW}(z_1)\odot\kappa_{\cW}(z_2)\odot\kappa_{\cW}(z_3)^{\odot2}\odot\kappa_{\cW}(z_4)^{\odot2}\big\rangle
\end{align}
Here $\odot$ denotes the symmetric tensor product, with its evaluation pairing normalized so that a pure sixth power evaluates as the product of the six linear functionals. The local choice of $b$ recovers \eqref{eq:section3_local_probe_conditional_factorization}. General mode tensors describe six-field backgrounds with the same determinant channel tensor.

The local $SU(6)$-invariant probe therefore produces only one independent spatial--channel structure. Within the full $SU(6)$-invariant six-field class, changing spatial or angular modes leaves a single channel tensor. Restricting invariance to the physical $SU(2)\times SU(3)$ row-column symmetry permits additional channel structures. The latter already have a ten-dimensional invariant space when $\dim H_L=2$, compared with seven dimensions under full $SU(6)$ invariance. Section~\ref{sec:duality_preparation_learning} constructs such angular charged preparation readouts and tests controlled approximations of both channels. This does not identify the original bare pointwise charge-six functional with an individual block.

\subsection{Reconstructing descendant propagation in the continuum completion}
\label{sec:architecture_descendant_reconstruction}

We reconstruct four-point propagation by sewing the descendant states above each channel seed, using their three-point couplings and bilinear pairings. The calculation takes place in the CAR/Fock completion of the limiting neural correlations.

Starting from a seed $s_\mu$, negative modes of the commuting row and column affine currents generate descendant spaces $\mathcal D_{\mu,n}^R$ and $\mathcal D_{\mu,n}^C$ at grade $n$. The current levels are $K$ and $N$, respectively, by the bilinear contraction established in Section~\ref{sec:explicit-fermionic-WZW}. For either sector, the incoming and outgoing dual probes specify the matrix elements to be computed from the limiting Cauchy contraction. 

Chiral  sewing pairs the intermediate module $\cH_\mu=\bigoplus_{n\geq 0}\cH_{\mu,n}$ with its restricted graded dual $\cH_\mu^\vee = \bigoplus_{n\geq 0}\cH_{\mu,n}^*$.  We use the evaluation pairing $B_{\rm sew}(\eta,u)=\eta(u)$ and the outgoing local coordinate $\zeta = 1/z$.  The contragradient current and Virasoro actions are fixed by
\begin{align}
    B_{\rm sew}(J_m^{a,\vee}\eta,u) = -B_{\rm sew}(\eta,J_{-m}^a u),\qquad B_{\rm sew}(L_m^\vee\eta,u) = B_{\rm sew}(\eta,L_{-m}u).
    \label{eq:sewing_radial_duality}
\end{align}
The current sign comes from the weight-one inversion transformation.  At grade zero, an orthonormal primary basis $\{e_{\mu,a}\}$ and its dual $\{e_\mu^a\}$ satisfy $B_{\rm sew}(e_{\mu}^a,e_{\mu,b}) = \delta^a{}_b$.  I particular the outgoing dual of a $\overline{\mathbf3}$ channel is a $\mathbf3$ channel, and that of a $\mathbf6$ channel is $\overline{\mathbf6}$.

The Fock inner product is antilinear in its first argument and has radial adjoints $J_m^a{}^\dagger = J_{-m}^a$ for Hermetian generators and $L_m^\dagger = L_{-m}$.  The map $\mathfrak r_\mu u= (u,\cdot)_{\rm F}$ identifies physical states antilinearly with their duals.  Let $\iota_{\mu,n}:\cD_{\mu,n}\to \cH_{\mu,n}$ evaluate the Fock space descendants, and set $\Xi_I^{(\mu,n),\vee} = \mathfrak r_\mu\iota_{\mu,n}\Xi_I^{(\mu,n)}$.  The sewing matrix and physical quotient are 
\begin{align}
    G^{(\mu,n)}_{IJ}=B_{\mathrm{sew}}\!\left(\Xi_I^{(\mu,n),\vee},\iota_{\mu,n}\Xi_J^{(\mu,n)}\right),\qquad
\overline{\mathcal D}_{\mu,n}=\mathcal D_{\mu,n}/\ker\iota_{\mu,n}.
\end{align}
In these bases $G^{(\mu,n)}$ is the Gram matrix on the Fock space.  Its null vectors are exactly the descendant combinations whose Fock norm vanishes, so removing them gives the same physical quotient.  We denote the resulting invertible matrix by $\overline G^{(\mu,n)}$. Sewing remains complex bilinear in the independent incoming and outgoing data.

The limiting current action determines the descendant couplings. Schematically, for a row current,
\begin{align}
    [R_m(A),V_q(v;u)] &=u^m V_q(\rho_q(A)v;u),
    \label{eq:charged_row_action}
\end{align}
with an analogous column relation. Once the external readouts, outgoing dual probes, charge-balancing prescription and seed normalization are chosen, this relation gives a recursion at each grade. On the physical quotient, the incoming functional $\mathcal A_{12}^{(\mu,n)}$ acts on the dual intermediate space, while the outgoing functional $\mathcal A_{34}^{(\mu,n)}$ acts on the intermediate space. Their components are
\begin{subequations}\label{eq:architecture_descendant_couplings}
\begin{align}
    \bigl(C_{\rm L}^{(\mu,n)}\bigr)_I&=\mathcal A_{12}^{(\mu,n)}\!\left(\Xi_I^{(\mu,n),\vee}\right),\\
\bigl(C_{\rm R}^{(\mu,n)}\bigr)_I&=\mathcal A_{34}^{(\mu,n)}\!\left(\overline\Xi_I^{(\mu,n)}\right),
\end{align}
\end{subequations}
where $\overline\Xi_I^{(\mu,n)}$ denotes the corresponding quotient basis vector. The channel seed in \eqref{eq:architecture_channel_seed} specifies the grade-zero channel dependence, with its scalar normalization set by the chosen probes and seed convention. Higher-grade couplings follow from current-mode recursion without fitting KZ coefficients.

We use $z_{ab}=z_a-z_b$ and the cross ratio
\begin{equation}
    x=\frac{z_{12}z_{34}}{z_{13}z_{24}},
    \label{eq:section3_cross_ratio_convention}
\end{equation}
with the standard conformal frame $(z_1,z_2,z_3,z_4)=(0,x,1,\infty)$ (the insertion at infinity is defined by the usual conformal limit). In this convention $x\to0$ is the $12$-channel degeneration and $x\to1$ the $23$-channel degeneration. A grade-$n$ descendant above channel $\mu$ has Sugawara $L_0$ eigenvalue $h_\mu+n$. Pairing the two coupling vectors through the inverse intermediate-state form and summing over grades gives
\begin{align}
    F_\mu^{(N_{\max})}(x) &=x^{h_\mu-h_1-h_2}\sum_{n=0}^{N_{\max}}x^n\bigl(C_{\rm R}^{(\mu,n)}\bigr)^{\mathsf T}\bigl(\overline G^{(\mu,n)}\bigr)^{-1}C_{\rm L}^{(\mu,n)}.
    \label{eq:architecture_four_point_series}
\end{align}
Each coefficient is independent of the basis chosen on the null quotient. The superscript $\mathsf T$ denotes ordinary transpose, as required by the complex-bilinear sewing form.

Two distinct resources enter the construction. The cutoff $N_{\max}$ counts descendant grades, while $L$ counts sampled neural features. Equation~\eqref{eq:architecture_four_point_series} is calculated in the limiting CAR/Fock theory. Obtaining it as a limit of \eqref{eq:transport_finite_neural_four_point} requires matching probes and convergence of their composite correlators.

\subsection{Comparing the reconstructed sectors with WZW transport}
\label{sec:architecture_four_point_transport}

The six-channel architecture gives paired row and column sectors. Their two-component sewn series, $\mathbf F_{2,3}(x)$ and $\mathbf F_{3,2}(x)$, inherit the channel correspondence 
\begin{align}
    j=1&\longleftrightarrow\overline{\mathbf3}, & j=0&\longleftrightarrow\mathbf6.
    \label{eq:wzw_channel_dictionary}
\end{align}
from \eqref{eq:charge_two_architecture_decomposition}. The limiting traceless row currents realize $\widehat{\mathfrak{su}}(2)_3$, and the traceless column currents realize $\widehat{\mathfrak{su}}(3)_2$. Together with the common number current, these belong to the six-complex-fermion parent $\widehat{\mathfrak{u}}(6)_1$. The two Sugawara central charges sum with the number-current central charge $1$ to give $6$.

For the charge-one and selected charge-two insertions, the external row and column representations are
\begin{align}
    \widehat{\mathfrak{su}}(2)_3:\quad&\left(\tfrac12,\tfrac12,1,1\right), & \widehat{\mathfrak{su}}(3)_2:\quad &\left(\mathbf3,\mathbf3, \overline{\mathbf3},\overline{\mathbf3}\right).
    \label{eq:architecture_frozen_externals}
\end{align}
The quadratic Casimirs of the limiting currents give row-channel exponents $1/10$ for $j=1$ and $-3/10$ for $j=0$, paired with column-channel exponents $-4/15$ for $\overline{\mathbf3}$ and $2/15$ for $\mathbf6$. These exponents follow from the selected channels and current operators before solving a WZW transport equation. Their relative splittings agree in magnitude, with the ordering reversed between the paired sectors.

For the specified external representations, WZW chiral block vectors satisfy the Knizhnik--Zamolodchikov system
\begin{align}
    (k_{\mathrm{aff}}+h^\vee)\frac{d}{dx}\mathbf F(x) &=\left(\frac{\Omega_{12}}{x}+\frac{\Omega_{23}}{x-1}\right)\mathbf F(x),
    \label{eq:architecture_kz_equation}
\end{align}
Here $k_{\mathrm{aff}}=K$ in the row theory and $k_{\mathrm{aff}}=N$ in the column theory, while $\Omega_{ij}=2\sum_a t_i^at_j^a$ acts on the appropriate invariant tensor space. We use the Hermitian generator normalization $\operatorname{tr}(t^at^b)=\delta^{ab}/2$ of Section~\ref{sec:background}. The KZ matrices are fixed by these WZW representation data and provide a comparison for the sewn series; they are not fitted to its coefficients.

Affine sewing performed independently of the KZ equations agrees with the KZ recurrence through grade eight for $\widehat{\mathfrak{su}}(2)_3$ and grade four for $\widehat{\mathfrak{su}}(3)_2$. At those grades, the descendant forms also agree with the independent Shapovalov/null-state calculation.

The shared number-current sector supplies an additional constraint. A lowest-grade charge-$q$ sector has fermionic weight $q/2$, of which $q^2/(2NK)$ belongs to the number current. For the charge-two intermediate sector of the six-fermion parent, the number-current weight relative to two charge-one insertions is $1/6$. The paired non-Abelian exponents consequently sum to $-1/6$. The compensating scalar factor is $x^{1/6}(1-x)^{1/3}$: its exponents are the number-current charge products $q_1q_2/(NK)$ and $q_2q_3/(NK)$ for $(q_1,q_2,q_3,q_4)=(1,1,2,2)$.

To combine the row and column block outputs, introduce a bilinear pairing $P$ between their block spaces. Write the two KZ systems as $d\mathbf F/dx=(R_0/x+R_1/(x-1))\mathbf F$. Requiring the bilinear pairing, multiplied by $x^{1/6}(1-x)^{1/3}$, to remain constant under KZ transport gives
\begin{subequations}
\begin{align}
    \frac16P+\bigl(R_0^{(2,3)}\bigr)^{\mathsf T}P+P R_0^{(3,2)} &=0,
    \label{eq:architecture_pairing_horizontality}\\
    \frac13P+\bigl(R_1^{(2,3)}\bigr)^{\mathsf T}P+P R_1^{(3,2)}  &=0.
    \label{eq:architecture_pairing_horizontality_x1}
\end{align}
\end{subequations}

The condition at $x=0$ leaves two channelwise normalization constants. The condition at $x=1$ fixes their ratio, leaving a one-dimensional solution space for $P$. Appendix~\ref{app:charged_level_rank_pairing} verifies these two conditions explicitly. The resulting $P$ was held fixed when the block series were separately reconstructed by affine sewing; consistency with their local expansions and monodromy was checked without refitting. 

This comparison separates what the channel architecture supplies from what the transport calculation determines. The architecture fixes the paired grade-zero channels, and the common number current fixes the scalar compensation. These $s$-channel data leave two channelwise constants; their ratio is fixed by the $t$-channel residue $R_1=\Omega_{23}/5$.

The NNFT construction thus gives explicit finite-neural mode and charged readouts, their parameter-space correlations, and fixed projectors onto the two lowest-grade channels. Its limiting mode correlations support a CAR/Fock reconstruction whose finite-grade outputs pass the stated WZW transport checks. Section~\ref{sec:duality_preparation_learning} and Appendix~\ref{app:lr_controlled_preparations} supply explicit charged angular probe families, physical null descent and controlled fixed-grade width/regulator limits. With specified compatible affine Ward data, a descendant-tail bound gives an angular approximation sequence for the selected blocks. This uses the stated sewing input; a direct identification of the original bare pointwise charge-six functional is separate.

\section{Finite-width presentations and regulator design}
\label{sec:finite_width_regulator_design}

In Section~\ref{sec:quantized_neural_transport}, the common fermionic parent related the charged channels of $SU(2)_3$ and $SU(3)_2$ and provided the operator construction used to compare their descendant expansions with WZW transport. We now use the associated coset construction to compare two neural presentations of the same limiting current sector. The target current algebra and stress-tensor correlations are held fixed, while the elementary neural fields and composite readouts used to represent them are varied.

We compare the two constructions at the same feature width using the same selected correlation functions and their common continuum reference. Within each construction, the frequency distribution, regulator and cutoffs provide further choices. We report approximation errors for specified observables and numerical error measures.\bigskip

For $N,K\geq2$, the conformal embedding
\begin{align}
    \widehat{\mathfrak{su}}(N)_K \oplus \widehat{\mathfrak{su}}(K)_N \subset \widehat{\mathfrak{su}}(NK)_1
    \label{eq:fw_conformal_embedding}
\end{align}
places two commuting current sectors inside a common parent. In the integrable WZW theory, their stress tensors add to the parent stress tensor, although their currents do not exhaust the parent algebra. The associated coset identification is
\begin{align}\label{eq:fw_same_sector_coset}
    \frac{SU(NK)_1}{SU(K)_N} \simeq SU(N)_K.
\end{align}
Here $SU(M)_k$ denotes the level-$k$ WZW chiral theory. The coset consists of the chiral fields of the parent that commute with the embedded $\widehat{\mathfrak{su}}(K)_N$ current algebra. For this embedding, the branching rules identify that chiral algebra with $SU(N)_K$ \cite{Ostrik:2014LevelRankTensorCategories}.

We focus on the direct $SU(2)_3$ presentation and the $SU(6)_1/SU(3)_2$ coset presentation, both constructed from the bosonic substrate of
Section~\ref{sec:bosonicsubstrate}. The direct presentation requires one scalar field and one commuting $\beta\gamma$ pair, while the coset presentation uses five scalar fields and fifteen ghost pairs.

We first specify the coset currents and the coset stress-energy readout. We then compare their finite-width correlation functions with those of the continuum limit under changes of the parameter distribution, regulator and architecture. General Wakimoto formulas and the parent-based denominator normal-product prescription are given in Appendix~\ref{app:sunk_wakimoto}.

\subsection{Neural coset construction}

The coset presentation of $SU(6)_1/SU(3)_2$ starts from the level-one $\mathfrak{su}(6)$ Wakimoto construction, whose elementary neural fields are
\begin{align}
    \phi_{\mathrm P}^{u}, \qquad u=1,\ldots,5, \qquad \bigl(\beta_{\mathrm P}^{pq},\gamma_{\mathrm P}^{pq}\bigr), \qquad 1\leq p<q\leq6.
    \label{eq:fw_parent_neural_fields}
\end{align}
These fields follow the architectures and parameter laws of Section~\ref{sec:bosonicsubstrate}, including the shared frequency samples across the scalar and ghost sectors.\footnote{The bosonic parent is $SU(6)_1$, rather than the $U(6)_1$ fermionic parent used in the preceding section. There is therefore no additional number-current sector in this construction. }

Let ${\mathcal{J}^{\mathrm P}}^{\mu}(z)$, $\mu=1,\ldots,35$, denote the current readouts obtained by evaluating the general Wakimoto formulas at $(N,K)=(6,1)$. Choose Hermitian generators $s^\mu$ of $\mathfrak{su}(6)$ with $\operatorname{tr}_6(s^\mu s^\nu)=\delta^{\mu\nu}/2$. The three currents retained by the coset can then be written as the following linear superposition of parent currents,
\begin{align}
    \mathfrak{J}_{\mathrm C}^a(z) := 2\sum_{\mu=1}^{35} \operatorname{tr}_6\!\left[ s^\mu(t^a\otimes\mathbf1_3)\right]{\mathcal{J}^{\mathrm P}}^{\mu}(z), \qquad a=1,2,3,
    \label{eq:fw_embedded_currents}
\end{align}
where $t^a=\sigma^a/2$ are the Hermitian generators of $\mathfrak{su}(2)$. To define the coset stress-energy subtraction, we also require the eight denominator currents, parametrizing the $\widehat{\mathfrak{su}}(3)_2$ algebra, 
\begin{align}
    \mathfrak{J}_{\mathrm{D}}^A(z) := 2\sum_{\mu=1}^{35}\operatorname{tr}_6\!\left[s^\mu(\mathbf1_2\otimes T^A)\right]{\mathcal{J}^{\mathrm P}}^{\mu}(z),\qquad A=1,\ldots,8,
    \label{eq:fw_denominator_currents}
\end{align}
where $T^A$ are Hermitian $\mathfrak{su}(3)$ generators normalized by $\operatorname{tr}_3(T^AT^B)=\delta^{AB}/2$. Both sets of currents use the same parent fields, with fixed coefficients given by the traces.

Using factorization of the tensor-product trace,
\begin{align}
    \operatorname{tr}_6\!\left[(t^a\otimes\mathbf1_3)(t^b\otimes\mathbf1_3)\right] = \operatorname{tr}_2(t^at^b)\operatorname{tr}_3(\mathbf1_3) = \frac32\delta^{ab},
    \label{eq:fw_level_three_trace}
\end{align}
and the level-one parent contraction therefore gives
\begin{align}
    \bE\!\left[\mathfrak{J}_{\mathrm C}^a(z)\mathfrak{J}_{\mathrm C}^b(w)\right] \longrightarrow\frac{3\delta^{ab}}{2(z-w)^2}.
    \label{eq:fw_coset_current_normalization}
\end{align}
Together with the $\mathfrak{su}(2)$ commutation relations, this shows that $\mathfrak{J}_{\mathrm C}^a$ generate a level-3 current algebra. Similar arguments hold for the denominator currents $\mathfrak{J}_{\mathrm{D}}^A$ and their level-2 algebra.

The fixed projections can be expanded directly in the parent neural fields. In the triangular-coordinate convention used for the general Wakimoto construction, the raising combination is
\begin{align}
    \mathfrak{J}_{\mathrm C}^1+i \mathfrak{J}_{\mathrm C}^2={}&\beta_{\mathrm P}^{14}+\beta_{\mathrm P}^{15}\gamma_{\mathrm P}^{45}+\beta_{\mathrm P}^{16}\gamma_{\mathrm P}^{46} + \beta_{\mathrm P}^{25}+\beta_{\mathrm P}^{26}\gamma_{\mathrm P}^{56}+\beta_{\mathrm P}^{36},
    \label{eq:fw_coset_raising_readout}
\end{align}
while the Cartan current is
\begin{align}
    \mathfrak{J}_{\mathrm C}^3={}&i\sqrt7\left(\frac{\sqrt3}{2}\partial\phi_{\mathrm P}^{3} +\frac{3}{2\sqrt5}\partial\phi_{\mathrm P}^{4} +\frac{3}{\sqrt{30}}\partial\phi_{\mathrm P}^{5}\right) + \sum_{p=1}^{3}\sum_{q=4}^{6}\beta_{\mathrm P}^{pq}\gamma_{\mathrm P}^{pq}.
    \label{eq:fw_coset_cartan_readout}
\end{align}
All fields on the right-hand sides are evaluated at the same input. As in the direct construction, the elementary Wick subtractions in these readouts vanish for the specified rotationally invariant ensemble.

The remaining combination, $\mathfrak{J}_{\mathrm C}^1-i \mathfrak{J}_{\mathrm C}^2$, contains higher ghost products, scalar-gradient couplings, and derivatives of the ghost fields. Its finite sum and the denominator-current expansions are determined by the parent-current projection formulas above and the general Wakimoto readout in Appendix~\ref{app:sunk_wakimoto}.\bigskip

The parent stress-energy neural field is
\begin{align}
    T_{\mathrm P}(z)={}&-\frac12\sum_{u=1}^{5}\bigl(\partial\phi_{\mathrm P}^{u}(z)\bigr)^2-\sum_{1\leq p<q}^6\beta_{\mathrm P}^{pq}(z)\partial\gamma_{\mathrm P}^{pq}(z)-\frac{i}{2\sqrt7}\sum_{u=1}^{5}\sqrt{u(u+1)}\,\partial^2\phi_{\mathrm P}^{u}(z).
    \label{eq:fw_parent_stress}
\end{align}
This is the Wakimoto stress readout at $(N,K)=(6,1)$, in the orthonormal scalar basis used above \cite{deBoer:1996ti}. In a similar way, the denominator stress-energy neural field is constructed from the eight currents \eqref{eq:fw_denominator_currents}:
\begin{align}
    T_{\mathrm{D}}(z):=\frac15\sum_{A=1}^{8}:\!\mathfrak{J}_{\mathrm{D}}^A\mathfrak{J}_{\mathrm{D}}^A\!:\!_{\mathrm{aff}}(z).
    \label{eq:fw_denominator_stresses}
\end{align}
The prefactor follows from the denominator level and dual Coxeter number. The affine products are defined through the elementary-field contractions and derivative corrections of these particular parent-based readouts. They are not ordinary squares of the denominator current.

We then define the coset stress-energy neural field by the difference
\begin{align}
    T_{\mathrm{C}}(z) := T_{\mathrm P}(z)-T_{\mathrm{D}}(z).
    \label{eq:fw_coset_stresses}
\end{align}
Every term is evaluated on the same sample of the parent neural fields. In particular, the denominator is not represented by a separately sampled $SU(3)_2$ network. This preserves the cross contractions between the parent and denominator stress readouts.

In the physical continuum current sector, the conformal embedding identifies the coset stress tensor with the Sugawara tensor of the three retained currents $\mathfrak{J}_{\mathrm C}^a$. Its central charge is
\begin{align}
    c_{\mathrm{C}} = c_{SU(6)_1}-c_{SU(3)_2} = 5-\frac{16}{5} = \frac95,
\end{align}
matching the direct $SU(2)_3$ target.

At finite width and fixed regulators, we evaluate \eqref{eq:fw_coset_stresses} directly, including all affine derivative corrections. We note that, despite projecting down to the coset currents, all five scalar fields and fifteen ghost pairs still contribute to the coset stress-tensor. This justifies a thorough comparison of the current and stress-tensor correlators of the two presentations, emphasizing their possible differences at finite width.\bigskip

The comparisons that follow concern the current and stress correlation functions of the direct $SU(2)_3$ presentation and the $SU(6)_1/SU(3)_2$ presentation. The coset identification specifies their common continuum reference, while their finite-width corrections and regulator dependence are calculated from their respective neural ensembles.

We continue to denote the sampled feature width by $L$. Equal values of $L$ do not imply equal computational resources: the direct presentation uses $3L$ Gaussian output amplitudes, whereas the parent-based presentation uses $35L$, in addition to the shared frequency and phase parameters. It will become clear that their composite readouts have different evaluation costs.

\subsection{Regulator choices and the infinite-width limit}
\label{sec:fw_regulators}

Having specified the direct and coset presentations, we now examine how their correlation functions approach the common continuum reference under different choices of frequency distribution. In Section~\ref{sec:bosonicsubstrate}, we chose $R_{\epsilon_B}(p)=e^{-\epsilon_B(p+p^{-1})}$ to control both endpoints of the frequency integrals when removing the annular cutoffs. We retain this choice as a reference, but broaden the comparison to include other regulator profiles.

On a finite annulus, the neural fields and polynomial correlators considered here are already well defined without the need for a regulator. The regulator, instead, controls their approach to the large-annulus limit, by introducing an exponential damping at large frequencies. We consider three representative profiles:
\begin{align}
    R^{1}_{\epsilon_B}(p) &=e^{-\epsilon_B p}, & R^{2}_{\epsilon_B}(p)&=(1-e^{-p/\epsilon_B})e^{-\epsilon_B p}, & R^{3}_{\epsilon_B}(p) &=e^{-\epsilon_B(p+p^{-1})}.
    \label{eq:fw_regulator_choices}
\end{align}
Each profile uses its corresponding normalization in the parameter distribution and neural amplitudes, 
\begin{align}
    Z_{\alpha,\eta}^{(\Lambda_B)} &= \int_{p_-}^{p_+}\frac{R_{\epsilon_B}^\eta(p)}{p^{1-2\alpha}}dp, & \eta&\in\{1,2,3\}.
\end{align}
The exponential profile $R^1$ has the same radial damping form as the fermionic prior with its normalization and regulator parameters defined separately.

These profiles illustrate a much broader freedom in choosing the parameter distribution. We use them to examine how this choice affects convergence to the same continuum reference at finite width and cutoffs. We fix the annular endpoints to $p_-=0.01$ and $p_+=1000$ throughout the numerical comparisons below. The damping parameter $\epsilon_B$ is selected separately for each regulator profile, but the same value is used for the direct and coset presentations. We first determine these values from the infinite-width stress-tensor correlator, then hold them fixed when varying the width $L$ and spectral exponent $\alpha$.

All regulator choices are evaluated on the same set of $64$ ordered insertion pairs $(z_s,w_s)$ in the unit disk, with $z_s\ne w_s$.
We denote the direct $SU(2)_3$ presentation by $\mathrm{W}$ and the $SU(6)_1/SU(3)_2$ presentation by $\mathrm{C}$.\bigskip

To select $\epsilon_B$, we first take the large-width limit analytically at fixed annular cutoffs and regulator profile. For $X\in\{\mathrm{W},\mathrm C\}$ and $\eta\in\{1,2,3\}$, write
\begin{align}
    \mu_{\infty,X,\eta}^{TT}(z,w;\epsilon_B) := \lim_{L\to\infty} \bE_{\eta,\Lambda_B} \!\left[T_X(z)T_X(w)\right], \qquad \Lambda_B=(0.01,1000,\epsilon_B).
    \label{eq:fw_gaussian_tt_reference}
\end{align}
The quantity $\mu_{\infty,X,\eta}^{TT}$ is the infinite-width correlator at the chosen finite regulator. The spectral exponent $\alpha$ drops out in this limit, so it does not enter in the selection of $\epsilon_B$.

The common continuum reference stress-tensor two-point function is $9/[10(z-w)^4]$. We therefore define the dimensionless ratio at each insertion pair to be the ratio between the measured two-point function and the continuum one,
\begin{align}
    Q_{X,\eta,s}(\epsilon_B) :=\frac{10}{9}(z_s-w_s)^4\mu_{\infty,X,\eta}^{TT}(z_s,w_s;\epsilon_B).
    \label{eq:fw_normalized_gaussian_tt}
\end{align}
The continuum reference is, therefore, $Q_{X,\eta,s}=1$.

To select a damping parameter that minimizes the discrepancy from the continuum reference, we report both the mean ratio and its relative RMS deviation from unity,
\begin{align}
    \overline Q_{X,\eta}(\epsilon_B) &=\frac1{64}\sum_{s=1}^{64}Q_{X,\eta,s}(\epsilon_B),\nonumber\\
    E_{X,\eta}(\epsilon_B) &=\left[\frac1{64}\sum_{s=1}^{64}\left|Q_{X,\eta,s}(\epsilon_B)-1\right|^2\right]^{1/2}.
    \label{eq:fw_gaussian_relative_error}
\end{align}
A mean close to one can conceal cancellations between errors at different insertion pairs. The quantity $E_{X,\eta}$ measures, instead, the deviations before averaging and is used to select the best regulator parameter.

For each profile, we choose a single matched value $\epsilon_{B,\eta}^{\star}$ by numerically minimizing the larger of the two presentation errors:
\begin{align}
    \mathcal E_\eta(\epsilon_B):=\max\left\{E_{\mathrm{W},\eta}(\epsilon_B),E_{\mathrm C,\eta}(\epsilon_B)\right\}.
    \label{eq:fw_regulator_minimax_objective}
\end{align}
Thus $\epsilon_{B,\eta}^{\star}$ is a relative-error minimax choice on the fixed set of points $\{(z_s, w_s)\}$. Separate values are not selected for the two presentations or for different $\alpha$.

The numerical search uses a positive logarithmic grid, followed by refinement of every detected local minimum and comparison with the $\epsilon_B=0$ boundary. The values in Table~\ref{tab:tt-relative-gaussian-main} are the numerical minima found by this procedure.

\begin{table}[t]
    \centering
    \small
    \setlength{\tabcolsep}{5pt}
    \begin{tabular}{lrrrrr}
        \toprule
        Regulator & $\epsilon_{B,\eta}^{\star}$ & $\operatorname{Re}\overline Q_{\mathrm{W}}$ & $E_{\mathrm{W}}$ & $\operatorname{Re}\overline Q_{\mathrm C}$ & $E_{\mathrm C}$ \\
        \midrule
        $R^1$ & $0.0206692504$ & $0.945072$ & $\mathbf{0.073951}$ & $0.818289$ & $0.236269$ \\
        $R^2$ & $0.0206496645$ & $0.945113$ & $\mathbf{0.073891}$ & $0.813471$ & $0.238665$ \\
        $R^3$ & $0.0198029579$ & $0.897404$ & $\mathbf{0.108578}$ & $0.457145$ & $0.561155$ \\
        \bottomrule
    \end{tabular}
    \caption{
    Regulated infinite-width $TT$ correlators at the matched damping parameters selected using \eqref{eq:fw_regulator_minimax_objective}. All rows use $p_-=0.01$, $p_+=1000$, and the same $64$ ordered unit-disk pairs without a minimum-separation cut. The continuum reference is $\overline Q=1$ and $E=0$. Bold entries identify the smaller relative RMS error within each row. The mean columns display real parts; the imaginary parts round to zero at the displayed precision, while the errors use the full complex ratios. }
    \label{tab:tt-relative-gaussian-main}
\end{table}

For all three profiles, the direct presentation has the smaller regulated Gaussian $TT$ error. This is a presentation-dependent approximation bias at the chosen regulator settings; both constructions realize the same target current sector in their ordered continuum limits. Including the $T\mathcal{J}^a$ and $\mathcal{J}^a\mathcal{J}^a$ correlators in the regulator search yields the same selected $\epsilon_B$ values: the $TT$ error dominates the combined objective at these settings.\bigskip

For the remainder of the numerical study, each profile uses the fixed parameter triple
\begin{align}
    \Lambda_{B,\eta}^{\star} = \left(0.01,1000,\epsilon_{B,\eta}^{\star}\right),\qquad \eta=1,2,3,
    \label{eq:fw_selected_regulator_parameters}
\end{align}
with the same choice applied to both presentations. We then vary $L$ and $\alpha$ without readjusting $\epsilon_B$. The subsequent search for a preferred spectral exponent is therefore distinct from the regulator selection above.

\subsection{Finite-width analysis and the optimal distribution}

With the regulator parameters fixed by \eqref{eq:fw_selected_regulator_parameters}, we now vary the feature width $L$ and spectral exponent $\alpha$. Changing $\alpha$ redistributes the sampled frequencies while preserving the large-width limit. As a global hyperparameter, selecting its value requires a numerical search, but doing so leaves the field content and number of parameters per neuron unchanged.
This allows the parameter distribution to be adjusted without increasing the width, the number of elementary fields, or the number of sampled parameters per neuron.

We focus on the stress-tensor correlator and extend the relative-error diagnostic used above to finite width. For $X\in\{\mathrm W,\mathrm C\}$, we define the two-point function and dimensionless ratio between it and the continuum limit,
\begin{align}
    \mu_{X,\eta}^{TT}(L,\alpha;z,w) &:= \bE_{\eta,\Lambda_{B,\eta}^{\star},\alpha}\!\left[T_X(z)T_X(w)\right],\nonumber\\
    Q_{X,\eta,s}(L,\alpha) &:= \frac{10}{9}(z_s-w_s)^4\mu_{X,\eta}^{TT}(L,\alpha;z_s,w_s),
    \label{eq:fw_finite_width_tt_ratio}
\end{align}
where the expectation uses the width-$L$ ensemble. The regulator parameters and the $64$ ordered insertion pairs are the same as in the preceding subsection. The population relative RMS error is similarly defined,
\begin{align}
    E_{X,\eta}^{TT}(L,\alpha) := \left[\frac1{64}\sum_{s=1}^{64}\left|Q_{X,\eta,s}(L,\alpha)-1\right|^2\right]^{1/2}.
    \label{eq:fw_finite_width_tt_error}
\end{align}
As before, the squared moduli of the complex deviations are averaged, so errors at different insertion pairs cannot cancel. This quantity, thus, measures the deviation of the ensemble correlation function from the continuum reference, and will serve as our main numerical probe throughout this subsection.\bigskip

\begin{figure}[!tp]
    \centering
    \begin{subfigure}[t]{0.48\textwidth}
        \centering
        \includegraphics[width=\linewidth]{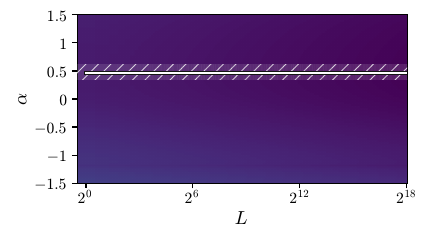}
        \caption{Direct presentation $\mathrm W$, regulator $R^1$.}
        \label{fig:fw_tt_W_R1}
    \end{subfigure}
    \hfill
    \begin{subfigure}[t]{0.48\textwidth}
        \centering
        \includegraphics[width=\linewidth]{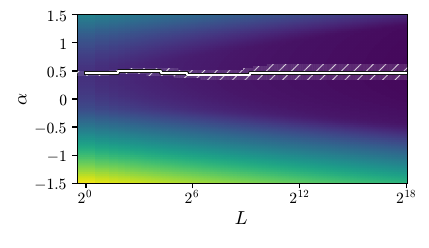}
        \caption{Coset presentation $\mathrm C$, regulator $R^1$.}
        \label{fig:fw_tt_C_R1}
    \end{subfigure}

    \medskip

    \begin{subfigure}[t]{0.48\textwidth}
        \centering
        \includegraphics[width=\linewidth]{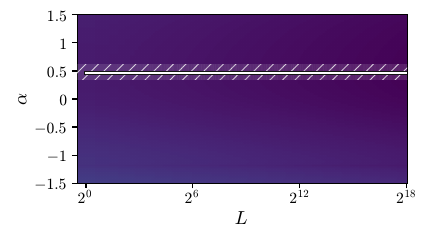}
        \caption{Direct presentation $\mathrm W$, regulator $R^2$.}
        \label{fig:fw_tt_W_R2}
    \end{subfigure}
    \hfill
    \begin{subfigure}[t]{0.48\textwidth}
        \centering
        \includegraphics[width=\linewidth]{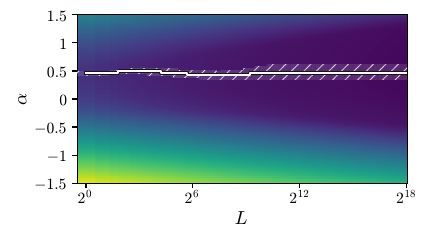}
        \caption{Coset presentation $\mathrm C$, regulator $R^2$.}
        \label{fig:fw_tt_C_R2}
    \end{subfigure}

    \medskip

    \begin{subfigure}[t]{0.48\textwidth}
        \centering
        \includegraphics[width=\linewidth]{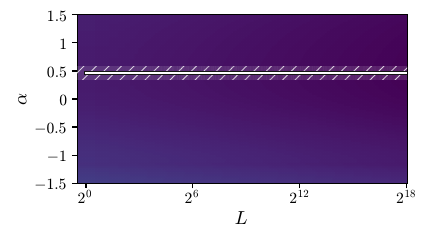}
        \caption{Direct presentation $\mathrm W$, regulator $R^3$.}
        \label{fig:fw_tt_W_R3}
    \end{subfigure}
    \hfill
    \begin{subfigure}[t]{0.48\textwidth}
        \centering
        \includegraphics[width=\linewidth]{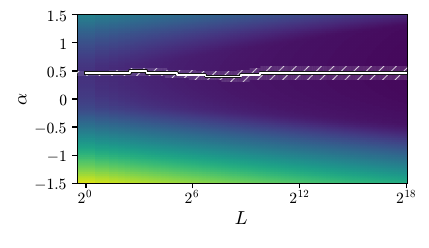}
        \caption{Coset presentation $\mathrm C$, regulator $R^3$.}
        \label{fig:fw_tt_C_R3}
    \end{subfigure}

    \medskip

    \input{figures/finite_width/tt_heatmap_scale.tex}

    \caption{    $\bE[T(z_1)T(z_2)]$: $\mathrm W=SU(2)_3$ versus $\mathrm C=SU(6)_1/SU(3)_2$. Relative RMS error of the finite-width $TT$ correlator, with color representing $\log_{10}E_{X,\eta}^{TT}$ on a common scale. The horizontal axis is the feature width $L$ on a logarithmic scale, and the vertical axis is the spectral exponent $\alpha$. Solid curves mark unique sampled minimizers at fixed $L$; the hatched regions contain exponents whose RMS error is within $5\%$ of the minimum. Tied minima are retained as sets. The evaluated grid has $301$ spectral exponents in $[-5,5]$ and $301$ distinct integer widths from $1$ to $2^{18}$. All panels use $p_-=0.01$, $p_+=1000$, and the same $64$ ordered insertion pairs. Each row uses the matched damping parameter from Table~\ref{tab:tt-relative-gaussian-main}, without retuning it at finite width.}
    \label{fig:fw_tt_alpha_width}
\end{figure}

Figure~\ref{fig:fw_tt_alpha_width} displays $\log_{10}E_{X,\eta}^{TT}$ as a function of $L$ and $\alpha$ for both the direct (W) and coset (C) presentations, for all three regulators. Reading a vertical slice compares frequency distributions at fixed width. Reading a horizontal slice shows how a fixed distribution approaches its regulated Gaussian limit as the number of sampled features increases. We also provide examples of higher-point functions in Figure~\ref{fig:fw_ttt_alpha_width}, and examples of $TT$ for other WZW theories in Figure~\ref{fig:fw_tt_alpha_width_su2_3_su_2_4_su_3_4}.

In each of these plots, we observe a local minimum at each fixed width $L$ determined by a choice of parameter $\alpha$ minimizing the error with respect to the continuum limit. Furthermore, this optimal point is not shared between regulator prescriptions and different $L$. This motivates defining $\alpha_{X,\eta}^{\star}(L)$ as an optimal spectral exponent for each choice of correlator, regulator and width $L$. Let $\mathcal A_{\mathrm{grid}}$ denote the sampled set of spectral exponents. At each fixed width, we define
\begin{align}
    E_{X,\eta}^{TT,\star}(L) &:= \min_{\alpha\in\mathcal A_{\mathrm{grid}}} E_{X,\eta}^{TT}(L,\alpha),\nonumber\\
    \mathcal A_{X,\eta}^{\star}(L) &:=\operatorname*{arg\,min}_{\alpha\in\mathcal A_{\mathrm{grid}}} E_{X,\eta}^{TT}(L,\alpha).
    \label{eq:fw_optimal_spectral_exponent}
\end{align}
When the minimizing exponent is unique, its values define the optimality curve $\alpha_{X,\eta}^{\star}(L)$. If several exponents attain the same minimum, we may retain the complete minimizing set rather than selecting one arbitrarily.

In those cases it becomes important to look at the near-optimal region
\begin{align}
    \mathcal A_{X,\eta}^{5\%}(L):= \left\{\alpha\in\mathcal A_{\mathrm{grid}}: E_{X,\eta}^{TT}(L,\alpha) \leq 1.05\,E_{X,\eta}^{TT,\star}(L)
    \right\}.
    \label{eq:fw_near_optimal_spectral_region}
\end{align}
This distinguishes a well-localized optimum from a range of distributions with comparable accuracy. The $5\%$ threshold applies to the RMS error itself, not to its square. Minima at the boundary of the sampled $\alpha$ interval are interpreted only within that interval.

Where the minima are numerically resolved, we show the optimality curves in Figures~\ref{fig:fw_tt_alpha_width}, \ref{fig:fw_ttt_alpha_width}, and \ref{fig:fw_tt_alpha_width_su2_3_su_2_4_su_3_4}, together with their near-optimal regions. Their width dependence shows that a spectral exponent selected at one width need not remain optimal at another. Increasing $L$ while keeping $\alpha$ fixed can therefore yield a larger correlation error than is attainable at the same width by adjusting the parameter distribution, without introducing additional neural fields or sampled parameters. The near-optimal regions indicate whether this adjustment gives a substantial improvement or only a small change in accuracy. A broader optimization could also vary $\epsilon_B$ jointly with $\alpha$ at each $L$, keeping the annular endpoints and error measure fixed. Here, we instead retain the damping parameters selected in the preceding subsection, so the displayed curves identify the preferred spectral distributions conditional on those regulator choices, the chosen correlator, and the insertion bank.

At fixed regulator parameters, all values of $\alpha$ approach the same Gaussian correlation function within a given presentation. The strict infinite-width error is consequently independent of $\alpha$, even when finite-width errors have a distinct minimum. It becomes natural to expect that a finite-width optimality curve does not select a uniquely preferred distribution in the Gaussian limit.

\paragraph{Continuous and discrete presentation choices.}

The comparison contains two different choices. Within either presentation, $\alpha$ continuously changes the finite neural ensemble without changing its field content. On the other hand, the coset identification provides a discrete alternative, $X\in\{\mathrm W,\mathrm C\}$, with the same continuum current sector but different elementary fields and composite readouts. The continuum identification fixes neither the preferred spectral distribution nor the relative finite-width errors of these two constructions.

For each regulator, the direct presentation has the smaller $TT$ error at all $301\times301$ matched $(L,\alpha)$ grid points. While not made explicit here, a similar numerical analysis on all other two-point functions between the currents and $T$ brings about a similar conclusion. Likewise for other level and rank choices and higher-point functions: the direct construction always has smaller errors than the coset construction.

The additional choice remains useful even when the direct construction is preferred: it shows that the same continuum target admits distinct finite approximation families whose accuracy must be assessed rather than inferred from their physical equivalence. Unlike varying $\alpha$, switching between $\mathrm W$ and $\mathrm C$ also changes the number of elementary fields and the cost of evaluating the readouts. The heatmaps compare equal feature widths, not equal total computational budgets. The discrete choice of presentation and the continuous spectral tilt therefore control approximation accuracy at a fixed continuum target. Tuning $\alpha$ leaves the network width unchanged, with an observable- and regulator-dependent preferred value.

We note that the larger $TT$ error of the coset presentation can be understood from its construction, $T_{\mathrm C}=T_{\mathrm P}-T_{\mathrm D}$. Its two-point function combines the parent and denominator correlators with their mixed contractions. At finite regulator, the deviations of these terms from their continuum values do not cancel completely in the cases studied here. The large-width limit calculation shows that the dominant discrepancy is already present in the $SU(6)_1$ parent. In particular, its ghost and scalar background-charge contributions cancel in the continuum but receive different regulator corrections. The coset construction therefore inherits a substantial bias even after finite-width effects have disappeared.

The denominator subtraction partially reduces the parent's absolute bias, rather than simply adding further error. However, it also reduces the continuum $TT$ coefficient from $5/2$ to $9/10$, making the remaining discrepancy larger relative to the target. For the three regulator choices and the insertion bank used here, the resulting relative error exceeds that of the direct $SU(2)_3$ construction. This explains a substantial part of the direct presentation's advantage: the parent-based readout depends on cancellations that are more sensitive to the regulator. Naturally, this observation only holds for stress-tensor correlators, and does not indicate that the coset construction is necessarily less accurate on all observables; it is only on the ones studied here.

\subsection{Local calibration and computational dependence}
\label{sec:local_calibration_dependence}

The analysis in the previous subsection compared different frequency distributions while keeping the prescribed elementary-field amplitudes fixed. We now focus on these amplitudes at finite width. Indeed, these provide another way to adjust the correlators---changing the relative scalar and ghost contributions can compensate part of the remaining discrepancy from the continuum. The additional fields in the coset presentation allow for more such adjustments, but need not make the approximation more accurate or the fitting problem easier.

We introduce one real amplitude parameter for each scalar field and coupled ghost pair:
\begin{align}
    \phi^A
    &\longmapsto e^{\theta_{\phi,A}/2}\phi^A,
    &
    \beta^a
    &\longmapsto e^{\theta_{\mathrm g,a}/2}\beta^a,
    &
    \gamma^a
    &\longmapsto e^{\theta_{\mathrm g,a}/2}\gamma^a.
    \label{eq:local_calibration_amplitudes}
\end{align}
The derivatives acquire the same factors because the amplitude parameters are independent of the insertion position. There are two adjustable parameters for $\mathrm W$ and twenty for $\mathrm C$. The normalizers and all current and stress-tensor coefficients remain fixed, and $\theta=0$ recovers the original presentation.

Unlike varying $\alpha$, these changes modify the elementary covariances, even at infinite width. They define approximate correlation functions near the original neural construction, rather than exact WZW theories with different levels. We refer to fitting these amplitudes against the fixed continuum reference as a calibration.

\paragraph{Correlation data and local response.}

An amplitude change generally affects several observables at once. Our experiment will therefore include correlations of types $\mathcal{J}\mathcal{J}$, $\mathcal{J}\mathcal{J}\mathcal{J}$, $\mathcal{J}\mathcal{J}\mathcal{J}\mathcal{J}$, $TT$, $T\mathcal{J}$, and $T\mathcal{J}\mathcal{J}$. Out of all possible combinations of these, we select thirteen correlators, and evaluate each at four separated configurations of points on the complex plane, giving $52$ complex values.

For both the direct and coset presentation, $X\in\{\mathrm W,\mathrm C\}$, we define
\begin{align}
    \mathcal F_{X,\lambda}(\theta)
    :=
    g_\lambda\,
    \bE_\theta[\mathcal O_{X,\lambda}],
    \qquad \lambda=1,\ldots,52,
    \label{eq:local_calibration_rescaled_correlators}
\end{align}
where $\mathcal O_{X,\lambda}$ is the specified product of currents and stress tensors evaluated at a given configuration of points. The fixed factor $g_\lambda$ is introduced to remove the leading dependence on those points. For example, it is $(z_1-z_2)^2$ for $\mathcal{J}\mathcal{J}$ and $(z_1-z_2)^4$ for $TT$. These factors are common to both presentations and are fixed throughout our experiment.

We collect the real and imaginary parts into a $52\times 2=104$-component vector
\begin{align}\label{eq:correlation_vector}
    y_X(\theta)
    =
    \bigl(
        \operatorname{Re}\mathcal F_{X,1},\ldots,
        \operatorname{Re}\mathcal F_{X,52},
        \operatorname{Im}\mathcal F_{X,1},\ldots,
        \operatorname{Im}\mathcal F_{X,52}
    \bigr)^{\mathsf T}.
\end{align}
The continuum vector $y_{\mathrm{ref}}$ uses the same factors and ordering, but its continuum correlation functions are evaluated at $\theta=0$.
The selected set of correlators, coordinate factors, and positions are all detailed in Appendix~\ref{app:local_calibration}.

Near $\theta=0$, we can characterize the response to these amplitude deformations by expanding the vector of correlators, $y_X(\theta)=y_X(0)+\left.\partial_\theta y_X\right|_0\theta+O(\|\theta\|^2)$. We then define the response to this deformation, $A_X$, and its associated symmetric matrix $H_X$, as
\begin{align}
    D_X &:= \left.\partial_\theta y_X\right|_0,
    &
    A_X &:= \frac{D_X}{\sqrt{104}},
    &
    H_X &:= A_XA_X^{\mathsf T}.
    \label{eq:local_calibration_response}
\end{align}
The range of $A_X$ encodes how the correlation functions change, to first order. Its singular values measure their response strengths. A residual outside this range cannot be removed by an infinitesimal amplitude adjustment.
We then call response rank the number of independent directions in which the correlation vector can change under infinitesimal amplitude adjustments, i.e. $\operatorname{rank}(A_X) = \dim\operatorname{range}(A_X)$.

In the continuum limit, one can show that the response ranks are $2$ and $8$ for $\mathrm W$ and $\mathrm C$, respectively. The response range of the direct presentation can be shown to be contained in that of the coset presentation, up to numerical tolerance. Consequently, we see that while their original correlators agree, their responses to amplitude changes do not.

With the original regulator, $R^3$, $\epsilon_B^\star$, $\alpha=0$, and $L=256$, the resolved ranks become $2$ and $20$, and the range of the direct presentation is no longer contained in that of the coset presentation.
In this finite-width regime, we can characterize those changes in the response, via the condition number. On the response space, the condition number of $H_X$ is the squared ratio of the largest and smallest singular values of $A_X$, and for both presentations those take the values
\begin{align}
    \operatorname{cond}(H_{\mathrm W})
    &\simeq731,
    &
    \operatorname{cond}(H_{\mathrm C})
    &\simeq7.77\times10^{12}.
\end{align}
The additional coset directions therefore have very unequal response strengths.
Within this local approximation, equally sized corrections along different response directions can therefore require very different amplitude changes. The greater local flexibility of the coset network does not by itself imply more efficient calibration toward the common continuum correlators.

\paragraph{Calibrating the parameters}

The response spectrum describes how infinitesimal amplitude changes affect the correlation data. We now test whether finite adjustments of those amplitudes can reduce discrepancies from the common continuum reference. For both presentations, we consider 255 distinct integer widths sampled log-uniformly on the interval $1\leq L\leq 10^6$, with regulator $R^3$, $\epsilon_B^\star$, and $\alpha=0$. At each width, we vary the two amplitude parameters of $\mathrm W$ and the twenty of $\mathrm C$, starting from $\theta = 0$, keeping the width, frequency distribution, regulator, and coefficients defining the currents and stress tensor fixed.

The amplitudes are fitted to a prescribed subset $F$ of correlation data using the bounded least-squares procedure described below. We then evaluate the same fitted amplitudes on the remaining original coordinates $H$ and on a separate set of validation configurations $V$. This tests whether improvements on the fitted correlations extend to the other evaluated observables and positions. These sets are further detailed in Appendix~\ref{app:local_calibration}.

For each set $S\in\{F,H,V\}$, let $y_{X,S}(\theta;L)$ contain the real and imaginary parts of the corresponding rescaled neural correlators, evaluated at width $L$, and let $y_{\mathrm{ref},S}$ contain their continuum values at the same positions. We measure the root-mean-square discrepancy
\begin{align}
    R_{X,S}(\theta;L) &:= \sqrt{\frac{1}{n_S} \left\| y_{X,S}(\theta;L)-y_{\mathrm{ref},S} \right\|^2},
    &
    n_F &= 40, & n_H &= 64, & n_V &= 104.
    \label{eq:local_calibration_subset_rms}
\end{align}
For validation, both the neural and continuum correlators, including their coordinate factors, are evaluated at a different set of positions from the fitted and held sets.

We determine the amplitude adjustments by reducing the fitted correlation error,
\begin{align}
    \mathcal L_{X,L}^{\mathrm{fit}}(\theta) = \frac{n_F}{104} \frac{(R_{X,F}(\theta;L))^2}{2} + \frac{10^{-6}}2\|\theta\|^2.
    \label{eq:local_calibration_loss}
\end{align}
The second term is a quadratic penalty which favors small departures from the original amplitudes $\theta=0$. It also serves to regulate parameter combinations that are weakly constrained by the fitted correlators in the local least-squares steps.

For each width $L$ and presentation $X$, we initialize the amplitude parameters at $\theta=0$ and perform at most twenty-four bounded local least-squares steps to reduce the fitting loss $\mathcal L_{X,L}^{\mathrm{fit}}(\theta)$. At each iteration, we evaluate the fitted correlation vector $y_{X,F}(\theta;L)$ and its Jacobian, and use them to compute a candidate update $\Delta\theta$ subject to the amplitude bounds $|\theta_i+\Delta\theta_i|\leq0.3$ and coordinate step limits $|\Delta\theta_i|\leq0.02$. Before accepting an update, we apply backtracking based on the actual fitting loss. The optimization terminates when the prescribed stopping criterion is satisfied or the iteration limit is reached, yielding the fitted parameters $\widehat\theta_{X,L}$. We then evaluate the initial and final RMS discrepancies, $R_{X,S}(0;L)$ and $R_{X,S}(\widehat\theta_{X,L};L)$, for each set $S\in\{F,H,V\}$, without further parameter adjustment. This comparison measures the improvement on the fitted data and tests whether it extends to the held data and separate validation configurations.

Writing $\widehat\theta_{X,L}$ for the amplitudes obtained after fitting, we define the RMS values, before and after the optimization steps, by
\begin{align}
    R_{X,S}^{\mathrm{before}}(L) &:= R_{X,S}(0;L),
    &
    R_{X,S}^{\mathrm{after}}(L) &:= R_{X,S}(\widehat\theta_{X,L};L).
\end{align}
Their ratio measures the change in agreement on each evaluation set: values below one indicate improvement, while values above one indicate deterioration.

\begin{figure}[t]
    \centering
    \includegraphics[width=0.75\linewidth]{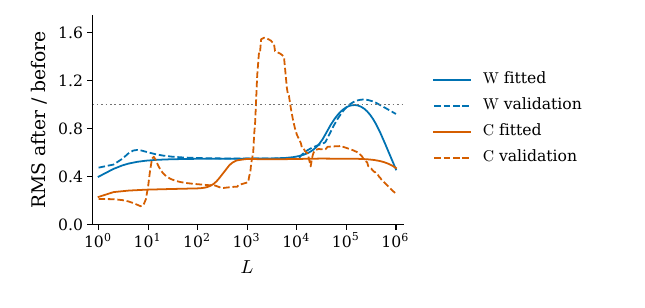}
    \caption{
    After/before RMS ratios for independent fits. Solid curves show $R_{X,F}^{\mathrm{after}}/R_{X,F}^{\mathrm{before}}$ on 40 fitted real coordinates, corresponding to 20 complex evaluations. Dashed curves show $R_{X,V}^{\mathrm{after}}/R_{X,V}^{\mathrm{before}}$ on the 104 validation real coordinates, corresponding to 52 complex evaluations. The gray reference line is one. Each ratio uses the same presentation's initial error at the corresponding width. Held-coordinate ratios are not plotted; more values are given in Table~\ref{tab:app_local_calibration_results}.
    }
    \label{fig:local_calibration}
\end{figure}

Figure~\ref{fig:local_calibration} compares the fitted and validation after/before RMS ratios across all 255 widths. Table~\ref{tab:app_local_calibration_results} summarizes the ratio extrema and the numbers of improved and worsened widths, while Table~\ref{tab:calibration_loggrid_endpoints} gives the absolute before and after residuals at $L=1$ and $L=10^6$. Fitted RMS decreases at all $255$ widths for both presentations. However, held RMS worsens at $35$ widths for $\mathrm W$ and $30$ for $\mathrm C$; validation worsens at $29$ and $35$, respectively. The largest validation after/before ratios are $1.04034$ for $\mathrm W$ at $L=218776$ and $1.55480$ for $\mathrm C$ at $L=2188$.

At $L=10^6$, validation ratios are $0.921068$ and $0.257934$, but the corresponding final absolute RMS values are $2.10865$ and $2.45236$ for $\mathrm W$ and $\mathrm C$. A larger fractional reduction therefore need not give smaller absolute error. The direct presentation has smaller final validation RMS at every sampled width in this fixed comparison.

The extra calibration freedom can therefore reduce selected discrepancies without necessarily improving the whole current sector. The non-uniform $L$-dependence is a good indicator for how complicated choosing a good finite-width presentation is, even when the continuum algebra is fixed. In our experiments, the direct presentation attains the smaller final validation RMS throughout the grid, and the fractional improvements vary with width in both presentations. These results show how the accuracy obtained through calibration depends on the chosen presentation and width, even when the limiting current algebra is shared.

\paragraph{Estimating correlation functions.}

The calibration experiment adjusts the field normalizations to reduce
differences from the continuum reference. We now consider how accurately
a correlation function of a fixed neural ensemble can be estimated
from a finite number of samples. We distinguish the difference between
the ensemble mean and the continuum reference from the uncertainty
introduced by estimating that mean.

For each presentation, we study the rescaled Cartan-current product
\begin{align}
    U_X(z,w)
    :=
    (z-w)^2\mathcal J_X^3(z)\mathcal J_X^3(w),
    \label{eq:calibration_cartan_observable}
\end{align}
at $z=1.3+0.1i$ and $w=0.3+0.05i$. We use the same $255$ widths
as in the calibration experiment, with regulator $R^3$,
$\epsilon_B^\star$, and $\alpha=0$, keeping the field normalizations
at $\theta=0$. At every width, we calculate the sampling uncertainty
corresponding to an average of $N_{\mathrm{samp}}=4096$ independent
estimates, using the integrated covariance of a single estimate.
Thus, $L$ controls the number of features in the neural construction,
while $N_{\mathrm{samp}}$ controls the number of independent estimates
used to approximate its mean.

Some averages over the random network parameters can be performed
analytically before sampling the remaining variables. For the estimator
used below, we first average over the Gaussian amplitudes, angles,
and phases, leaving an expression that depends on the sampled radii.
We then replace its scalar sample average by the known expectation
of that contribution. These operations preserve the finite-width
expectation, including the contributions from repeated features and
the correlations induced by shared frequencies and phases.
Appendix~\ref{app:local_sampling} gives the estimator definitions
and numerical checks.

Let $\mu_X$ contain the real and imaginary parts of $\bE[U_X]$,
and let $\widehat\mu_X$ be the average of $N_{\mathrm{samp}}$
independent unbiased estimates of $\mu_X$. If
$\Sigma_{\mathrm{est},X}$ is the covariance matrix of one such
estimate, the mean-square error per real component relative to
the continuum reference $\mu_{\mathrm{target}}=(3/2,0)^{\mathsf T}$ is
\begin{align}
    \frac12\bE\!\left[
        \|\widehat\mu_X-\mu_{\mathrm{target}}\|^2
    \right]
    &=
    \frac12\|\mu_X-\mu_{\mathrm{target}}\|^2
    +
    \frac{1}{2N_{\mathrm{samp}}}
    \operatorname{tr}\Sigma_{\mathrm{est},X}.
    \label{eq:cartan_bias_variance}
\end{align}
The first term measures the squared discrepancy of the regulated
ensemble mean, while the second measures the sampling uncertainty
of its estimate. We denote their square roots by $B_X(L)$ and
$\mathrm{SE}_X(L)$, respectively. Increasing $N_{\mathrm{samp}}$
reduces $\mathrm{SE}_X(L)$ while leaving $B_X(L)$ unchanged.

\begin{figure}[!tp]
    \centering
    \begin{subfigure}[t]{.48\textwidth}
        \centering
        \includegraphics[width=\linewidth]
        {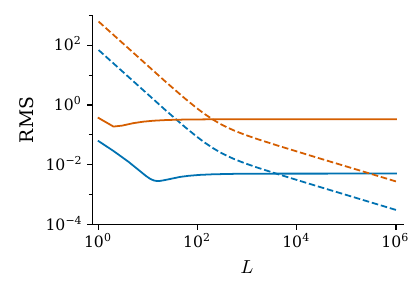}
        \caption{Error in the Cartan correlation.}
    \end{subfigure}
    \hfill
    \begin{subfigure}[t]{.48\textwidth}
        \centering
        \includegraphics[width=\linewidth]
        {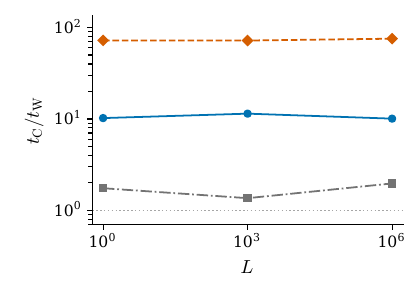}
        \caption{Coset/direct computation times.}
    \end{subfigure}
    \caption{
    Sampling uncertainty and computation time at $R^3$, $\epsilon_B^\star$, and $\alpha=0$. Left: blue curves denote $\mathrm W$ and orange curves denote $\mathrm C$. Solid curves show the RMS discrepancy $B_X(L)$ from the continuum reference, and dashed curves show the RMS sampling uncertainty $\mathrm{SE}_X(L)$ for $N_{\mathrm{samp}}=4096$ independent estimates. The calculation uses $\theta=0$ and the estimator with its scalar contribution averaged analytically, evaluated at all $255$ widths. Right: ratios of median coset and direct computation times from three repetitions at $L=1,10^3,10^6$. Blue solid curves with circles denote the full calculation; orange dashed curves with diamonds denote evaluations with adjustable parameters; gray dash-dotted curves with squares denote evaluations with fixed parameters. The two repeated-evaluation tasks each contain $1000$ calls. Lines connect the three measured widths, and the dotted horizontal reference marks a ratio of one. Both panels use logarithmic axes. Estimator definitions and timing details are given in Appendix~\ref{app:local_calibration}.
    }
    \label{fig:calibration_resources}
\end{figure}

Figure~\ref{fig:calibration_resources} compares these two contributions
across the width grid. For the chosen estimator, the sampling
uncertainty decreases with width in both presentations, and
$\mathrm{SE}_{\mathrm C}(L)=9\,\mathrm{SE}_{\mathrm W}(L)$ throughout
the grid. At $L=10^6$, we obtain
\begin{align}
    B_{\mathrm W}
    &\simeq 5.05189\times10^{-3},
    &
    \mathrm{SE}_{\mathrm W}
    &\simeq 3.06877\times10^{-4},
    \\
    B_{\mathrm C}
    &\simeq 0.334644,
    &
    \mathrm{SE}_{\mathrm C}
    &\simeq 2.76190\times10^{-3}.
\end{align}
At this width, the discrepancy from the continuum reference dominates
the error in both presentations, so further sampling would produce
little improvement in their total accuracy. The finite-width means
approach approximately $1.50714$ and $1.97326$ as $L$ increases at
the fixed regulator, whereas the continuum reference is $3/2$.
A discrepancy therefore remains even when the sampling uncertainty
and the finite-width correction become small.

\paragraph{Computational cost.}

We also compare the time required to evaluate the complete correlation vector used in the calibration experiment. We distinguish three tasks. The first constructs the correlation expressions, performs the bounded fit, and evaluates the fitted parameters at the validation positions. The second repeatedly evaluates previously integrated expressions while retaining their dependence on the adjustable amplitude parameters. The third fixes the amplitudes at $\theta=0$ and combines the corresponding coefficients before repeated evaluations.

These tasks permit different amounts of previously computed information to be reused. In the second task, each call recomputes the dependence on the amplitudes, using the same final fitted values throughout the timing measurement. In the third, the fixed amplitudes allow the correlations to be evaluated directly from their collected coefficients. Evaluations at new insertion positions require the corresponding integrals to be computed again.

We measure each task at $L=1,10^3,10^6$, using the same regulator and $\alpha=0$ as in the preceding experiments. Each measurement is repeated three times, and the two repeated-evaluation tasks each contain $1000$ calls. Figure~\ref{fig:calibration_resources} shows the ratio of the median coset time to the median direct time. Across the three widths, this ratio ranges from $10.08$ to $11.43$ for the full calculation, from $72.20$ to $75.61$ when the amplitude dependence is recomputed, and from $1.36$ to $1.97$ when the amplitudes are fixed and the coefficients have been combined in advance.

The relative cost therefore depends strongly on the expressions retained during evaluation and the preparatory work included in the measurement. These calculations evaluate ensemble correlations through finite-width formulas, which allow integrated coefficients to be reused as $L$ changes. The timing protocol and achieved validation errors are given in Appendix~\ref{app:local_timing}. Comparing the time required to reach a common continuum accuracy would additionally require both presentations to attain that accuracy.

\section{Finite-resource geometry, compression and transfer}
\label{sec:finite_resource_geometry}

The fermionic construction gives a common Fock-space description of the charged sectors of level--rank paired WZW theories. We use it to study two questions: how well a reduced collection of fermionic modes approximates sewn amplitudes, and how information learned from one current sector can be used in the other. We first consider finite-mode compression, then prediction of fusion coefficients, and finally the transfer of unknown charged-channel couplings measured through the finite neural fields. Numerical methods are given in Appendix~\ref{app:s5_upgrade_methods}.

\subsection{Polarized mode compression}
\label{sec:finite_mode_compression}

A finite collection of fermionic modes can be chosen to approximate the amplitudes of a larger Fock space. The choice must respect the separation between occupied sea modes and particle modes, which enters the current algebra through its polarization. We describe a retained  space by an isometry $Q=(Q_-,Q_+)$, with
\begin{align}
 Q^\dagger Q=1_m,\qquad  P_\pm=Q_\pm Q_\pm^\dagger,\qquad  m=m_-+m_+.
 \label{eq:finite_resource_isometry}
\end{align}
The projectors $P_-$ and $P_+$ specify the retained sea and particle spaces. Rotations within either space leave the compressed observables unchanged. Their relative orientation and their placement in the reference space affect the resulting amplitudes.

The retained fermionic operators satisfy
\begin{align}
 \{a(Qe_\mu),a^\dagger(Qe_\nu)\}  &=\delta_{\mu\nu}1,\nonumber\\  \{a(Qe_\mu),a(Qe_\nu)\}&=0.
 \label{eq:compressed_car}
\end{align}
The effect of compression on one-particle operators can be calculated explicitly. For $A_Q=Q^\dagger A Q$ and $P=P_-+P_+$, 
\begin{align}
 [A_Q,B_Q]-Q^\dagger[A,B]Q  =-Q^\dagger A(1-P)BQ  +Q^\dagger B(1-P)AQ .
 \label{eq:compressed_commutator_defect}
\end{align}
For Hermitian $A,B$, the norm of this difference obeys 
\begin{align}
 \|[A_Q,B_Q]-Q^\dagger[A,B]Q\|_{\rm op}  \leq 2\delta_Q(A)\delta_Q(B),  \qquad  \delta_Q(A)=\|(1-P)AQ\|_{\rm op}.
 \label{eq:compressed_commutator_bound}
\end{align}
The bound relates the commutator error to the components of $AQ$ and $BQ$ that leave the retained space. The affine central term also depends on the sea--particle polarization and is checked separately through the Schwinger trace.

\paragraph{Compression of sewn amplitudes.}
\label{sec:observable_dependent_compression}
\label{sec:finite_resource_experiments}

We test how well selected fermionic subspaces reproduce the sewn four-point amplitudes of $SU(2)_3$ and $SU(3)_2$. The reference one-particle space is $\mathbb C^8\otimes\mathbb C^6$, containing eight Neveu--Schwarz modes from $-7/2$ to $7/2$ and flavor space $\mathbb C^2\otimes\mathbb C^3$. Four modes lie on each side of the sea--particle polarization. The currents act by mode shifts tensored with the appropriate $SU(2)$ or $SU(3)$ generators.

For $SU(2)_3$, we retain $(m_-,m_+)=(6,6),(8,6),(10,6)$ and compute descendants through grade three. For $SU(3)_2$, we use $(6,6),(8,6)$ through grade two. The external sectors are $(\mathbf2,\mathbf2,\mathbf3,\mathbf3)$ and $(\mathbf3,\mathbf3,\overline{\mathbf3}, \overline{\mathbf3})$, respectively, with intermediate channels $j=0,1$ and $\overline{\mathbf3},\mathbf6$.

At each grade, the reference descendant Gram matrix determines the physical positive space after null vectors have been removed. We choose this reference space once and use it to evaluate the compressed descendant contractions. The same regularized inverse and fixed Ward couplings are used for every retained subspace. The regularization and numerical construction are given in Appendix~\ref{app:s5_upgrade_methods}.

The resulting finite-rank sewn amplitudes depend on the chosen reference quotient because compressed descendants can have different null directions. We therefore measure their accuracy against both the finite reference and independent KZ transport.

For a set $S$ of cross ratios and the two intermediate channels, the relative squared error is
\begin{align}
 \mathcal E_S(P_-,P_+)  =\frac{\sum_{x\in S,c}  \|F_c(Q;x)-F_c^{\rm ref}(x)\|_2^2}  {\sum_{x\in S,c}\|F_c^{\rm ref}(x)\|_2^2},  \qquad  R_S=\sqrt{\mathcal E_S}.
 \label{eq:compression_observable_objective}
\end{align}
The sums assign equal weight to each channel and its orthonormal singlet components.

We use $161$ equally spaced cross ratios in $[0.1,0.9]$, divided into $64$ fitting nodes, $32$ nodes for selecting the fitted solution, and $65$ evaluation nodes. Two methods select the retained subspaces. Direct optimization adjusts their coordinates explicitly; neural optimization produces the same coordinates using a fixed-input tanh network. The neural output layer can reach every subspace available to the direct method. Both methods use five matched initializations and $64$ gradient evaluations per initialization.

We compare the fitted subspaces with Hardy truncation, which retains the modes of smallest absolute frequency, and with randomly selected polarized subspaces. The evaluation uses the same cross ratios and error measure for all methods.

\paragraph{Compression results.}
\label{sec:finite_resource_experimental_results}

\begin{table}[t]
\centering
\small
\begin{tabular}{lrrrr}
\toprule
condition & Hardy & Random & Direct & Neural\\
\midrule
$SU(2)_3$ $(6,6)$ & 0.0458 & 7.0854 & 0.0520 & 0.0738\\
$SU(2)_3$ $(8,6)$ & 0.5792 & 6.5633 & 0.3828 & 0.4162\\
$SU(2)_3$ $(10,6)$ & 0.3871 & 6.2779 & 0.4054 & 0.4010\\
$SU(3)_2$ $(6,6)$ & 0.3865 & 9.8797 & 0.3387 & 0.3479\\
$SU(3)_2$ $(8,6)$ & 0.5470 & 9.8454 & 0.5492 & 0.5579\\
\bottomrule
\end{tabular}
\caption{Relative RMS error of the finite-rank sewn amplitudes against the reference calculation, evaluated on held-out cross ratios. Random, direct and neural results are averaged over five matched initializations. Each condition uses the indicated numbers of retained sea and particle modes.}
\label{tab:charged_projector_fits}
\end{table}

Table~\ref{tab:charged_projector_fits} shows substantial improvements over Hardy truncation at $SU(2)_3$ with $(8,6)$ retained modes and at $SU(3)_2$ with $(6,6)$ retained modes. The gains exceed $0.02$ for every initialization in both conditions. Hardy truncation performs as well as or better than the optimized methods in the remaining conditions.

The finite reference contributes substantial truncation error: its relative RMS discrepancies from KZ transport are $16.27\%$ for $SU(2)_3$ and $29.53\%$ for $SU(3)_2$. The improvements above concern the approximation of this finite reference. Appendix~\ref{app:s5_upgrade_methods} reports the errors against KZ, projector distances and computational costs.

We also examine projection of the physical reference descendants directly into a smaller particle--hole excitation space. Applying the projection to the physical Fock vectors preserves their null relations. Through grade two, the reference Fock and affine Gram matrices agree to $2.84\times10^{-14}$, with projected null residuals below $4.37\times10^{-15}$. A restriction from $1107$ to $360$ excitation states gives a physically accurate grade-two reference on a shorter cross-ratio domain. Optimization within a flavor-preserving radial family reaches Hardy-level accuracy without a further gain. The details of this projection and its admissible subspaces are included in Appendix~\ref{app:s5_upgrade_methods}.

\subsection{Fusion-label transfer}
\label{sec:duality_transfer}

Level--rank duality relates the integrable representation labels and fusion rules of the paired WZW theories. We first examine whether this relation improves the prediction of fusion coefficients from a limited set of labeled examples. The correspondence between the simple-current orbits can be calculated from Young-diagram transposition and the orbit data.

For orbit labels $a,b,c$, we use the orbit-averaged fusion multiplicity
\begin{align}
 \mathsf F_X(a,b;c)  =\frac{\sum_{\lambda\in[a],\,\mu\in[b],\,\nu\in[c]}  N_{\lambda\mu}^{\ \nu}}  {|[a]|\,|[b]|}.
 \label{eq:transfer_fusion_dataset_target}
\end{align}
The fusion coefficients are computed from the Verlinde formula, with numerical integrality checked to $10^{-6}$ before rounding.

We study two transfer directions: $SU(5)_2\to SU(2)_5$, containing three simple-current orbits and $27$ fusion entries, and $SU(4)_3\to SU(3)_4$, containing five orbits and $125$ entries. The first target contains $13$ zeros and $14$ ones; the second contains $72$ zeros, $52$ ones and one entry equal to two.

Each primary is described by its affine Dynkin labels, dimension, quadratic Casimir, conformal weight and index:
\begin{align}
 x_{N,K}(\lambda)  &=(\lambda_1,\ldots,\lambda_{N-1},\lambda_0,  \dim V_\lambda,C_2,h_\lambda,I_\lambda),  \label{eq:transfer_full_local_features}\\
 C_2&=\lambda^TA^{-1}(\lambda+2\rho),\qquad  h_\lambda=\frac{C_2}{2(K+N)},\qquad  I_\lambda=\frac{\dim V_\lambda C_2}{N^2-1}.  \nonumber
\end{align}
Here $C_2$ and $I_\lambda$ use twice the generator-normalization convention associated with $\operatorname{tr}(t^at^b)=\delta^{ab}/2$; the displayed conformal weight includes the corresponding factor.

A shared neural encoder maps the primary features to a $24$-component vector. The vectors are averaged over each simple-current orbit. The predictor combines the resulting orbit features as
\begin{align}
 \Phi(e_a,e_b,e_c)  =(e_a+e_b,\ e_a\circledast e_b,\ e_c), 
 \label{eq:transfer_symmetric_triple_map}
\end{align}
followed by a width-$32$ tanh layer. The first two inputs enter symmetrically, as required by the fusion coefficients. We also test a predictor symmetric under all three label permutations.

Table~\ref{tab:s5t2_predictions} compares six predictors. The Exact/source column label denotes the identical predictions obtained by supplying the exact orbit correspondence or computing it from public representation data. The Blind column uses the unaligned orbit labels, and the Wrong column uses a deliberately incorrect correspondence. The Kernel column is the source-trained Gaussian-kernel ridge predictor defined in Appendix~\ref{app:s5_upgrade_methods}; the Median column is the weighted median of the source fitting coefficients. The $S_3$ control is a neural predictor invariant under all three label permutations.

Training and evaluation use thirty group partitions and five initializations. All permutations and primary representatives of the same unordered orbit multiset remain in one partition. Held-out target coefficients are excluded from fitting, preprocessing and checkpoint selection. We report mean absolute error averaged over held-out orbit multisets. The data allocation and training settings appear in Appendix~\ref{app:s5_upgrade_methods}.

\begin{table}[t]
\centering
\small
\begin{tabular}{lrrrrrr}
\toprule
orbits & Exact/source & Blind & Wrong & Kernel & Median & $S_3$ control\\
\midrule
3 & 0.7577 & 0.6640 & 0.5912   & 0.5872 & 0.5667 & 0.6729\\
5 & 0.4510 & 0.6011 & 0.5250   & 0.5002 & 0.6032 & 0.4056\\
\bottomrule
\end{tabular}
\caption{Held-out fusion-prediction MAE for $SU(5)_2\to SU(2)_5$ (three orbits) and $SU(4)_3\to SU(3)_4$ (five orbits). Exact/source uses the supplied or algebraically computed correspondence; Blind leaves the labels unaligned; Wrong uses an incorrect correspondence. Kernel and Median use source fitting labels, and the $S_3$ control is invariant under all three label permutations. Results are averaged over partitions, initializations and relabelings.}
\label{tab:s5t2_predictions}
\end{table}

The exact-map and source-computed predictors have identical errors, as expected from their common label correspondence. In the three-orbit task, both kernel and median prediction outperform the aligned neural predictor. In the five-orbit task, the aligned neural predictor improves on those baselines. The fully $S_3$-symmetric control has lower error in both transfer directions.

We also allow the learner to request selected target fusion coefficients, with budgets of up to three or seven queries for the respective tasks. The public representation data already identify the correct label correspondence before any query is made. Neither task reaches the specified group-MAE target of $0.40$ within its query budget.

The fusion experiments show that the label correspondence can be calculated from the available representation data, giving the same predictions as a supplied exact map. We next use the pairing for a different purpose: transferring information about the relative couplings of charged WZW channels. These couplings vary with an external control and are learned from source-sector correlators.

\subsection{Learning charged-channel couplings across level--rank duality}
\label{sec:duality_preparation_learning}

The charged four-point functions studied in Section~\ref{sec:quantized_neural_transport} have two intermediate channels. Their contributions depend on the couplings of the external charged states to those channels. We consider a family in which these couplings vary with a control parameter $s$, and write
\[
 \ell(s)=\bigl(\ell_1(s),\ell_2(s)\bigr)^T.
\]
For $SU(2)_3$, the two entries refer to the $j=1$ and $j=0$ channels. Level--rank duality pairs them with the $\overline{\mathbf3}$ and $\mathbf6$ channels of $SU(3)_2$.

We take the dependence of $\ell(s)$ on $s$ to be unknown. The learning task is to use a limited set of source-sector correlators to predict the corresponding correlators in the paired sector, including values at controls and cross ratios absent from training. The WZW channel structure and level--rank pairing are known. The control-dependent couplings are learned from the source observations.

To implement these observables, we construct charged polynomials from the angular modes of the finite Fourier--Berezin fields. Their coefficients are fixed by the physical descendant states and Ward couplings of the two WZW theories.

\paragraph{Charged angular observables.}

At grade $n$ in intermediate channel $\lambda$, let $S_{\lambda n}$ contain the physical descendant states in the reference Fock basis, with Gram matrix $G_{\lambda n}=S_{\lambda n}S_{\lambda n}^T$. The corresponding Ward couplings define
\begin{align}
 u_{\lambda n a} &=S_{\lambda n}^TG_{\lambda n}^+C_{\lambda n a},\nonumber\\
 v_{\lambda n b} &=S_{\lambda n}^TG_{\lambda n}^+D_{\lambda n b}.
\end{align}
Expanding these vectors in the Fock basis and substituting the neural angular modes of \eqref{eq:fermionic_neural_mode_network} gives charged polynomials $\mathcal U_{\lambda n a,L}$ and $\mathcal V_{\lambda n b,L}$. They carry charges $+2$ and $-2$, respectively, and are Wick ordered at fixed sampled frequencies.

We assemble their matrix elements into the finite-grade response
\begin{align}
 (F_{X,L})_{i\lambda}(x)
 =\sum_{n=0}^{D}x^{\alpha_\lambda+n}
 \sum_{a,b}\Phi_{\lambda,i,ba}
 \mathbb E_{\Theta_F}
 [\mathcal V_{\lambda n b,L}\mathcal U_{\lambda n a,L}],
 \label{eq:capstone_realized_neural_response}
\end{align}
where $\Phi_{\lambda,i,ba}$ denotes the external invariant contraction and $\alpha_\lambda$ is the channel exponent. The expectation uses the original Berezin pairing, followed by averaging over the sampled frequencies. Contributions from coincident frequency samples are included.

The powers of $x$ assemble the descendant sewing series. Their coefficients are calculated from the neural angular polynomials with the specified Ward data. The construction therefore gives explicit finite-neural observables for the charged-channel responses.

Let $C_A,C_B$ denote the CAR reference response matrices at $x_0=0.2$, and define $A_L(x)=F_{A,L}(x)C_A^{-1}$ and $B_L(x)=F_{B,L}(x)C_B^{-1}$. These matrices describe how the known source and target observables depend on the two channel coefficients.

For fixed readout vectors $v,w$, we use
\begin{align}
 y_{A,L}(s,x;v)&=\ell(s)^TA_L(x)v,\nonumber\\
 y_{B,L}(s,x;w)&=w^TB_L(x)P_0^{-1}\ell(s),
 \qquad P_0=x_0^{1/6}(1-x_0)^{1/3}P .
 \label{eq:capstone_realized_paired_readouts}
\end{align}
The matrix $P$ is the continuum level--rank pairing derived in Appendix~\ref{app:charged_level_rank_pairing}.

At each $s$, the source response is a linear combination of two known channel responses, weighted by the unknown entries of $\ell(s)$. The target response uses the corresponding weights transformed by $P_0^{-1}$. Source observations provide the training data for this dependence. The sector response matrices and the duality pairing are calculated independently.

\paragraph{Accuracy of the charged observables.}

We first compare the neural angular construction with WZW transport. The complete charge-two Fock spaces contain $5379$ states through grade three and $21777$ through grade four. Their physical Gram matrices and Ward couplings give sewing coefficients that agree with independent KZ recursion to $2.3\times10^{-15}$. The grade-three and grade-four calculations also agree on their common lower grades.

At finite neural width, the ensemble calculation includes all contributions from repeated frequency samples. The grade-two, grade-three and grade-four calculations contain $8$, $17$ and $35$ distinct charged monomials, with maximum degrees $6$, $8$ and $10$, respectively. Their exact frequency contractions are described in Appendix~\ref{app:lr_controlled_preparations}.

For the accuracy comparison we use $\alpha=1/2$, $\epsilon_F=0.001$ and $L=2^{21}$, with each grade normalized at $x_0=0.2$. We evaluate the ensemble responses at $34$ cross ratios in $[0.1,0.4]$.

\begin{table}[!htbp]
\centering
\begin{tabular}{lrrrr}
\toprule
 & \multicolumn{2}{c}{RRMS versus KZ (\%)} & \multicolumn{2}{c}{maximum node error (\%)}\\
grade & $SU(2)_3$ & $SU(3)_2$ & $SU(2)_3$ & $SU(3)_2$\\
\midrule
two   & 0.546 & 0.808 & 1.347 & 1.928\\
three & 0.217 & 0.319 & 0.529 & 0.775\\
four  & 0.118 & 0.153 & 0.234 & 0.336\\
\bottomrule
\end{tabular}
\caption{Relative errors of the finite-neural ensemble responses against full KZ transport at $34$ cross ratios in $[0.1,0.4]$. The width and regulator are fixed across grades, with separate CAR normalizations.}
\label{tab:lr_physical_precision}
\end{table}

The error decreases with descendant grade in both theories. At grade four, the relative RMS errors are $0.118\%$ for $SU(2)_3$ and $0.153\%$ for $SU(3)_2$, with maximum errors below $0.34\%$ on the sampled domain. The calculations concern ensemble averages; the width--regulator dependence and sampling estimates are developed in the appendix.

\paragraph{Prediction from source measurements.}

We next examine how accurately the unknown channel couplings can be transferred using source observations. Ten neural predictors are trained on grade-two source responses at $\epsilon_F=0.02$ and $L=131072$. Each predictor receives $48$ fitting observations and $16$ validation observations.

The trained networks are evaluated at $201$ control values and $29$ unseen cross ratios per family. We compare their predictions with the refined grade-three finite-neural responses and independent KZ calculations through grades three and four. The predictors retain their source-trained parameters throughout these comparisons.

\begin{table}[!htbp]
\centering
\begin{tabular}{lrr}
\toprule
mean test RRMS (\%) & $SU(2)_3\to SU(3)_2$ & reverse\\
\midrule
grade-three neural response & 0.322 & 0.346\\
grade-three KZ reference & 0.466 & 0.400\\
grade-four KZ reference & 0.363 & 0.367\\
\bottomrule
\end{tabular}
\caption{Transfer errors for ten predictors trained on source-sector charged observables. Each direction contains five families of control-dependent channel couplings. The grade-three finite-neural response is the primary target, with independent KZ comparisons through grades three and four. No target observations are used for network fitting.}
\label{tab:lr_no_refit}
\end{table}

The mean errors are below $0.5\%$ against full KZ transport in both directions. All ten families have aggregate errors below $1\%$ against the grade-three neural responses and KZ references. The largest relative error for an individual finite-neural test response is $4.14\%$.

The transfer uses the source observations together with the independently computed response matrices and continuum pairing. The latter provide the target-sector operators; the learned dependence on $\ell(s)$ supplies the unknown channel information. For the public Fourier family, an exact source estimator is more accurate than the learned predictor.

\begin{figure}[!tp]
\centering
\begin{subfigure}[t]{.48\textwidth}
\centering
\includegraphics[width=\linewidth]{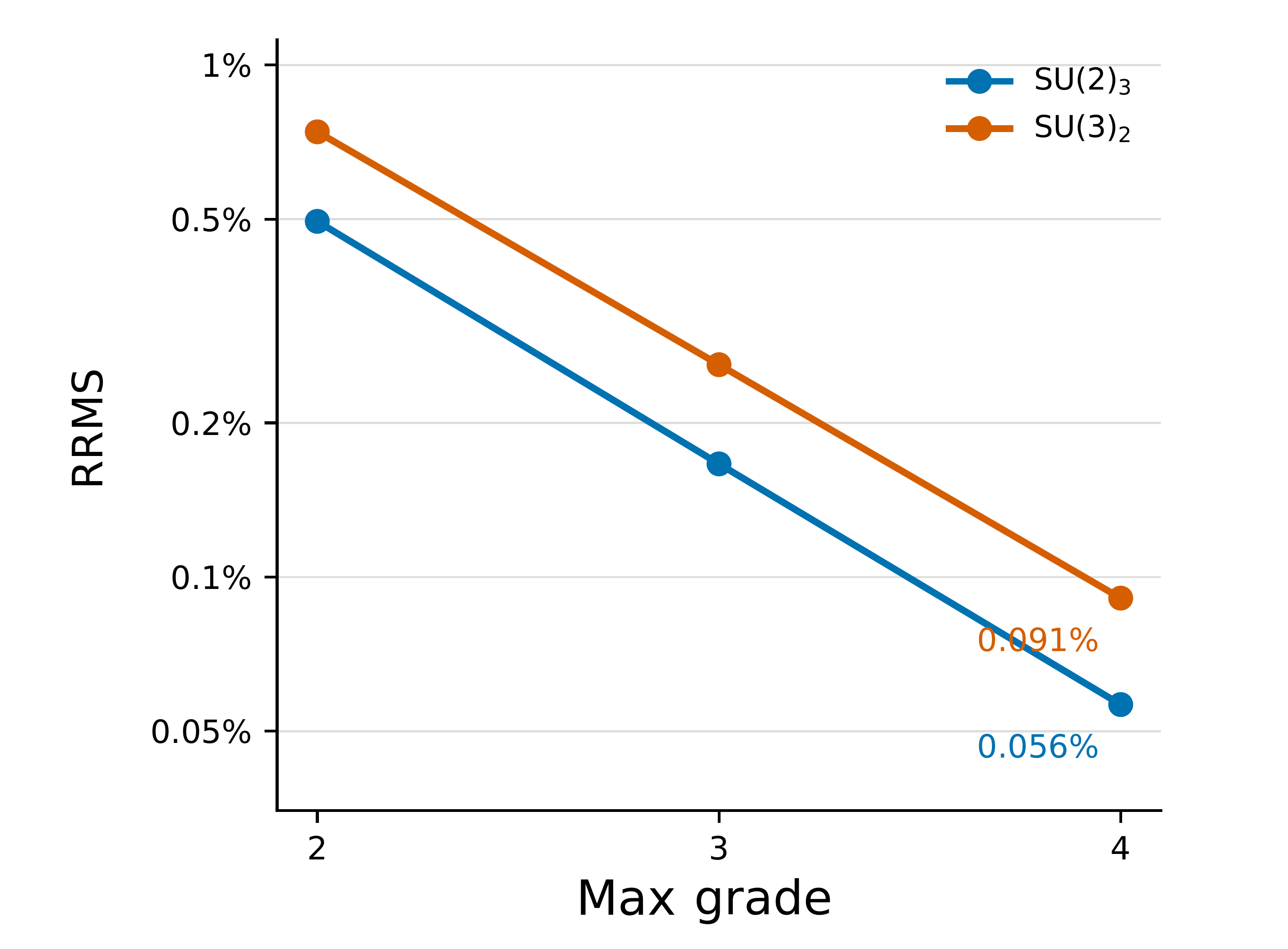}
\caption{Convergence with descendant grade.}
\end{subfigure}
\hfill
\begin{subfigure}[t]{.48\textwidth}
\centering
\includegraphics[width=\linewidth]{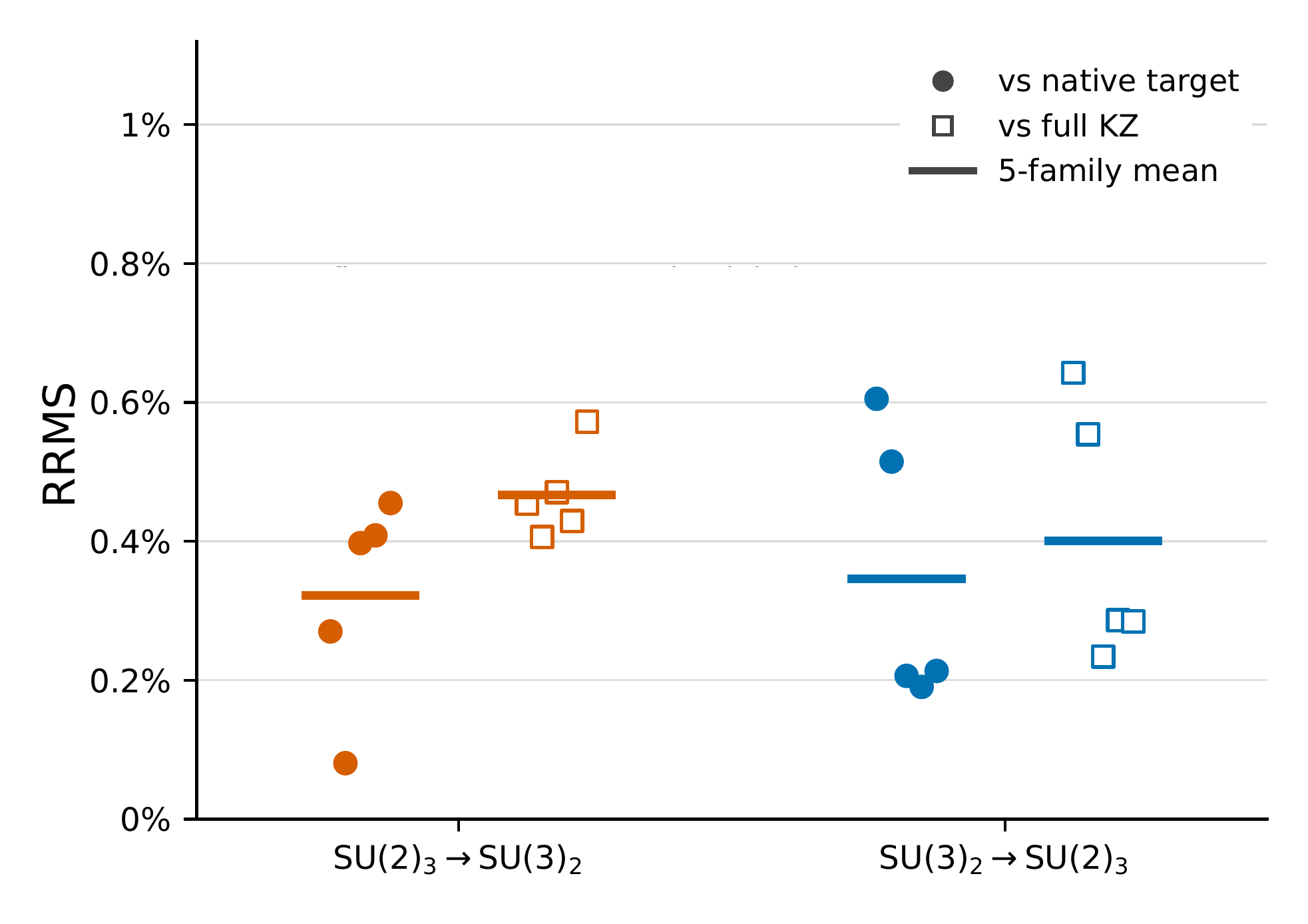}
\caption{Transfer errors across charged-state families.}
\end{subfigure}
\caption{Physical accuracy and transfer of charged-channel information. Left: CAR descendant truncations compared with KZ transport on $301$ cross ratios. Right: source-trained predictors evaluated against grade-three finite-neural responses on $201\times29$ control--cross-ratio combinations per family. Points give errors for individual families and horizontal bars the mean over five families. Each predictor uses $64$ source fitting and validation observations.}
\label{fig:lr_precision_transfer}
\end{figure}

\paragraph{Pure charged probes.}

We also implement the channel couplings through individual charged bra--ket matrix elements. The source uses a fixed charged ket and a control-dependent coherent bra. The target uses a fixed bra and a control-dependent incoming ket. For example, the source bra has the form $\mathcal B(s)=\ell_1(s)\mathcal B_1+\ell_2(s)\mathcal B_2$, where the two polynomials correspond to the intermediate channels. The coefficients of the basis probes are calculated from their physical Ward couplings. The complete construction is given in Appendix~\ref{app:lr_controlled_preparations}.

For these observables, the inner neural fields are evaluated at radius $S(x)=x$, including their weight-$1/2$ dilation factors. The radius changes the Bessel features entering the finite-width pairing. A grade-$n$ charged monomial has free energy $n+1$, and the channel and external-field factors produce the required sewing powers. The ensemble expectation includes all repeated-frequency contributions.

We test five nonlinear families of channel couplings in each direction, using grade three, $L=524288$, $\epsilon_F=0.02$ and radial frequency support $p\in[0.001,400]$. Each source predictor receives $48$ fitting observations and $16$ validation observations at $x=0.1,0.4$. The test uses new control values and cross ratios between those training geometries.

We repeat the calculation after adding independent Gaussian measurement noise with standard deviation equal to $1\%$ of the public operator RMS.

\begin{table}[!htbp]
\centering
\begin{tabular}{lrr}
\toprule
mean pure-probe RRMS (\%) & $SU(2)_3\to SU(3)_2$ & reverse\\
\midrule
no added noise & 1.35 & 2.49\\
$1\%$ measurement noise & 2.65 & 3.46\\
\bottomrule
\end{tabular}
\caption{Prediction errors for nonlinear charged-state families realized through single bra--ket matrix elements. The values are averaged over five families in each direction and measured against the finite-neural responses.}
\label{tab:lr_pure_learning}
\end{table}

The mean errors range from $1.35\%$ to $3.46\%$. All twenty combinations of family, transfer direction and noise condition have aggregate errors below $10\%$, with the largest family error below $7.32\%$. The finite-neural responses themselves differ from KZ transport by approximately $2\%$ at the selected regulator.

Kernel ridge regression on the source data gives smaller errors without added noise. Training directly on target observations with the same observation budget also improves accuracy. The experiment establishes transfer of nonlinear channel couplings through the realized charged observables, while these alternative estimators remain competitive.

\paragraph{Convergence of the angular observables.}

The angular construction approaches the physical WZW sewing amplitudes under controlled width, regulator and descendant limits. At fixed mode number and grade, the Fourier--Berezin angular moments converge under regulator removal when $L\epsilon_F^{2\alpha}\longrightarrow\infty$ for $0<\alpha<1$. The same moment estimates control conjugate products and sampling fluctuations.

With the specified affine Ward couplings and intertwiners, the limiting coefficients satisfy the KZ recurrence. A bound on the descendant tail then gives locally uniform convergence on compact subsets of $0<|x|<1$. The proof and normalization estimates are given in Appendix~\ref{app:lr_controlled_preparations}.

The angular projections are also important for the form of these observables. At finite regulator, polynomial expectations formed from finitely many point evaluations and derivatives are bounded and real analytic. Exact holomorphicity would force such an expectation to be constant. The corresponding obstruction and its assumptions are established in the appendix.

The finite-neural angular observables thus provide controlled approximations to the charged WZW responses. Their paired channel structure allows information learned from source correlators to be used in predicting the responses of the dual theory.

\section{Discussion and outlook}
\label{sec:discussion}

We have developed bosonic and fermionic neural constructions of positive-level WZW current algebras and used level--rank duality to investigate their finite approximations. The bosonic Fourier architecture realizes the scalar and first-order fields of the Wakimoto construction, while the fermionic Fourier--Berezin architecture produces commuting row and column current algebras through fermion bilinears. Both constructions recover the prescribed affine structures in controlled continuum limits. Their different elementary fields, parameter distributions and composite observables provide considerable freedom in how those limits are approached.

The finite-width analysis shows that this freedom has measurable consequences. Redistributing spectral weight between the frequency law and feature amplitudes preserves the elementary two-point functions while changing the higher moments of composite observables. The resulting corrections depend on the correlation function, insertion geometry and regulator. In particular, the spectral distributions that give the most accurate stress-tensor correlations can differ from those preferred by higher-point functions. The comparison between direct Wakimoto and coset constructions further reveals the sensitivity of composite observables to regulator corrections in the parent theory. Amplitude calibration can reduce these errors, although improvements in the fitted correlations do not always extend to other observables or insertion positions. The relative computational costs also depend strongly on the calculation being performed and the extent to which intermediate results can be reused. These findings show how architecture and parameter choice influence the accuracy and efficiency of neural realizations even when their limiting current algebras agree.

The fermionic construction provides a further connection between neural observables and the representation theory of WZW models. Angular projections of the Fourier--Berezin fields recover the limiting fermionic mode pairing, whose CAR/Fock completion supports charged sectors and affine descendants. For the six-channel construction, the decomposition of charge-two states identifies the paired intermediate channels of $SU(2)_3$ and $SU(3)_2$. Sewing their descendants reproduces the corresponding KZ expansions through the tested grades. We extend this construction to explicit charged angular polynomials of the finite neural fields, using physical Ward couplings to determine their coefficients. The resulting observables approximate the WZW sewing amplitudes with controlled width, regulator and descendant limits. The restricted channel content of local charge-balancing probes and the finite-regulator obstruction to exact holomorphic pointwise correlators further motivate the use of angular observables.

The finite-mode compression experiments examine how the fermionic state spaces can be approximated using fewer degrees of freedom. Isometric reduction preserves the canonical anticommutation relations of the retained modes, while the resulting current commutators and sewn amplitudes depend on the chosen subspaces. Optimizing these subspaces improves on conventional mode truncation in two of the five settings studied. Direct and neural optimization give comparable results, suggesting that the geometry of the retained modes matters more than the particular parameterization used to select them. The finite descendant reference introduces additional truncation error, which remains important when assessing the accuracy of the resulting conformal-block approximations.

Level--rank duality also provides a setting for transferring information between neural realizations. The fusion experiments show that the correspondence between simple-current orbits can be recovered directly from public representation data. Supplying this correspondence to a predictor gives the same results as computing it from the available algebraic information. A more interesting learning problem arises when the relative couplings of charged WZW channels are unknown. The charged angular observables allow these couplings to be inferred from source-sector measurements and used to predict responses in the paired theory. The source-trained predictors achieve subpercent mean errors against independent KZ calculations in both directions, without observations of the unknown target couplings. Pure charged probes extend the calculation to nonlinear coupling families with measurement noise. The transfer relies on independently computed WZW response operators and the continuum level--rank pairing, while the dependence on the unknown charged states is learned from data. Exact source estimators and direct target training remain competitive alternatives.

Several questions follow from these results. On the field-theory side, an important goal is to construct neural charged observables whose conformal-block structure emerges without prescribing the affine Ward couplings and sewing data. The present angular construction provides a controlled starting point, while the pointwise obstruction identifies restrictions on possible finite-regulator realizations. On the computational side, a fuller comparison should account jointly for approximation error, sampling uncertainty, mode truncation and the cost of evaluating the relevant observables. Such an analysis would help determine when the algebraic information supplied by duality leads to practical advantages in learning and computation. More broadly, the constructions developed here provide a setting in which exact relations between quantum field theories can inform the design and analysis of neural architectures.

\paragraph{Acknowledgments} The authors thank Christian Ferko and Jim Halverson for useful conversations during the development of this project. B.~R.~is supported by NWO Vidi grant 016.Vidi.189.182.
B.~S.~is supported by the National Research Foundation of Korea (NRF) grant RS-2025-00516583 and access to the ChatGPT for Academic Researchers program was provided by OpenAI.

\paragraph{AI usage statement} Codex (OpenAI) and Claude Code (Anthropic) assisted code development and numerical checking and ChatGPT assisted in manuscript editing. All scientific claims and outputs are the sole responsibility of the authors.

\appendix
\addtocontents{toc}{\protect\setcounter{tocdepth}{1}}
\section{Bosonic substrate in more detail}
\label{app:bosonic-substrate-more-detail}

\subsection{Neural Wakimoto fields for \texorpdfstring{$\widehat{\mathfrak{su}}(N)_K$}{su(N) level K}}
\label{app:sunk_wakimoto}

In Section~\ref{sec:bosonicsubstrate}, we introduced the bosonic substrate, consisting of scalar neural fields $\phi^A$ and commuting ghost pairs $(\beta^B,\gamma^B)$, and constructed the three currents of $\widehat{\mathfrak{su}}(2)_K$. We now give the corresponding construction for $\widehat{\mathfrak{su}}(N)_K$, with $N\geq2$ and $K\in\mathbb Z_{>0}$. The sampled feature width remains denoted by $L$.

Let $E_{jk}$ denote the numerical $N\times N$ matrix with a single nonzero entry equal to one in row $j$ and column $k$:
\begin{align}
    (E_{jk})_{rs} =\delta_{rj}\delta_{sk}.
\end{align}
For each pair $i<j$, corresponding to a positive root, introduce one commuting ghost pair $(\beta^{ij},\gamma^{ij})$. Thus the elementary neural fields are
\begin{align}
    \phi^A,    \qquad A=1,\ldots,N-1,    \qquad    (\beta^{jk},\gamma^{jk}),    \qquad 1\leq j<k\leq N.
    \label{eq:fw_general_wakimoto_fields}
\end{align}
The double superscript $(j,k)$ replaces the single ghost-species label $a$ used in the main text; it is not a neuron index. There are $N-1$ scalar fields and $N(N-1)/2$ ghost pairs.

All fields use the architectures and parameter laws of Section~\ref{sec:bosonicsubstrate}. In particular, the frequency samples remain shared between the scalar and ghost sectors. The construction below changes only the number of elementary fields and the composite readouts assembled from them.

We choose the Hermitian generators of $\mathfrak{su}(N)$ as follows,
\begin{subequations}
\begin{align}
    t^{S,jk} &= \frac12(E_{jk}+E_{kj}), & t^{Y,jk} &= \frac1{2i}(E_{jk}-E_{kj}),\qquad 1\leq j<k\leq N,
\end{align}
\begin{align}
    t^{D,A} &= \frac{\sum_{i=1}^{A}E_{ii}-A\,E_{A+1,A+1}}{\sqrt{2A(A+1)}}, \qquad A=1,\ldots,N-1.
\end{align}
\end{subequations}
The superscripts $S$, $Y$, and $D$ distinguish the symmetric, imaginary antisymmetric, and diagonal generators, respectively. Collectively, these $N^2-1$ matrices are denoted by $t^a$, with
\begin{align}
    \operatorname{tr}_N(t^at^b)
    &=    \frac{\delta^{ab}}2,    &    [t^a,t^b] &= i f^{ab}{}_{c}t^c.
    \label{eq:fw_general_generator_normalization}
\end{align}
For $N=2$, this gives $t^a=\sigma^a/2$. For $N=3$, it gives the usual Gell-Mann basis.

To express all current components in one finite sum, extend the notation for the ghost coordinates by
\begin{align}
    \gamma^{ii}=1, \qquad \gamma^{jk}=0\quad\text{for }j>k.
    \label{eq:fw_extended_ghost_coordinates}
\end{align}
These are constant assignments, not additional fields, and all their input derivatives vanish.

With $\partial=\partial_z$, the current associated with $t^a$ takes the compact form

\begin{align}\label{eq:fw_general_wakimoto_readout}
    \mathcal{J}^a(z) ={}& \sum_{\ell=0}^{N-1} \sum_{1\leq s_0<\cdots<s_\ell}^N \sum_{r=1}^{N}(-1)^\ell(t^a)_{s_\ell r} :\!\Bigg[\sum_{u=1}^{s_0}\sum_{v=s_0+1}^{N}\beta^{uv}\gamma^{rv}\gamma^{u s_0} \\\nonumber
    &\hspace{8mm}
        +i\sqrt{2(K+N)}\,\gamma^{r s_0}\sum_{A=1}^{N-1}(t^{D,A})_{s_0s_0}\partial\phi^A \\\nonumber
    &\hspace{8mm}-(K+N-s_0)\partial\gamma^{r s_0}\Bigg]\prod_{\nu=0}^{\ell-1}\gamma^{s_\nu s_{\nu+1}}\!:(z),
\end{align}
where $s_j< s_i$ for $j<i$. All elementary fields on the right-hand side are evaluated at $z$ and their antiholomorphic arguments are suppressed. The current index $a$ is not summed.

The index $\ell$ counts the number of ghost factors in the final product, not the number of sampled neurons. For $\ell=0$, the sequence contains only $s_0$, and the empty product is defined to be one. For $\ell=1$, the factor is $-\gamma^{s_0s_1}$; for $\ell=2$, it is $\gamma^{s_0s_1}\gamma^{s_1s_2}$. The indices increase strictly, so $\ell\leq N-1$ and the expansion terminates. These products are the finite inverse expansion associated with the triangular ghost coordinates. For example, the two sequences connecting indices $1$ and $3$ contribute $-\gamma^{13}+\gamma^{12}\gamma^{23}$.

The three terms in square brackets have distinct roles. The first contains the ghost products, the second couples the ghost coordinates to the scalar gradients, and the third supplies the ghost-derivative terms. The restriction $u\leq s_0<v$ guarantees that every $\beta^{uv}$ appearing in the first term belongs to an existing ghost pair. The constant extension \eqref{eq:fw_extended_ghost_coordinates} determines which remaining factors vanish or reduce to one.

The factor $K+N$ contains the usual shift by the dual Coxeter number $h^\vee=N$. The coefficients $K+N-s_0$ also include the normal-ordering correction appropriate to the chosen ghost coordinates. They are fixed by the Wakimoto construction and are not adjustable network parameters \cite{deBoer:1996ti}.

Every $\partial$ acts only on the elementary field immediately following it. In particular, $\partial\gamma^{r s_0}$ does not differentiate the final product of ghost coordinates. The normal-ordering prescription is applied after expanding the brackets and products: each term containing a $\beta^{uv}$ is ordered with that field to the left of its accompanying polynomial in the $\gamma$ fields. The colons denote the elementary Wick subtractions defined in Section~\ref{sec:bosonicsubstrate}, not a sequence of nested normal products.

For the regulated ensemble used here, the coincident elementary contractions entering these current readouts vanish. The readouts can therefore be evaluated by ordinary multiplication of the displayed neural outputs, while retaining the explicit derivative terms. The normal-ordering symbols specify the associated free-field operators in the continuum limit.

In the ordered infinite-width and regulator-removal limits, the elementary contractions and the Wakimoto prescription give the target current relation \eqref{eq:target-current-ope} \cite{deBoer:1996ti}. At finite width and fixed regulators, \eqref{eq:fw_general_wakimoto_readout} instead defines the composite neural outputs whose correlation functions are evaluated in the parameter ensemble.

In this basis, the stress-energy neural field is
\begin{align}
    T(z) ={}&-\frac12\sum_{A=1}^{N-1}:\!\bigl(\partial\phi^A\bigr)^2\!:(z) - \sum_{1\leq j<k\leq N}:\!\beta^{jk}\partial\gamma^{jk}\!:(z)-\frac{i}{\sqrt{K+N}}\sum_{A=1}^{N-1}\boldsymbol\rho^A\partial^2\phi^A(z).
    \label{eq:fw_general_stress_readout}
\end{align}
The first two terms sum the scalar and first-order-field contributions. The last term is the scalar background-charge contribution, with $\boldsymbol\rho$ the Weyl vector, defined as half the sum of the positive roots. Consequently, in our basis,
\begin{align}
    \boldsymbol\rho^A &= \frac{\sqrt{A(A+1)}}2.
\end{align}
Thus the improvement term is completely specified by the numerical Cartan basis; it introduces no additional parameter. For $SU(2)$, $\boldsymbol{\rho}^1=1/\sqrt2$, while for $SU(3)$, $\boldsymbol\rho=\left(\frac1{\sqrt2},\sqrt{\frac32}\right)$.

The elementary normal products in \eqref{eq:fw_general_stress_readout} are distinct from the affine products of the composite currents. With the affine prescription, the continuum Sugawara relation is
\begin{align}
    T(z) = \frac1{K+N}\sum_{a=1}^{N^2-1}:\!\mathcal{J}^a\mathcal{J}^a\!:\!_{\mathrm{aff}}(z)
    \label{eq:fw_general_sugawara_relation}
\end{align}
\cite{deBoer:1996ti}.
The elementary-field expression \eqref{eq:fw_general_stress_readout} is used as the corresponding finite neural readout. It must not be replaced by the ordinary quadratic sum of the current outputs.

For the rotationally invariant regulated ensemble, the elementary coincident contractions in \eqref{eq:fw_general_stress_readout} vanish, so its elementary normal products require no additional numerical subtraction. Its continuum central charge is
\begin{align}
    c &= (N-1)+N(N-1) - \frac{12}{K+N}\sum_{A=1}^{N-1}(\rho^A)^2 = \frac{K(N^2-1)}{K+N}.
\end{align}

The direct $SU(2)_3$ construction and the $SU(6)_1$ parent used in the coset comparison in Section~\ref{sec:finite_width_regulator_design} are obtained by setting $(N,K)=(2,3)$ and $(N,K)=(6,1)$, respectively. The same formulas therefore specify both presentations without changing the elementary architecture or the normal-ordering convention.

\paragraph{Parent-based affine normal products.}
For completeness we specify the finite differential polynomial used for the denominator products in  \eqref{eq:fw_denominator_stresses}. Let $P,Q$ be the elementary normally ordered parent-current polynomials obtained from \eqref{eq:fw_general_wakimoto_readout} and \eqref{eq:fw_denominator_currents}. Write the elementary continuum contraction as $\langle u(z+t)v(z)\rangle_0=c_{uv}t^{-m_{uv}}$. The nonzero entries are $c_{\partial\phi^A,\partial\phi^B}=-\delta^{AB}$ with $m=2$, and $c_{\beta^a,\gamma^b}=-\delta^{ab}$, $c_{\gamma^a,\beta^b}=\delta^{ab}$ with $m=1$, as in \eqref{eq:phibetagamma_ope_definition}. Derivative jets contract by
\[
 H_{(u,d),(v,e)}(t)=c_{uv}(-1)^d(m_{uv})_{d+e}\,t^{-m_{uv}-d-e},
\]
where $(m)_r=m(m+1)\cdots(m+r-1)$ and $(m)_0=1$.

Expand $P(z+t)Q(z)$ by the bosonic Wick rule over all cross-contraction matchings $\mathcal C$ between its two factors, including the empty matching. Taylor expand only the remaining left fields about $z$. If $\mathcal C$ has total pole order $p(\mathcal C)$, retain the total Taylor degree $p(\mathcal C)$ of those left fields. Equivalently, define
\[
 \mathcal N_{\rm aff}(P,Q)(z) =[t^0]\sum_{\mathcal C}  \left(\prod_{(a,b)\in\mathcal C}H_{ab}(t)\right)  :\!\prod_{a\notin\mathcal C}u_a(z+t)        \prod_{b\notin\mathcal C}v_b(z)\!:_{\rm el}.
\]
Here $[t^0]$ selects the Laurent constant term after the Taylor expansion; the colons on the right denote elementary normal ordering. The matching and Taylor sums are finite. The empty matching gives the ordinary product, while the other matchings supply the affine derivative corrections. This convention also reproduces \eqref{eq:su2-neural-affine-products}.

We use $:\!\mathfrak J_{\mathrm D}^A\mathfrak J_{\mathrm D}^A\!:\!_{\rm aff} =\mathcal N_{\rm aff}(\mathfrak J_{\mathrm D}^A,\mathfrak J_{\mathrm D}^A)$, then form $T_{\mathrm D}=\tfrac15\sum_A\mathcal N_{\rm aff} (\mathfrak J_{\mathrm D}^A,\mathfrak J_{\mathrm D}^A)$ and $T_{\mathrm C}=T_{\mathrm P}-T_{\mathrm D}$. The continuum contractions specify the coefficients of this readout polynomial once. The resulting polynomial is then evaluated on the regulated finite neural fields, and its correlations use their full finite-width parameter law. No continuum identity, discarded null polynomial, or Gaussian contraction at the mean kernel replaces that finite expectation.

\subsection{Finite-width stress-tensor two-point function}
\label{app:finite_width_TT}

In general, computing the exact finite-width expressions for all correlators in the neural representation of the affine algebra $\widehat{\mathfrak{su}}(N)_K$ is technically demanding. However, in simpler cases, we can go through the derivation exactly. That is the situation chosen here, where we evaluate the stress-tensor correlator of the direct $\widehat{\mathfrak{su}}(2)_K$ presentation at arbitrary width $L$. Throughout this calculation, we use the symmetric regulator and radial normalization introduced in Section~\ref{sec:bosonicsubstrate},
\begin{align}
    R_{\epsilon_B}(p)
    &=e^{-\epsilon_B(p+p^{-1})},
    &
    Z_\alpha
    &:=Z_\alpha^{(\Lambda_B)}
      =\int_{p_-}^{p_+}
        R_{\epsilon_B}(p)p^{2\alpha-1}\,dp,
    \qquad
    \Lambda_B=(p_-,p_+,\epsilon_B).
\end{align}
Every elementary feature carries the compensating factor $p^{-\alpha}$, with
\begin{align}
    C_\phi^2=\frac{4Z_\alpha}{\sigma^2},
    \qquad
    C_\beta=-i\sqrt{\frac{Z_\alpha}{\sigma^2}},
    \qquad
    C_\gamma=\sqrt{\frac{Z_\alpha}{\sigma^2}}.
\end{align}
The scalar and ghost fields share their frequency samples, while their Gaussian output weights are independent.

The elementary coincident contractions vanish, so the stress readout is
\begin{align}
    T(z)
    =
    -\frac12(\partial\phi(z))^2
    -\beta(z)\partial\gamma(z)
    -\frac{i}{\sqrt{2(K+2)}}\partial^2\phi(z),
    \qquad \bE[T(z)]=0.
\end{align}
Expanding its two-point function and removing terms with odd Gaussian-amplitude moments gives
\begin{align}
    \bE[T(z)T(w)]
    ={}&
    \frac14\bE[(\partial\phi(z))^2(\partial\phi(w))^2]
    \nonumber\\
    &+\frac12\bE[(\partial\phi(z))^2
                         \beta(w)\partial\gamma(w)]
    +\frac12\bE[\beta(z)\partial\gamma(z)
                         (\partial\phi(w))^2]
    \nonumber\\
    &+\bE[\beta(z)\partial\gamma(z)
                  \beta(w)\partial\gamma(w)]
    -\frac{1}{2(K+2)}
       \bE[\partial^2\phi(z)\partial^2\phi(w)].
    \label{eq:app_TT_decomposition}
\end{align}
We contract the Gaussian output weights first and then average the scalar phases and frequencies. Gaussian factorization is not assumed for the finite-width neural fields themselves.

\paragraph{Radial integrals.}

Write $z-w=re^{i\theta}$ and $z+w=\tilde{r}e^{i\tilde{\theta}}$. The elementary derivative kernels are
\begin{align}
    \mathcal C(r,\theta) &:= \bE[\partial\phi(z)\partial\phi(w)]
    =
    -2e^{-2i\theta} \int_{p_-}^{p_+} R_{\epsilon_B}(p)p\mathrm{J}_2(2pr)\,dp, \nonumber\\
    \mathcal D(r,\theta) &:= \bE[\partial^2\phi(z)\partial^2\phi(w)]
    =
    2e^{-4i\theta} \int_{p_-}^{p_+} R_{\epsilon_B}(p)p^3\mathrm{J}_4(2pr)\,dp.
    \label{eq:app_TT_derivative_kernels}
\end{align}
Both are independent of $L$ and $\alpha$, and $\mathcal D(r,\theta)=\partial_z\partial_w\mathcal C(r,\theta)$. The repeated-frequency contributions are expressed through
\begin{align}
    \mathcal H_\alpha(r,\theta) := Z_\alpha e^{-4i\theta} \int_{p_-}^{p_+} R_{\epsilon_B}(p)p^{3-2\alpha} \mathrm{J}_4(4pr)\,dp,
    \qquad
    \mathcal H_\alpha(0,\theta):=0.
    \label{eq:app_TT_repeated_frequency_integral}
\end{align}
The dependence of these functions on $\Lambda_B$ is left implicit. At fixed $\epsilon_B>0$, the symmetric regulator makes all these integrals finite after annulus removal for every real $\alpha$.

\paragraph{Mixed and scalar contributions.}

The mixed terms vanish even though the frequencies are shared. Averaging the output weights and scalar phases gives
\begin{align}
    \bE[(\partial\phi(z))^2 \beta(w)\partial\gamma(w)] &= \frac{4Z_\alpha^2}{L^2} \sum_{j,k=1}^{L} \bE_W\!\left[ \frac{W_j^2W_k^2}{p_j^{2\alpha}p_k^{2\alpha}} \right]
    =0.
    \label{eq:app_TT_mixed_zero}
\end{align}
For $j\ne k$, the two angular averages vanish separately; for $j=k$, the remaining angular factor is $e^{4i\varphi_j}$, with $W_j=p_je^{i\varphi_j}$, and also averages to zero. The term with $z$ and $w$ exchanged vanishes by the same argument.

For the scalar fourth moment, the Gaussian weight contractions separate into contributions with two distinct neuron indices and contributions with one repeated index. The former give the factor $L(L-1)/L^2$. For the latter, the fourth amplitude moment is $3\sigma^4$, and the scalar-phase average is
\begin{align}
    \int_0^{2\pi}\frac{d\delta}{2\pi} \sin^2(a+\delta)\sin^2(b+\delta) = \frac14+\frac18\cos\bigl(2(a-b)\bigr).
\end{align}
The constant term vanishes upon averaging the remaining factor $W^4$ over its angle. Consequently,
\begin{align}
    \frac14 \bE[(\partial\phi(z))^2(\partial\phi(w))^2]
    =
    \frac12\left(1-\frac1L\right)\mathcal C(r,\theta)^2 + \frac{3}{2L}\mathcal H_\alpha(r,\theta).
    \label{eq:app_TT_scalar_four_point}
\end{align}
The background-charge contribution already follows from the second line of \eqref{eq:app_TT_derivative_kernels} and has no finite-width correction.

\paragraph{Ghost contribution.}

The four Gaussian ghost amplitudes admit three pairings. After contracting them, the remaining double sum again separates into equal and unequal neuron indices. For unequal indices, the frequency averages leave the product of the two cross $\beta\partial\gamma$ contractions, giving
\begin{align}
    \bE[\beta(z)\partial\gamma(z) \beta(w)\partial\gamma(w)]_{\mathrm{off}} = \left(1-\frac1L\right)\mathcal C(r,\theta)^2.
    \label{eq:app_TT_ghost_off_diagonal}
\end{align}

For equal indices, the frequency average must be taken after combining the three amplitude pairings. Writing $\vartheta_W(x)=Wx+\bar W\bar x$, their one-neuron contribution before frequency averaging is
\begin{align}
    \frac{4Z_\alpha^2W^4}{p^{4\alpha}}
    \left[
        1+\cos^2\vartheta_W(r,\theta)
         +\cos^2\vartheta_W(\tilde{r},\tilde\theta)
    \right].
\end{align}
Using $\cos^2u=(1+\cos2u)/2$, the terms independent of the insertion positions vanish under the angular integral. The remaining contribution is
\begin{align}
    \bE[\beta(z)\partial\gamma(z) \beta(w)\partial\gamma(w)]_{\mathrm{diag}}
    =
    \frac2L\left[
        \mathcal H_\alpha(r,\theta)+\mathcal H_\alpha(\tilde{r},\tilde\theta)
    \right].
    \label{eq:app_TT_ghost_diagonal}
\end{align}
In particular, the same-frequency $\beta\beta$ and $\partial\gamma\partial\gamma$ pairing contributes before frequency averaging, even though each of those elementary two-point functions vanishes after averaging.

\paragraph{Exact result and continuum limit.}

Combining \eqref{eq:app_TT_mixed_zero}-- \eqref{eq:app_TT_ghost_diagonal} gives
\begin{equation}
    \begin{aligned}
        \bE[T(z)T(w)] ={}&
        \frac32\mathcal C(r,\theta)^2 -\frac{\mathcal D(r,\theta)}{2(K+2)} + 
        \frac1L\left[
            \frac72\mathcal H_\alpha(r,\theta)
            +2\mathcal H_\alpha(\tilde{r},\tilde\theta)
            -\frac32\mathcal C(r,\theta)^2
        \right].
    \end{aligned}
    \label{eq:app_TT_exact_finite_width}
\end{equation}
This expression is exact at finite width and fixed $\Lambda_B$: there are no higher powers of $L^{-1}$. The $\alpha$ dependence is confined to the finite-width coefficient. Its dependence on $z+w=\tilde{r}e^{i\tilde\theta}$ also displays the finite-width violation of translation invariance associated with the ghost features.

We first take $L\to\infty$, then remove the annular cutoffs at fixed $\epsilon_B>0$, and finally take $\epsilon_B\to0^+$. At fixed $z\ne w$, the regulated derivative kernels approach
\begin{align}
    \mathcal C(r,\theta)\longrightarrow-\frac1{(z-w)^2},
    \qquad
    \mathcal D(r,\theta)\longrightarrow\frac6{(z-w)^4}.
\end{align}
It follows that
\begin{align}
    \lim_{\epsilon_B\to0^+}
    \lim_{\substack{p_-\to0\\p_+\to\infty}}
    \lim_{L\to\infty}
    \bE[T(z)T(w)]
    &=
    \left(\frac12+1-\frac3{K+2}\right)
    \frac1{(z-w)^4}
    \nonumber\\
    &=
    \frac{3K}{2(K+2)(z-w)^4}.
    \label{eq:app_TT_continuum_limit}
\end{align}
Thus the scalar, ghost, and background-charge contributions reproduce the stress-tensor two-point function with central charge $c_K=3K/(K+2)$. The finite-width coefficient need not remain bounded as the regulator is removed, so the exact $L^{-1}$ dependence at fixed $\Lambda_B$ does not justify exchanging these limits.

\section{Finite-width readouts and parameter sharing}
\label{app:optimal_distribution}

The sharing prescription of Section~\ref{sec:bosonicsubstrate} is not unique. Alternatively, each scalar and ghost pair can use independent frequencies $W_j^{\phi,A}$ and $W_j^{\mathrm{gh},B}$. The scalar phases $\delta_j^A$ and Gaussian amplitudes $a_j^A$ can also be shared or independent across species. We label these choices by $(F,\delta,a)$, with $\mathrm S$ and $\mathrm I$ denoting shared and independent variables. The main convention is $\mathrm{SSI}$. Different neuron indices remain independent, and each $\beta\gamma$ pair always retains its common frequency and coupled amplitudes. All marginal parameter laws and current coefficients are unchanged.

\paragraph{Continuum limit.}

Independent scalar amplitudes or independent scalar phases preserve the complete elementary covariance at fixed regulator. These prescriptions therefore have the same Gaussian limit. In particular, common amplitudes $a_j^A=a_j$ are allowed when the phases are independent. Sharing all three families instead makes the five coset scalars identical and fails to reproduce the target current sector.

When the frequencies are independent but the scalar phases and amplitudes are shared ($\mathrm{ISS}$), different scalar fields have nonzero mixed covariances at fixed regulator. For $R^3$ and $\alpha\in[-1.5,1.5]$, their differentiated contributions vanish in the ordered continuum limit at separated nonzero insertion positions. The convergence is logarithmic at $\alpha=0$. This recovers the required current and stress sector, but does not identify the regulated Gaussian fields or their undifferentiated scalar zero modes. The prescribed elementary Wick subtractions are retained, including the mixed coincident contractions present in this case.

\paragraph{Finite-width accuracy.}

We compare the admissible prescriptions using the relative $TT$ error \eqref{eq:fw_finite_width_tt_error}, with $p_-=0.01$ and $p_+=1000$. Unlike the main-text comparison, each prescription selects its own $(\alpha,\epsilon_B)$ from $[-1.5,1.5]\times[10^{-4},0.2]$. The original $64$ insertion pairs determine the selection; the reserved $16$ pairs are evaluated without retuning. Table~\ref{tab:sharing_choices} gives representative coset results.

\begin{table}[t]
    \centering
    \small
    \setlength{\tabcolsep}{5pt}
    \begin{tabular}{lrrrrr}
        \toprule
        $(F,\delta,a)$
        & Variables/neuron
        & $\alpha$
        & $\epsilon_B$
        & $E_{\rm select}^{TT}$
        & $E_{\rm valid}^{TT}$ \\
        \midrule
        $\mathrm{SSI}$ & 38 & 0.8633 & 0.15310
        & 2.4652 & 3.0647 \\
        $\mathrm{SIS}$ & 38 & 1.4414 & $0.20000^\dagger$
        & 2.3759 & 2.2965 \\
        $\mathrm{IIS}$ & 76 & 0.8906 & 0.14427
        & 2.3858 & 2.9708 \\
        $\mathrm{III}$ & 80 & 0.8906 & 0.14427
        & 2.3870 & 2.9718 \\
        \bottomrule
    \end{tabular}
    \caption{
    Coset $TT$ relative RMS errors at $R^3$ and $L=4096$, after joint parameter selection. Counts include all independent real sampling variables per neuron. These are the best values found in the stated search domain, not certified global minima. The dagger marks its upper $\epsilon_B$ boundary.
    }
    \label{tab:sharing_choices}
\end{table}

The preferred prescription depends on width and regulator. For the coset, $\mathrm{IIS}$ gives the smallest selected $TT$ error at $L=64,256,1024$ with $R^3$, and at both checked widths, $L=256,4096$, with $R^1$ and $R^2$. At $R^3,L=4096$, $\mathrm{SIS}$ is preferred. At their respective $TT$-selected parameters, changing from $\mathrm{SSI}$ to $\mathrm{SIS}$ also reduces the scaled $TTT$ error from $4.18$ to $2.18$, but increases the $\mathcal{J}^1\mathcal{J}^2\mathcal{J}^3$ error from $0.374$ to $0.667$. Thus the improvement is not uniform across observables. For the direct presentation, shared and independent frequencies give exactly the same $TT$ correlator, although their three-point functions differ.

\paragraph{Number of independent parameters.}

The main convention uses $38L$ independent real variables for the coset, compared with $80L$ when all families are independent. For the direct presentation, the corresponding counts are $6L$ and $8L$. The alternative $\mathrm{SIS}$ also uses $38L$ variables. It replaces five independent scalar amplitudes and one common phase with one common scalar amplitude and five independent phases. No admissible prescription tested here uses fewer sampled variables.%

We retain $\mathrm{SSI}$ to minimize the independent frequency and phase variables while preserving independent scalar amplitudes. This is a choice of a smaller parameter ensemble, not a claim of uniformly optimal accuracy or runtime. Independent frequency integrals can factorize, so the number of sampled variables alone does not determine the cost of evaluating correlators.

\subsection{Supplementary stress-tensor comparisons}
\begin{figure}[t]
    \centering
    \begin{subfigure}[t]{0.48\textwidth}
        \centering
        \includegraphics[width=\linewidth]{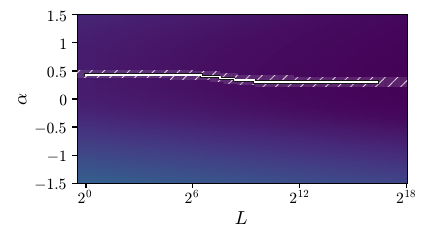}
        \caption{Direct presentation $\mathrm W$, regulator $R^1$.}
        \label{fig:fw_ttt_W_R1}
    \end{subfigure}
    \hfill
    \begin{subfigure}[t]{0.48\textwidth}
        \centering
        \includegraphics[width=\linewidth]{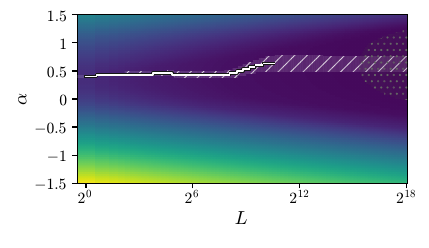}
        \caption{Coset presentation $\mathrm C$, regulator $R^1$.}
        \label{fig:fw_ttt_C_R1}
    \end{subfigure}

    \medskip

    \begin{subfigure}[t]{0.48\textwidth}
        \centering
        \includegraphics[width=\linewidth]{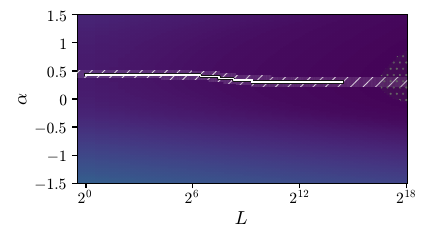}
        \caption{Direct presentation $\mathrm W$, regulator $R^2$.}
        \label{fig:fw_ttt_W_R2}
    \end{subfigure}
    \hfill
    \begin{subfigure}[t]{0.48\textwidth}
        \centering
        \includegraphics[width=\linewidth]{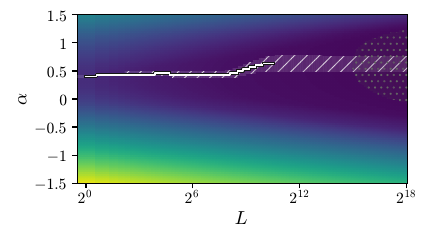}
        \caption{Coset presentation $\mathrm C$, regulator $R^2$.}
        \label{fig:fw_ttt_C_R2}
    \end{subfigure}

    \medskip

    \begin{subfigure}[t]{0.48\textwidth}
        \centering
        \includegraphics[width=\linewidth]{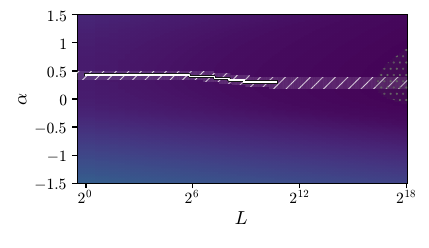}
        \caption{Direct presentation $\mathrm W$, regulator $R^3$.}
        \label{fig:fw_ttt_W_R3}
    \end{subfigure}
    \hfill
    \begin{subfigure}[t]{0.48\textwidth}
        \centering
        \includegraphics[width=\linewidth]{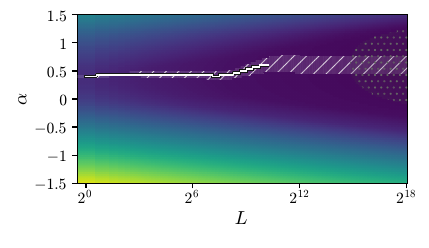}
        \caption{Coset presentation $\mathrm C$, regulator $R^3$.}
        \label{fig:fw_ttt_C_R3}
    \end{subfigure}

    \medskip

    \input{figures/finite_width/ttt_su2_k3_scale}

    \caption{
    $\bE[T(z_1)T(z_2)T(z_3)]$: $\mathrm W=SU(2)_3$ versus $\mathrm C=SU(6)_1/SU(3)_2$. Relative population RMS $E=\operatorname{RMS}_s|\mu_s/\mu_{\rm CFT}-1|$. Original 64 ordered unit-disk pairs, augmented by fixed independent third points; no separation cut. Shared frequencies and scalar phases; the asymmetric ghost architecture of \eqref{eq:bosonic-ghost-architecture}; full same-parent affine stress subtraction. $p_-=.01$, $p_+=1000$; fixed TT-selected damping for $R^1,R^2,R^3$: 0.0206692503922, 0.0206496645453, 0.0198029578588. 91 sampled $\alpha\in[-1.5,1.5]$, 301 distinct integer $L\in[1,2^{18}]$. Viridis shows $\log_{10}E$ on the common displayed scale from $-1$ to $115$; white denotes resolved unique sampled minima and hatching the strict 5\%-RMS region. Stippling flags a numerical-sensitivity indicator exceeding 1\% of $E$. }
    \label{fig:fw_ttt_alpha_width}
\end{figure}

\begin{figure}[t]
    \centering
    \begin{subfigure}[t]{0.48\textwidth}
        \centering
        \includegraphics[width=\linewidth]{figures/finite_width/tt_W_R1.pdf}
        \caption{Direct presentation $\mathrm W$, $SU(2)_3$.}
        \label{fig:fw_tt_W_su2_k3_comparison}
    \end{subfigure}
    \hfill
    \begin{subfigure}[t]{0.48\textwidth}
        \centering
        \includegraphics[width=\linewidth]{figures/finite_width/tt_C_R1.pdf}
        \caption{Coset presentation $\mathrm C$, $SU(6)_1/SU(3)_2$.}
        \label{fig:fw_tt_C_su2_k3_comparison}
    \end{subfigure}
    \smallskip
    \input{figures/finite_width/tt_heatmap_scale.tex}

    \smallskip

    \begin{subfigure}[t]{0.48\textwidth}
        \centering
        \includegraphics[width=\linewidth]{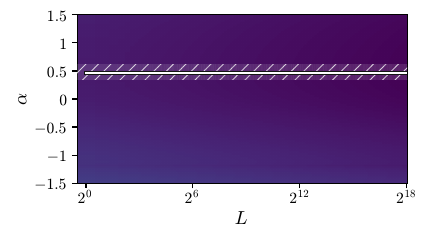}
        \caption{Direct presentation $\mathrm W$, $SU(2)_4$.}
        \label{fig:fw_su2_k4_tt_W_R1}
    \end{subfigure}
    \hfill
    \begin{subfigure}[t]{0.48\textwidth}
        \centering
        \includegraphics[width=\linewidth]{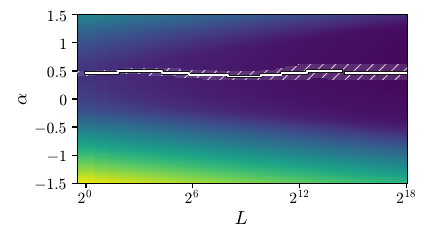}
        \caption{Coset presentation $\mathrm C$, $SU(8)_1/SU(4)_2$.}
        \label{fig:fw_su2_k4_tt_C_R1}
    \end{subfigure}
    \smallskip
    \input{figures/finite_width/tt_su2_k4_scale.tex}

    \smallskip

    \begin{subfigure}[t]{0.48\textwidth}
        \centering
        \includegraphics[width=\linewidth]{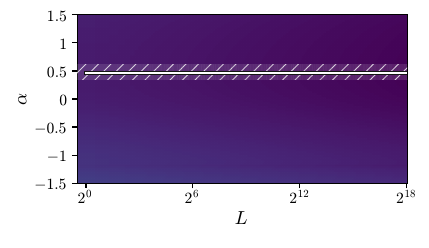}
        \caption{Direct presentation $\mathrm W$, $SU(3)_4$.}
        \label{fig:fw_su3_k4_tt_W_R1}
    \end{subfigure}
    \hfill
    \begin{subfigure}[t]{0.48\textwidth}
        \centering
        \includegraphics[width=\linewidth]{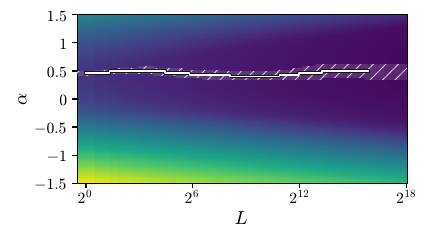}
        \caption{Coset presentation $\mathrm C$, $SU(12)_1/SU(4)_3$.}
        \label{fig:fw_su3_k4_tt_C_R1}
    \end{subfigure}

    \smallskip

    \input{figures/finite_width/tt_su3_k4_scale.tex}

    \caption{
    Stress-correlator comparisons for the indicated direct and coset theories at $R^1$. Axes, minima, near-optimal regions and sweep conventions are as in Figure~\ref{fig:fw_tt_alpha_width}, with displayed $\alpha\in[-1.5,1.5]$. Each row has its own color scale, ending at $80$, $81$ and $81$. All panels use $p_-=0.01$, $p_+=1000$ and the same $64$ ordered insertion pairs; damping is selected by the matched infinite-width procedure of \eqref{eq:fw_regulator_minimax_objective}. Equal feature width does not imply equal computational cost.}
    \label{fig:fw_tt_alpha_width_su2_3_su_2_4_su_3_4}
\end{figure}

\clearpage
\section{Response geometry, calibration and computational resources}
\label{app:local_calibration}
\label{app:stress_comparison_scope}

This appendix specifies the correlation data and evaluation sets used in Section~\ref{sec:local_calibration_dependence}, and gives the numerical details of the calibration, sampling, and timing calculations.

Unless stated otherwise, the finite-width calculations use
\begin{align}
    R^3_{\epsilon_B}(p)
    &=e^{-\epsilon_B(p+p^{-1})},
    &
    \epsilon_B^\star&=0.019802957858766932,
    &
    p_-&=0.01,
    &
    p_+&=1000.
    \label{eq:app_local_regulator}
\end{align}
The scalar and ghost fields share their frequency samples, and the scalar species share their additive phases. Each ghost pair retains the coupled construction of Section~\ref{sec:bosonicsubstrate}. For the coset presentation, the stress readout remains $T_{\mathrm C}=T_{\mathrm P}-T_{\mathrm D}$, with both terms evaluated on the same parent fields. It is not replaced by the Sugawara readout of the three retained currents.

\subsection{Correlation data and evaluation sets}
\label{app:local_calibration_data}

The comparison uses the thirteen ordered operator patterns listed in Table~\ref{tab:app_local_patterns}. Each pattern is evaluated at four configurations of separated points. A pattern containing $q$ fields uses the first $q$ positions of its configuration. This gives $52$ complex values for each complete set of four configurations.

\begin{table}[t]
    \centering
    \small
    \begin{tabular}{clcl}
        \toprule
        $\ell$ & Operator pattern
        & $\ell$ & Operator pattern \\
        \midrule
        0 & $\mathcal{J}^1\mathcal{J}^1$       & 7  & $\mathcal{J}^1\mathcal{J}^1\mathcal{J}^3\mathcal{J}^3$ \\
        1 & $\mathcal{J}^2\mathcal{J}^2$       & 8  & $\mathcal{J}^1\mathcal{J}^2\mathcal{J}^1\mathcal{J}^3$ \\
        2 & $\mathcal{J}^3\mathcal{J}^3$       & 9  & $\mathcal{J}^3\mathcal{J}^3\mathcal{J}^3\mathcal{J}^3$ \\
        3 & $\mathcal{J}^1\mathcal{J}^2$       & 10 & $TT$ \\
        4 & $\mathcal{J}^1\mathcal{J}^3$       & 11 & $T\mathcal{J}^3$ \\
        5 & $\mathcal{J}^1\mathcal{J}^2\mathcal{J}^3$    & 12 & $T\mathcal{J}^1\mathcal{J}^1$ \\
        6 & $\mathcal{J}^2\mathcal{J}^1\mathcal{J}^3$    &    & \\
        \bottomrule
    \end{tabular}
    \caption{
    Current and stress-tensor products used in the correlation comparisons of Section~\ref{sec:local_calibration_dependence}.
    }
    \label{tab:app_local_patterns}
\end{table}

The original configurations are denoted by $\mathbf h_c$, and those used for validation by $\mathbf v_c$, with $c=0,\ldots,3$. Their entries are given in Table~\ref{tab:app_local_positions}.

\begin{table}[t]
    \centering
    \small
    \setlength{\tabcolsep}{4pt}
    \begin{tabular}{crrrr}
        \toprule
        Configuration & $z_1$ & $z_2$ & $z_3$ & $z_4$ \\
        \midrule
        $\mathbf h_0$
        & $1.30+0.10i$ & $0.80+0.20i$
        & $0.30+0.05i$ & $-0.10+0.10i$ \\
        $\mathbf h_1$
        & $1.55+0.30i$ & $0.91-0.10i$
        & $0.42+0.12i$ & $-0.13-0.07i$ \\
        $\mathbf h_2$
        & $1.60+0.15i$ & $0.61+0.21i$
        & $0.53+0.23i$ & $-0.11+0.02i$ \\
        $\mathbf h_3$
        & $1.42-0.23i$ & $0.67+0.08i$
        & $0.18-0.09i$ & $-0.12+0.03i$ \\
        \midrule
        $\mathbf v_0$
        & $1.67+0.22i$ & $0.83-0.19i$
        & $0.77-0.21i$ & $-0.04+0.02i$ \\
        $\mathbf v_1$
        & $1.33-0.14i$ & $0.79+0.24i$
        & $0.36+0.08i$ & $-0.11-0.05i$ \\
        $\mathbf v_2$
        & $1.51+0.13i$ & $1.02-0.16i$
        & $0.25-0.13i$ & $0.19-0.12i$ \\
        $\mathbf v_3$
        & $1.28-0.27i$ & $0.68+0.18i$
        & $0.21+0.04i$ & $-0.08+0.07i$ \\
        \bottomrule
    \end{tabular}
    \caption{
    Original and validation-position configurations. A pattern of length $q$ uses the first $q$ entries of the corresponding row.
    }
    \label{tab:app_local_positions}
\end{table}

We multiply each correlation function by a fixed coordinate factor $g_\lambda$ before forming the error, as mentioned in Section~\ref{sec:local_calibration_dependence}. Writing $z_{ij}=z_i-z_j$, these factors are detailed in Table~\ref{tab:calibration_factors}.
\begin{table}
\centering
\begin{tabular}{lc}
    \toprule
    Correlator type & Coordinate factor \\
    \midrule
    $\mathcal{J}\mathcal{J}$   & $z_{12}^2$ \\
    $\mathcal{J}\mathcal{J}\mathcal{J}$  & $z_{12}z_{23}z_{31}$ \\
    $\mathcal{J}\mathcal{J}\mathcal{J}\mathcal{J}$ & $z_{12}^2z_{34}^2$ \\
    $TT$   & $z_{12}^4$ \\
    $T\mathcal{J}$   & $z_{12}^3$ \\
    $T\mathcal{J}\mathcal{J}$  & $z_{12}^2z_{13}^2$ \\
    \bottomrule
\end{tabular}
\caption{Coordinate factors and their associated correlators.}
\label{tab:calibration_factors}
\end{table}
The factors are complex, including the sign of $z_{31}$ in the three-current expression. The same factor multiplies the neural prediction and its continuum reference at the positions being evaluated. There is no division by a reference component and no fitted rescaling. Components with zero continuum reference therefore use absolute errors.

The fitted set $F$ contains ten operator patterns, each evaluated at configurations $\mathbf h_0$ and $\mathbf h_1$. The three excluded patterns are precisely $\mathcal{J}^2\mathcal{J}^2$, $\mathcal{J}^2\mathcal{J}^1\mathcal{J}^3$, and $\mathcal{J}^1\mathcal{J}^2\mathcal{J}^1\mathcal{J}^3$. Splitting each fitted correlator into real and imaginary parts gives a 40-real-component vector, $n_F=|F|=40$.

The held set $H$ is the complement within the original $104$-component vector. Equivalently, it contains all thirteen patterns at $\mathbf h_2,\mathbf h_3$, together with the three previously excluded patterns at $\mathbf h_0,\mathbf h_1$. Thus, $n_H=|H|=64$. The validation set $V$ uses all thirteen patterns at $\mathbf v_0,\mathbf v_1,\mathbf v_2,\mathbf v_3$, giving $104$ real coordinates.

Together, the fitted, held and validation sets form the basis for our numerical optimization experiment, detailed in Section~\ref{sec:local_calibration_dependence}.

\subsection{Bounded calibration and validation}
\label{app:local_bounded_calibration}

The calibration experiment introduced in Section~\ref{sec:local_calibration_dependence} probes $255$ different width values, in the set $\mathcal G_L=\{\operatorname{round}(10^{j/50}):j=0,\ldots,300\}$. For each width, we look at both the direct and coset presentations, $\mathrm{W}, \mathrm{C}$, start from $\theta=0$ and retain the same continuum target, regulator $R^3$, $\epsilon_B^\star$, and $\alpha=0$ law. The fitting objective is
\begin{align}
    \mathcal L_X^{\mathrm{fit}}(\theta)
    =\frac{1}{2\cdot104}\sum_{i\in F}
       \bigl(y_{X,i}(\theta) - y_{\mathrm{ref},i}\bigr)^2
      +\frac{10^{-6}}{2}\|\theta\|^2.
    \label{eq:app_local_fit_loss}
\end{align}

Every fit allows at most twenty-four bounded local least-squares steps, with $|\theta_i|\leq0.3$, physical coordinate step cap $0.02$, penalty $10^{-6}$, original gradient tolerance $10^{-11}$ and actual-loss backtracking. The full original vector and Jacobian are evaluated, but $H$ is masked out of the loss, gradient, bounded solve, acceptance and stopping test. The validation vector is never passed to the optimizer. There is no restart selection, bias matching, weight search or transfer of gains across widths.

\begin{table}[t]
    \centering\small\setlength{\tabcolsep}{4pt}
    \begin{tabular}{llrrrrrr}
    \toprule
    $X$ & Set & $\min\frac{R^{\rm after}}{R^{\rm before}}$ & $L_{\min}$ & $\max \frac{R^{\rm after}}{R^{\rm before}}$ & $L_{\max}$ & Improved & Worsened\\
    \midrule
    $\mathrm W$ & $F$ & 0.398214 & 1 & 0.994654 & 144544 & 255 & 0\\
    $\mathrm W$ & $H$ & 0.418142 & 1 & 1.049563 & 263027 & 220 & 35\\
    $\mathrm W$ & $V$ & 0.473206 & 1 & 1.040337 & 218776 & 226 & 29\\
    $\mathrm C$ & $F$ & 0.229583 & 1 & 0.548762 & 21878 & 255 & 0\\
    $\mathrm C$ & $H$ & 0.170406 & 7 & 1.365613 & 2188 & 225 & 30\\
    $\mathrm C$ & $V$ & 0.152453 & 7 & 1.554804 & 2188 & 220 & 35\\
    \bottomrule\end{tabular}
    \caption{Calibration on the set of $255$ distinct widths.
    Each presentation is newly fitted independently from $\theta=0$ at every width, with fixed $R^3$, $\epsilon_B^\star$, and $\alpha=0$. The smallest and largest values of the ratio $R^{\rm after}/R^{\rm before}$ are reported, together with the width at which that value is reached $L_{\min}$, $L_{\max}$.
    Improved/worsened counts use deviations from one exceeding $10^{-9}$ plus ten times the order-$24/40$ ratio change; other cases are unresolved at this reporting tolerance. $F,H,V$ retain their $40,64,104$ real-coordinate RMS definitions.}
    \label{tab:app_local_calibration_results}
\end{table}

\begin{table}[t]
    \centering\small\setlength{\tabcolsep}{4pt}
    \begin{tabular}{lrlrrr}
    \toprule
    $X$ & $L$ & Set & Before RMS & After RMS & After/before\\
    \midrule
    $\mathrm W$ & 1 & $F$ & $67571.9$ & $26908.1$ & 0.398214\\
    $\mathrm W$ & 1 & $H$ & $1.8445\times10^{6}$ & $7.7125\times10^{5}$ & 0.418142\\
    $\mathrm W$ & 1 & $V$ & $3.8166\times10^{6}$ & $1.8060\times10^{6}$ & 0.473206\\
    $\mathrm W$ & $10^6$ & $F$ & $0.0586106$ & $0.0268786$ & 0.458596\\
    $\mathrm W$ & $10^6$ & $H$ & $1.04758$ & $0.98285$ & 0.938208\\
    $\mathrm W$ & $10^6$ & $V$ & $2.28936$ & $2.10865$ & 0.921068\\
    $\mathrm C$ & 1 & $F$ & $9.5293\times10^{10}$ & $2.1878\times10^{10}$ & 0.229583\\
    $\mathrm C$ & 1 & $H$ & $2.3037\times10^{13}$ & $4.8181\times10^{12}$ & 0.209150\\
    $\mathrm C$ & 1 & $V$ & $5.9930\times10^{13}$ & $1.2715\times10^{13}$ & 0.212165\\
    $\mathrm C$ & $10^6$ & $F$ & $0.549793$ & $0.258519$ & 0.470212\\
    $\mathrm C$ & $10^6$ & $H$ & $5.41738$ & $1.48797$ & 0.274665\\
    $\mathrm C$ & $10^6$ & $V$ & $9.50771$ & $2.45236$ & 0.257934\\
    \bottomrule\end{tabular}
    \caption{Endpoint RMS values from the calibration experiment.}
    \label{tab:calibration_loggrid_endpoints}
\end{table}

Fitted RMS decreases in all $510$ fits. Validation RMS worsens at $29/255$ widths for $\mathrm W$ and $35/255$ for $\mathrm C$; held-coordinate RMS worsens at $35/255$ and $30/255$. For $\mathrm W$, the held and validation ratios reach their maxima
at $L=263027$ and $L=218776$, respectively; for $\mathrm C$, both
maxima occur at $L=2188$.

\subsection{Sampling uncertainty}
\label{app:local_sampling}

For the Cartan observable \eqref{eq:calibration_cartan_observable}, we consider three estimators obtained by performing different parts of the ensemble average analytically. The first averages over the Gaussian amplitudes, retaining the random radii, angles, and scalar phases. The second also averages over the angles and phases, conditional on the $L$ sampled radii. The estimator plotted in Figure~\ref{fig:calibration_resources} then replaces the remaining scalar sample average by its finite-ensemble expectation. All three estimators preserve the same mean within each presentation. Their construction retains the joint frequency and phase law and the contributions in which feature indices coincide.

The finite-width means have the form $\bE[U_X]=c_{0,X}+c_{1,X}/L$. With the numerical coefficients rounded, they are
\begin{align}
    \bE[U_{\mathrm W}]
    &\simeq
    1.50714452
    -\frac{0.07686623+0.05234317i}{L},
    \\
    \bE[U_{\mathrm C}]
    &\simeq
    1.97325899
    -\frac{0.69179610+0.47108855i}{L}.
\end{align}
Thus, the large-width RMS discrepancies at the fixed regulator are approximately $0.00505194$ and $0.334645$, respectively. At finite width, the $1/L$ contribution can partially cancel the remaining discrepancy, so these limiting values need not be the smallest errors on the width grid.

The estimator covariances are obtained from integrated single-feature moments, with their finite-width dependence retained through the terms in $L^{-1}$, $L^{-2}$, and $L^{-3}$. For the estimator with the scalar contribution averaged analytically, the centered coset estimate is nine times the centered direct estimate when evaluated on the same sampled radii. Consequently,
\begin{align}
    \Sigma_{\mathrm{est},\mathrm C}
    =
    81\,\Sigma_{\mathrm{est},\mathrm W}.
\end{align}
The computed covariance matrices satisfy this relation to a maximum relative discrepancy of $1.5\times10^{-15}$ across the width grid.

The radial moments were evaluated at quadrature orders $24$, $40$, and $56$, with order $56$ used for the reported results. The largest relative covariance change between orders $40$ and $56$ was below $8\times10^{-15}$. For the estimator averaged only over Gaussian amplitudes, the remaining moments were evaluated using four independently scrambled Sobol integrations at $2^{16}$ and $2^{18}$ points; the largest relative covariance change between these resolutions was $1.37\%$. These comparisons assess the numerical sensitivity of the moments used to calculate the sampling uncertainty.

\subsection{Computation times}
\label{app:local_timing}

The timing comparison uses three sequential repetitions for each presentation at $L=1,10^3,10^6$. Each repetition starts a new process and uses one computational thread on the same Intel processor (family $6$, model $197$), under Windows~11 with Python~3.14.3, NumPy~2.4.3, and double-precision arithmetic.

For the full calculation, we sum the times required to construct the presentation and its finite-width correlation expressions, integrate at the original configurations, perform the bounded fit, and integrate and evaluate at the validation configurations. The common continuum references are generated separately, and process startup is outside these timed phases.

The repeated evaluations reuse the correlation expressions and integrated coefficients at the original configurations. For the adjustable-parameter calculation, three preliminary calls are followed by $1000$ timed calls at the final fitted parameters. Each call recomputes the amplitude dependence. For the fixed-parameter calculation, we first set $\theta=0$ and collect the coefficients of the resulting finite polynomial in $1/L$, then time $1000$ evaluations of that polynomial. Coefficient collection is completed before this timing begins.

\begin{table}[t]
    \centering
    \small
    \setlength{\tabcolsep}{3pt}
    \begin{tabular}{rlccc}
        \toprule
        $L$ & Task
        & $\mathrm W$ time (s)
        & $\mathrm C$ time (s)
        & Ratio \\
        \midrule
        $1$
        & Full calculation & $24.78\,[22.76,32.85]$ & $253.39\,[244.31,311.95]$ & $10.224$ \\
        & Adjustable parameters & $0.1390\,[0.1375,0.2053]$ & $10.06\,[10.00,11.82]$ & $72.371$ \\
        & Fixed parameters & $0.00437\,[0.00436,0.00583]$ & $0.00763\,[0.00537,0.00884]$ & $1.746$ \\
        \midrule
        $10^3$
        & Full calculation & $24.55\,[23.36,29.15]$ & $280.71\,[230.55,288.96]$ & $11.433$ \\
        & Adjustable parameters & $0.1434\,[0.1390,0.1470]$ & $10.36\,[9.28,10.42]$ & $72.197$ \\
        & Fixed parameters & $0.00408\,[0.00404,0.00488]$ & $0.00554\,[0.00548,0.00635]$ & $1.358$ \\
        \midrule
        $10^6$
        & Full calculation & $25.39\,[22.27,35.13]$ & $255.92\,[248.62,377.87]$ & $10.081$ \\
        & Adjustable parameters & $0.1369\,[0.1222,0.2077]$ & $10.35\,[10.34,15.28]$ & $75.608$ \\
        & Fixed parameters & $0.00441\,[0.00373,0.00503]$ & $0.00869\,[0.00768,0.01094]$ & $1.970$ \\
        \bottomrule
    \end{tabular}
    \caption{
    Computation times for the three tasks. Entries give the median and the observed minimum and maximum across three repetitions. The adjustable-parameter and fixed-parameter times each cover $1000$ evaluations. Ratios are calculated from the unrounded median coset and direct times.
    }
    \label{tab:app_local_timings}
\end{table}

The timed fits reproduce the endpoint validation errors in Table~\ref{tab:calibration_loggrid_endpoints}. At $L=10^3$, their final validation RMS values are $583.408$ for $\mathrm W$ and $5064.13$ for $\mathrm C$. These errors are identical across the three repetitions of each calculation, while the elapsed times vary over the ranges reported in Table~\ref{tab:app_local_timings}. The comparison therefore records the computational cost of the specified procedure together with the accuracy it achieves.

\section{Finite-width fermionic correlations}
\label{app:finite_width_fermionic_corrections}

The finite-width corrections of the Fourier--Berezin fields arise from repeated frequency samples in their Berezin contractions. We derive bounds for fixed collections of modes, followed by exact corrections to four-point functions and their behavior under regulator removal.

\subsection{Moment bounds at fixed regulator}
\label{app:neural_fixed_grade_limit}

Fix a regulator $\Lambda$ and a finite collection of fermionic or angular readouts. Their empirical covariance is $C_L=L^{-1}\sum_{j=1}^L c(W_j)$, where the frequencies are independent, $C_\Lambda=\mathbb E_Wc(W)$, and $|c_{ab}(W)|\leq B_\Lambda$. The conditional $2r$-point Berezin functional is $\mathcal P_{L,r}=\operatorname{Pf}(C_L)$.

\begin{proposition}[Fixed-grade moment bounds]
\label{prop:neural_fixed_grade_bound}
Writing $M_r=(2r-1)!!$, one has
\begin{align}
\left|\mathbb E_W\mathcal P_{L,r}-\operatorname{Pf}(C_\Lambda)\right|&\leq\frac{2M_r\binom r2B_\Lambda^r}{L},
\label{eq:fg_bias}\\
\left(\mathbb E_W\left|\mathcal P_{L,r}-\mathbb E_W\mathcal P_{L,r}\right|^2\right)^{1/2}&\leq\frac{\sqrt2\,rM_rB_\Lambda^r}{\sqrt L}.
\label{eq:fg_variance}
\end{align}
\end{proposition}

\noindent\emph{Proof.} The Pfaffian contains $M_r$ products of $r$ covariance entries. Expanding their empirical averages gives $L^r$ frequency-index tuples, of which $(L)_r$ have distinct indices. These contribute the corresponding products of entries of $C_\Lambda$. The remaining fraction is at most $\binom r2/L$, with each product bounded by $B_\Lambda^r$, proving \eqref{eq:fg_bias}. Replacing one frequency changes the Pfaffian by at most $2rM_rB_\Lambda^r/L$. The Efron--Stein inequality gives \eqref{eq:fg_variance}.\hfill$\square$

For the Fourier features, $|J_n|\leq1$ gives $B_\Lambda\leq2Z_\alpha\max_{p\in[p_-,p_+]}p^{1-2\alpha}$, with additional fixed factors for radially rescaled modes. The bound is finite on a regulated annulus for a fixed mode bank.

\subsection{Exact corrections from shared frequencies}
\label{app:finite_width_cft_constraints}

For one output channel, each sampled frequency contributes a rank-two covariance $c_{ab}(W)=u_a(W)v_b(W)-v_a(W)u_b(W)$. This structure fixes the dependence of its higher moments on the number of sampled features.

\begin{proposition}[Same-channel exclusion factor]
\label{prop:fw_exclusion_factor}
For $2r$ fields in one output channel,
\begin{align}
\mathbb E_\Theta\left[\prod_{a=1}^{2r}\chi_L^A(z_a)\right]
&=s_r(L)\operatorname{Pf}(C_\Lambda),\nonumber\\
s_r(L)&=\frac{L^{(r)}}{L^r}=\prod_{j=0}^{r-1}\left(1-\frac jL\right).
\label{eq:fw_exclusion_factor}
\end{align}
\end{proposition}

\noindent\emph{Proof.} Let $\omega_j=\sum_{a<b}c_{ab}(W_j)e_a\wedge e_b=u_j\wedge v_j$. Since $\omega_j\wedge\omega_j=0$, the exterior expansion of the conditional Pfaffian contains only products with distinct frequency indices. Independence gives $L^{(r)}/L^r$ times the Pfaffian of the mean covariance.\hfill$\square$

In particular, the same-channel four-point function has the factor $s_2(L)=1-L^{-1}$, while its two-point function is independent of $L$.

For distinct channels $A\ne B$ sharing their sampled frequencies, let $c_W(d)$ be the one-feature contraction at separation $d=z-w$, and set $e=z'-w'$. The mixed four-point function is
\begin{align}
&\mathbb E_{\Theta_F}\left[\chi_L^A(z)\chi_L^A(w)\chi_L^B(z')\chi_L^B(w')\right]\nonumber\\
&\qquad=C_\Lambda(d)C_\Lambda(e)+\frac1L\operatorname{Cov}_W\bigl(c_W(d),c_W(e)\bigr),
\label{eq:fw_mixed_fourpoint}
\end{align}
where $\operatorname{Cov}_W(X,Y)=\mathbb E_W[XY]-\mathbb E_W[X]\mathbb E_W[Y]$ is complex bilinear. The shared-frequency term depends on both insertion separations.

\subsection{Regulator dependence and the ultraviolet limit}
\label{app:neural_width_projective_uv}

Consider the radial law with damping $e^{-\epsilon p}$, endpoints $p_-=0$, $p_+=\infty$, and $0<\alpha<1$. The mean kernel is $C_\epsilon(d)=d^{-1}(1-\epsilon/\sqrt{\epsilon^2+4|d|^2})$, which approaches $1/d$ at nonzero separation. For $d,e,d+e,d-e\ne0$, the mixed-channel covariance satisfies
\begin{align}
\lim_{\epsilon\downarrow0}\epsilon^{2\alpha}\operatorname{Cov}_\epsilon\bigl(c_W(d),c_W(e)\bigr)&=K_\alpha(d,e),\nonumber\\
K_\alpha(d,e)&=\frac{\Gamma(2\alpha)\Gamma(2-\alpha)}{\Gamma(1+\alpha)}\left[\frac{|d+e|^{2\alpha}}{(d+e)^2}-\frac{|d-e|^{2\alpha}}{(d-e)^2}\right].
\label{eq:neural_width_nonuniform_uv}
\end{align}
The angular Bessel integral and its Abel limit give this expression. The resulting correction scales as $(L\epsilon^{2\alpha})^{-1}$ and vanishes along the joint limit $L\epsilon^{2\alpha}\to\infty$.

The correction also changes the relative invariant-channel coefficients of the fermionic parent. For external representations $(F,F,\bar F,\bar F)$, let $U=u_N\otimes u_K$ and $V=v_N\otimes v_K$ be the direct and exchanged invariant tensors. Write $d_{ij}=z_i-z_j$ and $P_{L,\epsilon}(d,e)=C_\epsilon(d)C_\epsilon(e)+L^{-1}\operatorname{Cov}_\epsilon(c_W(d),c_W(e))$. The parent four-point tensor is
\begin{align}
D_{L,\epsilon}=-P_{L,\epsilon}(d_{13},d_{24})U+P_{L,\epsilon}(d_{14},d_{23})V.
\label{eq:neural_width_parent_tensor}
\end{align}
The leading difference between the relative channel corrections is
\begin{align}
\mathcal Q_\alpha(z_i)=\frac{K_\alpha(d_{13},d_{24})}{1/(d_{13}d_{24})}-\frac{K_\alpha(d_{14},d_{23})}{1/(d_{14}d_{23})}.
\label{eq:neural_width_projective_difference}
\end{align}
This expression is generically nonzero and nonholomorphic. The same correction contributes to the dilation response. With $\mathscr W_{\rm dil}=\sum_i(z_i\partial_{z_i}+\bar z_i\partial_{\bar z_i})+2$, homogeneity gives
\begin{align}
\lim_{\epsilon\downarrow0}L\epsilon^{2\alpha}\mathscr W_{\rm dil}(D_{L,\epsilon}-D_{\infty,\epsilon})
=2\alpha\left[-K_\alpha(d_{13},d_{24})U+K_\alpha(d_{14},d_{23})V\right].
\label{eq:neural_width_dilation_defect}
\end{align}
The shared-frequency corrections thus deform the relative invariant-channel coefficients and their insertion dependence. Their scaling agrees with the fixed-grade angular moment estimates in Appendix~\ref{app:lr_controlled_preparations}.

\section{Numerical methods for compression and fusion transfer}
\label{app:s5_upgrade_methods}

\paragraph{Polarized subspace parameterization.}

The compression experiments use real orthonormal frames in the $48$-dimensional reference one-particle space. For sea and particle ranks $(m_-,m_+)$, set $m=m_-+m_+$ and let $E_m$ embed the first $m$ coordinate vectors into $\mathbb R^{48}$. We parameterize retained frames by
\begin{align}
Q(z)=Q_0e^{\Omega(z)}E_m,
\label{eq:app_compression_flag_chart}
\end{align}
where $Q_0$ is orthogonal and $\Omega(z)$ is skew-symmetric. Its independent blocks mix sea, particle and unused coordinates, giving $d=m_-m_++m(48-m)$ real parameters. Rotations within the retained sea and particle spaces are omitted. The exponential coordinates cover the polarized-flag manifold, with possible nonuniqueness at the cut locus.

Direct optimization adjusts $z\in\mathbb R^d$. Neural optimization produces these coordinates through a fixed-input network of widths $8$, $32$ and $d$, with tanh hidden activation and a linear output. Its unrestricted output bias gives both methods the same family of retained subspaces. They use five matched initializations, generated with skew-generator Frobenius norm $0.2$ and seeds $9200+i$, $i=0,\ldots,4$. Random flags are obtained by Haar QR decomposition with seeds $55200+i$.

Both optimizers use Adam for $64$ gradient evaluations, clipping gradients at norm $10$ and selecting checkpoints every eight updates. The learning rates are $0.01$ for direct optimization and $0.001$ for neural optimization. Checkpoints retain the parameters, frame, optimizer state and objective value.

\paragraph{Reference descendant contractions.}

Let $\Psi_n(Q)$ contain the compressed descendants at grade $n$, with Gram matrix $G_n(Q)=\Psi_n(Q)\Psi_n(Q)^T$. The reference affine Gram matrix is $G_n^{\rm ref}$. At each grade we fix a positive reference quotient, retaining normalized-Gram eigenvalues above $10^{-9}\max(1,\max|\lambda|)$, and denote its lift by $L_n$. For reference Ward covectors $C_n,D_n$, define
\begin{align}
\widetilde G_n&=L_n^TG_n(Q)L_n,& H_n&=L_n^TG_n^{\rm ref}L_n,\nonumber\\
\widetilde C_n&=L_n^TC_n,&\widetilde D_n&=L_n^TD_n.
\end{align}
Writing $A_n=H_n^{-1/2}\widetilde G_nH_n^{-1/2}$, the inverse used in the sewing contraction is
\begin{align}
\mathcal I_n=H_n^{-1/2}A_n(A_n^2+10^{-4}\mathbf1)^{-1}H_n^{-1/2}.
\label{eq:app_compression_soft_inverse}
\end{align}
The contraction $\widetilde D_n^T\mathcal I_n\widetilde C_n$ is projected onto the prescribed external singlet component and summed with channel powers $x^{\alpha_\lambda+n}$. The same lift, inverse regularization, Ward data and channel exponents define the finite reference. Primary vectors are normalized by $\bigl(t_\lambda/\sum_v\|\psi_{\lambda,v}(Q)\|^2\bigr)^{1/2}$, where $t_\lambda$ is the reference primary trace; raw norms at most $10^{-16}$ are inadmissible.

The regularization has a simple stability bound. For Hermitian $G,H$, let $\mathcal R_\tau(G)=G(G^2+\tau^2\mathbf1)^{-1}$. Then
\begin{align}
\|\mathcal R_\tau(G)\|_{\rm op}&\leq(2\tau)^{-1},&
\|\mathcal R_\tau(G)-\mathcal R_\tau(H)\|_{\rm op}&\leq\tau^{-2}\|G-H\|_{\rm op}.
\label{eq:fg_stable_resolvent}
\end{align}
We use $\tau=10^{-2}$ in the reference-whitened Gram matrix. The fixed reference metric also gives covariance under changes $L_n\mapsto L_nR$ by invertible matrices $R$. We check nonorthogonal basis changes, rotations within each retained polarization, and the complex-bilinear sewing convention independently.

\paragraph{Evaluation and physical accuracy.}

The $161$ cross ratios in $[0.1,0.9]$ are partitioned into five index blocks: $B_0=0,\ldots,32$, $B_1=33,\ldots,63$, $B_2=64,\ldots,96$, $B_3=97,\ldots,128$, and $B_4=129,\ldots,160$. Fitting uses $B_0\cup B_1$, checkpoint selection uses $B_3$, and evaluation uses $B_2\cup B_4$. Each subset has its own reference denominator in the relative error \eqref{eq:compression_observable_objective}.

Structural checks include sea and particle projector distances in Frobenius and operator norms, variations of the retained subspaces, compressed null-vector overlaps, and mixed-mode current commutator defects. Relative commutator defects exclude reference denominators below $10^{-12}$, with exclusions recorded separately. The polarized Schwinger trace is compared against $-Kp\,\operatorname{tr}(t^at^b)$ for the relevant mode index $p$. The numerical excitation cutoff is applied separately from the isometric one-particle embedding.

\begin{table}[t]
\centering\small
\begin{tabular}{lrrr}
\toprule
condition & reference/KZ & direct/KZ & neural/KZ\\
\midrule
$SU(2)_3$ $(6,6)$ & 0.1627 & 0.1765 & 0.1896\\
$SU(2)_3$ $(8,6)$ & 0.1627 & 0.2206 & 0.2464\\
$SU(2)_3$ $(10,6)$ & 0.1627 & 0.2577 & 0.2519\\
$SU(3)_2$ $(6,6)$ & 0.2953 & 0.1945 & 0.2050\\
$SU(3)_2$ $(8,6)$ & 0.2953 & 0.2599 & 0.2650\\
\bottomrule
\end{tabular}
\caption{Relative RMS errors against independent KZ transport on the final evaluation nodes. Direct and neural errors are averaged over five matched initializations. The reference column gives the physical error of the finite sewing calculation used in the compression objective.}
\label{tab:s5c2_physical_errors}
\end{table}

\paragraph{Projection of physical reference states.}

The complementary reference-state calculation projects the physical Fock descendants directly into a retained particle--hole excitation space. With reference states arranged as rows of $S_n$ and $G_n=S_nS_n^T$, define $u_n=S_n^TG_n^+C_n$ and $v_n=S_n^TG_n^+D_n$. A contraction $T_U$ on the excitation space gives the projected sewing coefficient $\phi_n[(T_Uv_n)^T(T_Uu_n)]$, preserving the physical reference null relations. The grade-two calculation retains $360$ of $1107$ excitation states, with six holes and six particles. The three-start optimization uses a flavor-preserving radial contraction family; unrestricted projection onto the Ward ket-support supplies a fixed-coupling comparison at ranks at most four or nine.

\paragraph{Fusion-data partitions and training.}

Each fusion-data group contains the ordered triples belonging to one unordered multiset of simple-current orbit labels. The three-orbit and five-orbit tasks contain $10$ and $35$ groups, respectively. Their calibration pools reserve one and seven groups. The first is the all-distinct multiset; the second is selected by a fixed permutation with seed $440211+5$. The remaining groups are divided into fitting, selection and test sets of sizes $(5,2,2)$ and $(16,5,7)$.

Thirty partitions are generated using \texttt{default\_rng}$(52100+s)$, $s=0,\ldots,29$, keeping every permutation of a multiset in one group. Labeled source fitting and selection exclude the true counterparts of target selection, calibration and test groups. Source-teacher predictions supply the target fitting and selection values. Teacher and transferred-predictor initialization seeds are $710000+100s+i$ and $910000+100s+i$, with $i=0,\ldots,4$. Both use Adam for $500$ epochs at learning rate $0.02$, with checkpoint selection every $25$ epochs.

Primary features are centered and scaled using distinct primaries in the fitting groups; scales below $10^{-12}$ are replaced by one. The primary encoder precedes orbit averaging. If $\mathcal G$ is the collection of fitting groups and group $g$ contains $n_g$ ordered triples, each triple has weight $1/(|\mathcal G|n_g)$, giving equal weight to each orbit multiset.

\paragraph{Baselines, queries and uncertainty.}

The source kernel baseline averages $32$ fixed cosine random features over each orbit, using seed $4152026$, and applies the same symmetric input map as the neural predictor. Gaussian-kernel ridge regression uses the median positive mean-square fitting-feature distance as its scale and ridge parameter $0.01$. The weighted-median baseline minimizes the source fitting absolute loss. Neither baseline receives target coefficients.

Target-query budgets are $0,1,2,3$ for the three-orbit task and $0,1,2,4,7$ for the five-orbit task. Queries are chosen from the fixed calibration pool by source-information-gain or random selection. The public orbit data already determine a unique correspondence, so the information-gain policy follows its fixed tie rule. Repeated requests and symmetry-equivalent entries are counted, and queried groups remain in evaluation.

The primary MAE gives equal weight to held-out orbit multisets. Under equal ordered-triple weighting, aligned-neural and kernel MAE are $0.7622$ and $0.5974$ in the three-orbit task, and $0.4446$ and $0.4932$ in the five-orbit task. Uncertainty is evaluated by averaging initializations and relabelings within each partition, followed by $10000$ paired bootstrap resamples of the thirty partitions (seed $609401$) and a simultaneous maximum-deviation bound for the prescribed contrasts.

\paragraph{Reproducibility and computational costs.}

The numerical archive retains model parameters, compression frames, optimizer states, partitions, query sequences, predictions and evaluation results with file hashes. Separate numerical implementations reconstruct the neural outputs, frames, descendant contractions and error measures. Compression timings include setup, parameter selection, optimization, evaluation and reconstruction, with wall and CPU times and artifact sizes recorded. The fusion experiments retain worker wall times; their comparisons concern predictive accuracy and target-observation budgets.

\section{Charged angular observables and controlled limits}
\label{app:lr_controlled_preparations}

\subsection{Level--rank pairing of the charged sectors}
\label{app:charged_level_rank_pairing}

The charged four-point functions have external sectors $(\frac12,\frac12,1,1)$ for $SU(2)_3$ and $(\mathbf3,\mathbf3,\overline{\mathbf3},\overline{\mathbf3})$ for $SU(3)_2$. We denote these theories by $A$ and $B$, with intermediate channels ordered as $(j=1,j=0)$ and $(\overline{\mathbf3},\mathbf6)$. In the corresponding $x=0$ eigenbases, their KZ connections are $\mathscr A_X(x)=R_0^X/x+R_1^X/(x-1)$, with residues
\begin{align}
R_0^A&=\begin{pmatrix}1/10&0\\0&-3/10\end{pmatrix},
&R_1^A&=\begin{pmatrix}-1/5&1/5\\2/5&0\end{pmatrix},\nonumber\\
R_0^B&=\begin{pmatrix}-4/15&0\\0&2/15\end{pmatrix},
&R_1^B&=\begin{pmatrix}-2/15&2/5\\1/5&-1/3\end{pmatrix}.
\label{eq:architecture_six_channel_residue_matrices}
\end{align}
Both theories have $k+h^\vee=5$. The bilinear level--rank pairing is a constant nondegenerate matrix $P$ satisfying
\begin{align}
(R_0^A)^TP+PR_0^B&=-\frac16P,\nonumber\\
(R_1^A)^TP+PR_1^B&=-\frac13P.
\label{eq:app_charged_pairing_conditions}
\end{align}
The $x=0$ equation permits only the paired-channel eigenvalue combinations and makes $P$ diagonal. The $x=1$ equation requires its diagonal entries to sum to zero. Hence the pairing is unique up to an overall scalar; in the displayed invariant bases we take
\begin{align}
P=\begin{pmatrix}1&0\\0&-1\end{pmatrix}.
\label{eq:app_charged_pairing_matrix}
\end{align}

On a fixed branch, let $f(x)=x^{1/6}(1-x)^{1/3}$. The residue conditions imply
\begin{align}
\frac{d}{dx}\bigl(f(x)P\bigr)+\mathscr A_A(x)^Tf(x)P+f(x)P\mathscr A_B(x)=0.
\label{eq:app_charged_pairing_horizontality}
\end{align}
Thus $f(x)P$ pairs solutions of the two KZ systems horizontally. The transfer calculation uses the reference pairing $P_0=f(x_0)P$ at $x_0=0.2$. Its overall scalar is fixed by the chosen external invariant tensors and Ward normalizations; the residue equations determine the relative channel coefficients and cross-ratio dependence.

\subsection{Finite-neural realization and moment estimates}

\paragraph{Physical descendant polynomials.}

Fix a descendant cutoff $D$ and a finite collection of radially ordered angular modes. At grade $n$, use ordered products of negative affine current modes, with total mode number $n$, acting on the primary multiplet. These form a Poincar\'e--Birkhoff--Witt (PBW) spanning set before null relations are removed. Let $S_{\lambda n}$ contain their coefficient rows in the six-flavor CAR excitation space, using the real coefficient bases of the descendant calculation, and set $G_{\lambda n}=S_{\lambda n}S_{\lambda n}^T$. The specified Ward couplings determine the Riesz vectors in Section~\ref{sec:duality_preparation_learning} and their charged polynomials.

The reference null space is $N_{\lambda n}=\ker G_{\lambda n}$. Since $S_{\lambda n}^TN_{\lambda n}=0$, substitution of the actual neural angular modes by the linear map $Q_{L,\theta}$ preserves these relations: $Q_{L,\theta}S_{\lambda n}^TN_{\lambda n}=0$. The charged polynomials therefore descend from physical reference states.

At fixed grade, conditional Berezin integration gives a finite polynomial $P_D(\Gamma_{L,\theta})$ in the empirical angular-mode pairings. Let $\mu_\theta=\mathbb E\Gamma_{L,\theta}$. For a fixed regulator, the pairings converge to $\mu_\theta$ in $L^2$ as $L\to\infty$. The prescribed radial normalization gives $\mu_\theta\to I$ under regulator removal, and hence
\begin{align}
\lim_{\theta\to\mathrm{CAR}}\lim_{L\to\infty}\mathbb E P_D(\Gamma_{L,\theta})=P_D(I).
\label{eq:app_charged_ordered_limit}
\end{align}
Here $\theta\to\mathrm{CAR}$ denotes the fixed-mode limit of the mean pairing. The value $P_D(I)$ is the physical grade-$D$ sewing truncation. With compatible affine-primary Ward couplings and intertwiners, the Sugawara identity identifies its coefficients with the corresponding KZ Frobenius series \cite{Knizhnik:1984nr,TsuchiyaKanie1988,TUY1989}.

\paragraph{Width and regulator estimates.}

Each empirical mode pairing has the form $\Gamma_{L,j}=L^{-1}\sum_{a=1}^Lg_j(W_a)$. For the radial density $Z_\theta^{-1}p^{2\alpha-1}e^{-\epsilon p}$, outer modes $n\geq1$, and inner harmonics $1-k\leq0$, the elementary feature, up to its fixed normalization, is
\begin{align}
g_{n,k}(p,\phi)=Z_\theta p^{1-2\alpha}J_n(2pR)J_{1-k}(2pS)e^{i(k-n)\phi}.
\label{eq:app_mode_feature}
\end{align}
For $0<\alpha<1$, fixed positive radii and fixed mode indices, its joint moments satisfy
\begin{align}
\mathbb E\prod_{j=1}^r|g_{i_j}|\leq C_r Z_\theta^{r-1},\qquad r\geq2.
\label{eq:app_feature_moment_bound}
\end{align}
The fixed-order Bessel estimate $|J_m(p)|=O(p^{-1/2})$ bounds the radial integrand at infinity by a constant times $p^{-1-2\alpha(r-1)}$ after removing $Z_\theta^{r-1}$. At the origin, integrability follows from $n+|1-k|\geq1$.

A product of $r$ empirical pairings is summed over set partitions $\pi$ of its feature slots. A partition with $|\pi|$ blocks carries the factor $L^{(|\pi|)}/L^r$, with one joint feature moment per block. The bound \eqref{eq:app_feature_moment_bound} makes each collision contribution $O((Z_\theta/L)^{r-|\pi|})$. Applying the same argument to conjugate products gives $L^2$ convergence whenever $Z_\theta/L\to0$ and the first mode moments converge to CAR. For the untruncated exponential radial law, $Z_\theta\asymp\epsilon^{-2\alpha}$, so a sufficient joint limit is
\begin{align}
L\epsilon^{2\alpha}\longrightarrow\infty,\qquad 0<\alpha<1.
\label{eq:app_joint_width_regulator}
\end{align}
The estimates apply at fixed descendant grade and fixed mode bank.

\paragraph{Exact repeated-frequency evaluation.}

The ensemble calculations retain all coincidences among sampled frequencies. Let $\mathbf n$ be a nonzero monomial multi-index, choose $i$ with $n_i>0$, and write $\mu_{\mathbf b}=\mathbb E\prod_jg_j^{b_j}$. Distinguishing one feature slot gives the recursion
\begin{align}
M_{\mathbf n}(L)=\sum_{\substack{0\leq\mathbf b\leq\mathbf n\\b_i\geq1}}
\binom{\mathbf n-e_i}{\mathbf b-e_i}\mu_{\mathbf b}
L^{1-|\mathbf b|}\left(1-\frac1L\right)^{|\mathbf n|-|\mathbf b|}
M_{\mathbf n-\mathbf b}(L-1).
\label{eq:app_shared_feature_recursion}
\end{align}
The initial conditions are $M_{\mathbf0}(L)=1$ and $M_{\mathbf n}(0)=0$ for $\mathbf n\ne\mathbf0$. Conjugate modes have reversed angular harmonics. An independent multivariate generating-function calculation and direct frequency sampling check the numerical evaluation.

\subsection{Descendant convergence and normalization}

\paragraph{KZ descendant tail.}

The reference conformal blocks satisfy $F'=(R_0/x+R_1/(x-1))F$. On a fixed branch near $x=0$, write $F_\lambda(x)=x^a\sum_{n\geq0}v_nx^n$. The coefficients obey
\begin{align}
((a+n)I-R_0)v_n=-R_1\sum_{j<n}v_j.
\label{eq:app_kz_frobenius}
\end{align}

For the operator-norm estimates we use symmetric channel bases obtained from the rational residue matrices \eqref{eq:architecture_six_channel_residue_matrices} by diagonal similarities $S_A=\operatorname{diag}(1,\sqrt2)$ and $S_B=\operatorname{diag}(1,1/\sqrt2)$, acting as $R_i^X\mapsto S_X^{-1}R_i^XS_X$. In these bases $R_0$ is real symmetric with eigenvalue separation $2/5$, while $\|R_1\|_2=2/5$ for $SU(2)_3$ and $8/15$ for $SU(3)_2$. The pairing remains $S_A^TPS_B=P$. The nonresonant inverse bound therefore gives
\begin{align}
\|v_n\|_2\leq\frac{\beta}{n}\left\|\sum_{j<n}v_j\right\|_2,\qquad
\beta=\begin{cases}2/3,&SU(2)_3,\\8/9,&SU(3)_2.\end{cases}
\label{eq:app_coefficient_bound}
\end{align}
With $b_n=(\beta)_n/n!$, induction yields $\|v_n\|_2\leq\|v_0\|_2b_n\leq\|v_0\|_2$. On a fixed branch with $0<\delta\leq|x|\leq\rho<1$, the descendant tail through grade $D$ satisfies
\begin{align}
\left\|F_\lambda(x)-x^a\sum_{n=0}^Dv_nx^n\right\|_2
\leq |x^a|\|v_0\|_2\sum_{n>D}b_n\rho^n
\leq |x^a|\|v_0\|_2\frac{\rho^{D+1}}{1-\rho}.
\label{eq:app_kz_tail_bound}
\end{align}
The intermediate sum equals $(1-\rho)^{-\beta}I_\rho(D+1,\beta)$, where $I_\rho$ is the regularized incomplete beta function. A sharper estimate replaces $\|v_0\|_2$ by $\|\sum_{n=0}^Dv_n\|_2/\sum_{n=0}^Db_n$.

\paragraph{Normalization error.}

Let $U(x)=F(x)C^{-1}$ and $U_D(x)=F_D(x)C_D^{-1}$, with $C=F(x_0)$ and $C_D=F_D(x_0)$. Suppose $\eta_x\geq\|F(x)-F_D(x)\|_2$, $\eta_0\geq\|C-C_D\|_2$, and $s_D=\sigma_{\min}(C_D)>\eta_0$. Then
\begin{align}
\|U-U_D\|_2\leq
\|C_D^{-1}\|_2\eta_x+
(\|F_D(x)\|_2+\eta_x)\|C_D^{-1}\|_2\frac{\eta_0}{s_D-\eta_0}.
\label{eq:app_normalized_tail_bound}
\end{align}
Together, the mode-moment estimates, descendant tail and normalization bound give locally uniform convergence of the charged angular approximation on compact subsets of $0<|x|<1$, with width and regulator limits taken at fixed mode number and grade before the descendant limit. A compact family of channel coefficients contributes its corresponding uniform coefficient-norm factor. These are absolute matrix-norm estimates.

\begin{figure}[!ht]
\centering
\includegraphics[width=0.7\textwidth]{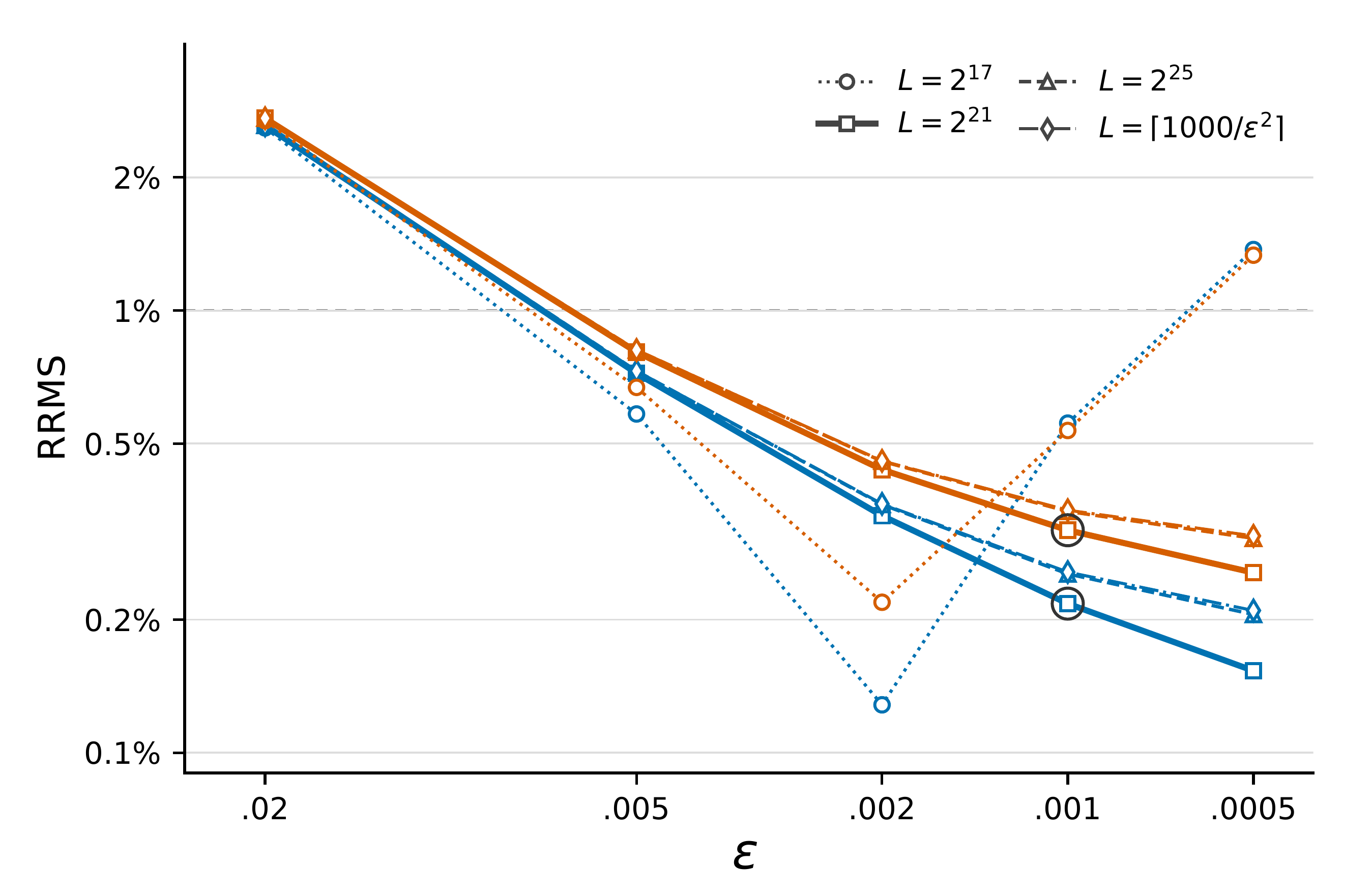}
\caption{Width and regulator dependence of the grade-three charged neural responses for $SU(2)_3$ and $SU(3)_2$. Each point is evaluated against KZ transport on the same $34$ cross ratios using the full finite-width ensemble expectation.}
\label{fig:lr_regulator_width}
\end{figure}

\subsection{Pure charged probes and pointwise readouts}

\paragraph{Pure charged-probe construction.}

For two intermediate channels and grades $0\leq n\leq D$, set $d=2(D+1)$ and choose unit physical reference vectors $e_{\lambda n}$ with duals $e_{\lambda n}^{\vee}$. The fixed source ket and charged bra polynomials are
\begin{align}
\mathcal K&=d^{-1/2}\sum_{\lambda,n}e_{\lambda n},\nonumber\\
\mathcal B_i(v)&=d^{1/2}\sum_{\lambda,n}c^A_{i\lambda n}(C_A^{-1}v)_\lambda e_{\lambda n}^{\vee}.
\label{eq:app_pure_source_bra}
\end{align}
Here $c^A_{i\lambda n}$ are the physical sewing coefficients. The factors $d^{\pm1/2}$ cancel in the matrix element, and $\mathcal K$ has unit reference norm.

In the target sector, choose unit vectors $q_{\lambda n}$ and their duals. The fixed bra and incoming charged kets are
\begin{align}
\mathcal Q&=d^{-1/2}\sum_{\lambda,n}q_{\lambda n}^{\vee},\nonumber\\
\mathcal T_j(w)&=d^{1/2}\sum_{\lambda,n}(w^Tc^B_{\lambda n})(C_B^{-1}P_0^{-1})_{\lambda j}q_{\lambda n}.
\label{eq:app_pure_target_probes}
\end{align}
Coherent linear combinations of these probes implement the control-dependent charged-channel couplings. The variable probes carry the norms determined by their coefficients.

At finite width, the physical vectors are replaced by their charged angular polynomials and contracted using the original Berezin pairing. The channel spectator factors enforce intermediate-channel selection, while angular isotropy removes cross-grade terms from the ensemble expectation. The radial experiment evaluates the inner modes at $S(x)=x$ with weight-$1/2$ dilation. A charge-two monomial at grade $n$ has free energy $n+1$; its radial dependence and the external channel factors produce $x^{\alpha_\lambda+n}$. A seven-rotation average implements the grade-three symmetry projection for sampled frequency banks. All frequency-collision contributions are retained.

\paragraph{Pointwise obstruction.}

Fix finite neural width and compact frequency support $0<p_-<p_+<\infty$. With the other insertion positions fixed, consider a finite polynomial in point-evaluation fields and finitely many derivatives, with coefficients independent of the moving point $z$. Its Berezin expectation and frequency average give a bounded real-analytic function $F(z,\bar z)$ on the plane: the Fourier features and their derivatives are uniformly bounded, and compact frequency support permits termwise analytic expansion.

If $F$ is holomorphic on a nonempty open set, real-analytic continuation gives $\partial_{\bar z}F=0$ throughout the plane. Liouville's theorem then makes $F$ constant. The $R_0$ residues of the two KZ systems are invertible, so their only constant solution is zero. The stated pointwise polynomial class therefore cannot give an exact nonconstant chiral block at finite regulator. The charged angular observables use the mode projections and controlled limits developed above.

\bibliographystyle{JHEP-compat}
\bibliography{bibliography}

\end{document}

%% file: figures/finite_width/tt_heatmap_scale.tex
\begingroup\begin{minipage}{0.8\linewidth}\centering
\includegraphics[width=\linewidth,height=2mm]{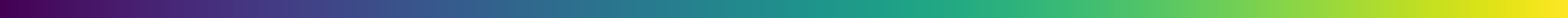}\par
\setlength{\unitlength}{0.01\linewidth}\begin{picture}(100,4)
\put(0,1){\makebox(0,0){\scriptsize$-1$}}
\put(20,1){\makebox(0,0){\scriptsize$15.2$}}
\put(40,1){\makebox(0,0){\scriptsize$31.4$}}
\put(60,1){\makebox(0,0){\scriptsize$47.6$}}
\put(80,1){\makebox(0,0){\scriptsize$63.8$}}
\put(100,1){\makebox(0,0){\scriptsize$80$}}
\end{picture}\par\vspace{1mm}
{\small$\log_{10}E^{TT}_{X,\eta}$}\end{minipage}\endgroup

%% file: figures/finite_width/ttt_su2_k3_scale.tex
\begingroup\begin{minipage}{0.8\linewidth}\centering
\includegraphics[width=\linewidth,height=2mm]{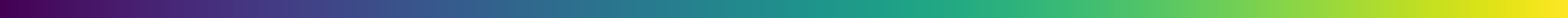}\par
\setlength{\unitlength}{0.01\linewidth}\begin{picture}(100,4)
\put(0,1){\makebox(0,0){\scriptsize$-1$}}
\put(20,1){\makebox(0,0){\scriptsize$22.2$}}
\put(40,1){\makebox(0,0){\scriptsize$45.4$}}
\put(60,1){\makebox(0,0){\scriptsize$68.6$}}
\put(80,1){\makebox(0,0){\scriptsize$91.8$}}
\put(100,1){\makebox(0,0){\scriptsize$115$}}
\end{picture}\par\vspace{1mm}
{\small$\log_{10}E^{TTT}_{X,\eta}$}\end{minipage}\endgroup

%% file: figures/finite_width/tt_su2_k4_scale.tex
\begingroup\begin{minipage}{0.8\linewidth}\centering
\includegraphics[width=\linewidth,height=2mm]{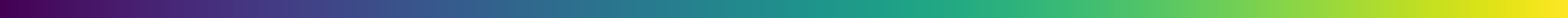}\par
\setlength{\unitlength}{0.01\linewidth}\begin{picture}(100,4)
\put(0,1){\makebox(0,0){\scriptsize$-1$}}
\put(20,1){\makebox(0,0){\scriptsize$15.4$}}
\put(40,1){\makebox(0,0){\scriptsize$31.8$}}
\put(60,1){\makebox(0,0){\scriptsize$48.2$}}
\put(80,1){\makebox(0,0){\scriptsize$64.6$}}
\put(100,1){\makebox(0,0){\scriptsize$81$}}
\end{picture}\par\vspace{1mm}
{\small$\log_{10}E^{TT}_{X,\eta}$}\end{minipage}\endgroup

%% file: figures/finite_width/tt_su3_k4_scale.tex
\begingroup\begin{minipage}{0.8\linewidth}\centering
\includegraphics[width=\linewidth,height=2mm]{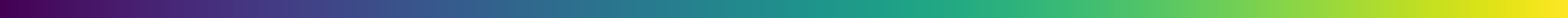}\par
\setlength{\unitlength}{0.01\linewidth}\begin{picture}(100,4)
\put(0,1){\makebox(0,0){\scriptsize$-1$}}
\put(20,1){\makebox(0,0){\scriptsize$15.4$}}
\put(40,1){\makebox(0,0){\scriptsize$31.8$}}
\put(60,1){\makebox(0,0){\scriptsize$48.2$}}
\put(80,1){\makebox(0,0){\scriptsize$64.6$}}
\put(100,1){\makebox(0,0){\scriptsize$81$}}
\end{picture}\par\vspace{1mm}
{\small$\log_{10}E^{TT}_{X,\eta}$}\end{minipage}\endgroup